\documentclass[11pt]{article}
\usepackage[a4paper,left=2.2cm,right=2.2cm,top=2cm,bottom=2cm]{geometry}

\pdfoutput=1
\usepackage[english]{babel}
\usepackage{graphicx}
\usepackage{setspace} 
\usepackage{rotating}
\usepackage{colortbl}
\usepackage{multirow}
\usepackage{booktabs} 
\usepackage{longtable}
\usepackage{pdflscape}   
\usepackage{calc}
\usepackage{float}
\usepackage{indentfirst}
\usepackage{epstopdf}
\usepackage[toc,page]{appendix}
\usepackage{amsmath,amsfonts,bm,amssymb}
\usepackage{amsthm}
\usepackage{latexsym}
\usepackage{soul}
\usepackage[flushleft]{threeparttable}

\usepackage[authoryear]{natbib}
\bibpunct{(}{)}{;}{}{,}{,}

\makeatletter

\definecolor{quotemark}{gray}{0.7}
\makeatletter
\def\fquote{%
    \@ifnextchar[{\fquote@i}{\fquote@i[]}
           }%
\def\fquote@i[#1]{%
    \def\tempa{#1}%
    \@ifnextchar[{\fquote@ii}{\fquote@ii[]}
                 }%
\def\fquote@ii[#1]{%
    \def\tempb{#1}%
    \@ifnextchar[{\fquote@iii}{\fquote@iii[]}
                      }%
\def\fquote@iii[#1]{%
    \def\tempc{#1}%
    \vspace{1em}%
    \noindent%
    \begin{list}{}{%
         \setlength{\leftmargin}{0.1\textwidth}%
         \setlength{\rightmargin}{0.1\textwidth}%
                  }%
         \item[]%
         \begin{picture}(0,0)%
         \put(-15,-5){\makebox(0,0){\scalebox{3}{\textcolor{quotemark}{``}}}}%
         \end{picture}%
         \begingroup\itshape}%

 \def\endfquote{%
 \endgroup\par%
 \makebox[0pt][l]{%
 \hspace{0.8\textwidth}%
 \begin{picture}(0,0)(0,0)%
 \put(15,15){\makebox(0,0){%
 \scalebox{3}{\color{quotemark}''}}}%
 \end{picture}}%
 \ifx\tempa\empty%
 \else%
    \ifx\tempc\empty%
       \hfill\rule{100pt}{0.5pt}\\\mbox{}\hfill\tempa,\ \emph{\tempb}%
   \else%
       \hfill\rule{100pt}{0.5pt}\\\mbox{}\hfill\tempa,\ \emph{\tempb},\ \tempc%
   \fi\fi\par%
   \vspace{0.5em}%
 \end{list}%
 }%
 \makeatother

\title{Deep Learning in Science \footnote{The research leading to the results of this paper has received financial support from the CNRS through the MITI interdisciplinary programs [reference: Artificial intelligence in the science system (ARISE)] and the French National Research Agency [reference: DInnAMICS -ANR-18-CE26-0017-01].}} 
\author{Stefano Bianchini, Moritz M\"uller, and Pierre Pelletier \footnote{Email: \texttt{s.bianchini@unistra.fr} ; \texttt{mueller@unistra.fr} ; \texttt{p.pelletier@unistra.fr}} \\ \\ \textit{BETA - Universit\'{e} de Strasbourg, France}}
\date{} 

\begin{document}
\maketitle

\begin{abstract}
Much of the recent success of Artificial Intelligence (AI) has been spurred on by impressive achievements within a broader family of machine learning methods, commonly referred to as Deep Learning (DL). This paper provides insights on the diffusion and impact of DL in science. Through a Natural Language Processing (NLP) approach on the arXiv.org publication corpus, we delineate the emerging DL technology and identify a list of relevant search terms. These search terms allow us to retrieve DL-related publications from Web of Science across all sciences. Based on that sample, we document the DL diffusion process in the scientific system. We find i) an exponential growth in the adoption of DL as a research tool across all sciences and all over the world, ii) regional differentiation in DL application domains, and iii) a transition from interdisciplinary DL applications to disciplinary research within application domains. In a second step, we investigate how the adoption of DL methods affects scientific development. Therefore, we empirically assess how DL adoption relates to re-combinatorial novelty and scientific impact in the health sciences. We find that DL adoption is negatively correlated with re-combinatorial novelty, but positively correlated with expectation as well as variance of citation performance. Our findings suggest that DL does not (yet?) work as an autopilot to navigate complex knowledge landscapes and overthrow their structure. However, the `DL principle' qualifies for its versatility as the nucleus of a general scientific method that advances science in a measurable way.

\end{abstract}

\begin{fquote}[Nils Nilsson][The Quest for Artificial Intelligence][2009]In today's world, the magic of AI is everywhere -- maybe it's not full AI but there are significant parts.
 \end{fquote}

\section{Introduction}


Most economic and policy analyses on the new wave of technological changes triggered by Artificial Intelligence (AI) and robotization have looked at the effects these technologies can have on economic growth \citep{aghion_2017}, labour market and productivity dynamics \citep{furman_2019, acemoglu_2020, vanroy_2020}, changes in skills \citep{graetz_2018}, and inequality and discrimination \citep{oneil_2016}. The paper at hand deals with the diffusion of Deep Learning (DL) in science and its consequences on scientific development. Our overarching goal is to add some empirical insights into the broader question of how AI shapes the process of knowledge creation.

The theory of re-combinatorial knowledge creation holds that new knowledge predominantly results from the recombination of existing pieces of knowledge
\citep{weitzman_1998, uzzi_2013, wang_2017}. The recombination principle opens the possibility of exponential knowledge growth. Indeed, measurable research outputs such as papers, patents, or innovations have been subject to high enduring growth rates over the last century. Yet, recent empirical evidence suggests that research productivity is ever falling and new ideas are increasingly getting harder to find \citep{bloom_2020}.

Several reasons may account for this decline in research productivity. There may be a `fishing-out' effect whereby the number of useful recombinations is inherently limited and low-hanging fruits are harvested first. Technological opportunities may thus naturally decline, until a new principle or natural phenomenon is discovered, which in turn opens up a plethora of possible combinations. Another possibility is that the potential for useful recombinations is ever increasing, but we have more and more difficulties in realizing the existence of those recombinations due to cognitive and social limitations \citep{fleming_2001}. Using existing knowledge effectively involves the challenge of searching for potentially relevant bits of knowledge, properly assessing their quality, ensuring their relevance for a given context and legitimizing their use in the absence of a universal canon. All of this becomes increasingly difficult within an expanding knowledge landscape that is not only becoming richer, but also more rugged. The `knowledge burden' translates into increased specialization and fragmentation in science where sub-disciplines are flourishing and researchers are working on fewer and fewer topics \citep{jones_2009}. Interdisciplinary research and teamwork can only partially recover the potential of cross-fertilization that is lost. Finally, even within the most narrow specialization, researchers are increasingly confronted with needle-in-the-haystack problems \citep{agrawal_2018}. For example, discoveries in the pharmaceutical sector have become progressively more difficult to achieve due to the proliferation of plausible targets for therapeutic innovation \citep{pammolli_2011}. Similarly, in molecular biology microarrays assess the individual activity of thousands of genes, among which a few of interest must be identified \citep{leung_2015}. 

Expectations are high that AI may resolve at least some of these issues. In particular because of the numerous breakthroughs and rapid improvements in predictions achieved with DL. The principle idea behind DL is that any mapping from input to output, even the most complex, can be arbitrarily approximated through a (deep) chain of very simple mappings that can be fitted with data on exemplary projections. This idea gave rise to a broader family of (data and computing intensive) machine learning methods that are used to discover representations, invariance and laws, unusual and interesting patterns that are somehow hidden in high-dimensional data \citep{hey_2009, brynjolfsson_2014, lecun_2015, agrawal_book_2018}.

The diffusion of AI in general and DL in particular across scientific disciplines has been documented in two previous empirical studies: i) \cite{cockburn_2018} and ii) \cite{klinger_2020}. This paper complements these studies by providing a more fine-grained identification of deep learning research (compared to .i) and evidence on a larger multidisciplinary Web of Science sample (compared to .ii). Our results corroborate the idea that DL serves as a general method of invention across the sciences and around the globe.

Given its diffusion, how does DL affect scientific development? \cite{agrawal_2018}, in a theoretical growth model, propose that AI may alter the knowledge production function in combinatorial-type research problems by affecting either `search' (i.e., knowledge access) or `discovery' (i.e., combining existing knowledge to produce new knowledge). AI in search makes potentially relevant existing knowledge available to the researcher. AI in discovery helps identifying valuable combinations among the available knowledge elements. In a needle-in-a-haystack problem, search would arrange the haystack and discovery would then find the needle. DL search can enhance haystack quality by yielding more, and probably more relevant, components; whereas DL discovery can increase the chances and speed of finding needles. The distinction between search and discovery is certainly relevant and fruitful. Yet, it tells us little about the direction of knowledge development, because it only deals with one body (or one haystack to stay in the picture) of pre-existing knowledge elements. However, knowledge explosion has two sides: increasing knowledge within each domain (larger haystacks) and increasing number of domains (more haystacks). A fundamental question is therefore whether DL deals with knowledge explosion \textit{within} a domain, or facilitates knowledge creation \textit{across} domains. 

There are many examples in science where DL search and discovery remain within the boundaries of established research areas -- e.g., protein-protein interactions (PPIs), nanoscale material properties, or power grid energy supply -- augmenting human scientific reasoning and finding unusual and interesting patterns in vast datasets. However, DL can also facilitate cross-fertilization across topics or sub-disciplines. Neural models for information retrieval (IR), for example, can establish connections that go beyond word similarities, but act on similarities of concepts, lexicons, semantic relations and ontologies \citep{mitra_2018}. Cross domain recommender system can assist target domain recommendation with the knowledge learned from other domains \citep{zhang_2019}. Also, DL is well suited to transferring methods that perform well in constrained, well-structured problem spaces (such as game playing or image analysis) to noisy, flawed and partially observed scientific problems. This is typically achieved by integrating previously unrelated data -- not only in the form of numerical measurements but also unstructured heterogeneous information such as text or images -- from different realms, and perform analysis on them \citep{zhang_2018}.

This leads us to investigate empirically how DL contributes to science in terms of \textit{re-combinatorial novelty} and \textit{impact}. Our analysis here is confined to the \textit{health sciences}. Although much of the discussion around DL is on whether DL qualifies as GPT and its potential to leverage re-combinatorial knowledge creation \citep{agrawal_2018, cockburn_2018, hain_2020, klinger_2020}, our paper is, to the best of our knowledge, the first to empirically assess the effects of DL as a research tool for knowledge creation.\footnote{In a similar vein, \cite{furman_2020} investigate the impacts of the Microsoft Kinect gaming system (fuelled by AI pattern recognition software) on the rate and type of knowledge production in the domains of computer science and electrical and electronics engineering. Kinect automatizes several tasks required to track, collect, and analyze complex 3D motion data in real-time; as such, it can influence knowledge workers' behaviour. The study shows that AI research technology in knowledge production leads to an increase in research output, an increase in research diversity, and a shift in research trajectories.} 

In this study, the concept of re-combinatorial novelty refers to novel re-combinations across domains, as proxied by scientific journals, whereas the concept of impact refers to the the relative importance of a work in the scientific community, as proxied by citation indices. Overall, we find that DL adoption is negatively associated with re-combinatorial novelty, yet shows the `high risk/high gain' profile of breakthrough research, reflected by a higher variance in citation performance. Our results suggest that researchers are using DL as a research tool primarily to cope with the explosion of knowledge within domains rather than across domains. Thus, DL seems to be currently deepening existing knowledge structures rather than overthrowing them. 

The remainder of this paper is structured as follows. Section \ref{sec:DeepLearning} retraces the evolution of deep learning and its alleged effects on the process of knowledge creation. Section \ref{sec:DLIdentification} provides the method for identifying search terms through Natural Language Processing (NLP), our DL search terms, and sample construction. Section \ref{sec:DiffusionInScience} documents some aspects of the diffusion of DL in science. In Section \ref{sec:DLHealth} we present the analysis on the contribution of DL to the health sciences. Section \ref{sec:concl} provides a discussion and concludes by indicating some areas for policy considerations.

\section{What is deep learning?}
\label{sec:DeepLearning}

Artificial Intelligence is at the heart of the current technological paradigm. This paradigm shares several similarities, in scale and scope, with previous technological revolutions that have shaped and fueled long-term cycles of economic growth and structural change. The term `Artificial Intelligence' was coined by the computer scientist John McCarthy in 1955 in the proposal for the Dartmouth Summer Research Project on Artificial Intelligence, which took place in 1956. This workshop was a seminal event for artificial intelligence as a field, whose objective would have been to ``\textit{[m]ake machines use language, form abstractions and concepts, solve kinds of problems now reserved for humans, and improve themselves}'' \citep{mccarthy_1955}. Since its inception, AI has suffered from shifting definitions, although most definitions have revolved around the simulation of intelligent behavior by machines, whereby intelligence generally implies the ability to perform complex tasks in the real-world environment and learn from experience.\footnote{Some definitions were more oriented towards describing the operational characteristics of an intelligent machine, while others focused on the objectives of AI research. We witness a collective effort to establish definitions that are understandable, technically accurate, technology-neutral and applicable to short- and long-term horizons. For instance, the European Commission recently refers to AI as ``\textit{[m]achines or agents that are capable of observing their environment, learning, and based on the knowledge and experience gained, taking intelligent action or proposing decisions}'' \cite[p.19]{annoni_2018}. According to the OECD, ``\textit{An AI system is a machine-based system that can, for a given set of human-defined objectives, make predictions, recommendations or decisions influencing real or virtual environments}'' \cite[p.23]{oecd_2019}. WIPO defines AI systems ``\textit{[a]s learning systems, that is, machines that can become better at a task typically performed by humans with limited or no human intervention}'' \cite[p.19]{wipo_2019}.} Often the terms machine learning, deep learning, and artificial intelligence are used interchangeably. This Section aims to briefly retrace the history of AI research, emphasizing the different approaches to machine intelligence and especially the approach based on deep neural networks.

\subsection{Approaches to machine intelligence}
\label{sec:machine_intell}

In the early days, AI has tackled and (often) solved problems that could be described by a list of formal mathematical rules. Problems of this kind are intellectually difficult for humans but relatively straightforward for computers because real-world knowledge can be hard-coded into formal languages and logical inference rules can be used to achieve solutions. This approach to machine intelligence is commonly refer to as `knowledge-based' approach. The typical architecture of a knowledge-based system includes a knowledge base and an inference engine (i.e., inference rules): the knowledge base contains a collection of real-world information and the inference engine enables the machine to deduce insights from the information stored in the knowledge base. This approach has been the dominant practice for the first decades. Applications, such as expert systems, were introduced in the 1970s and were aimed at simulating the judgement and behaviour of a human being who has knowledge and experience in a particular field. These applications proved to be very effective for solving certain types of problems (e.g., recommend antibiotics and dosages to clinicians based on the presence of bacteria and patient symptoms) but terribly scarce for problems that require a great deal of subjective and intuitive knowledge of the world, as well as the perceptual capabilities of the environment (e.g., performing sensory tasks such as recognizing a face in the midst of a large crowd). Much of these problems are indeed easy for humans to perform, but hard for humans to articulate formally and through mathematics \citep{nilsson_2009}.

During the same period, an alternative approach to machine intelligence began to take hold in the scientific community. This approach soon became known as `machine learning', and consisted in designing intelligent systems with the ability to acquire their own knowledge by extracting patterns from raw data. In other words, machine learning methods construct hypotheses directly from the data through inductive inference. Here is a classic example: if a large data set contains several instances of white swans and no instances of swans of other colors, a machine learning algorithm may infer that `all swans are white'. Inductive inferences consists of hypotheses which are always subject to falsification by additional data; for instance, there may still be an undiscovered island of black swans. Machine learning soon proved to be a valid alternative to knowledge-based systems. Machines could tackle problems involving real-world knowledge and reach certain human abilities, such as recognizing simple objects. From the 1980s, machine learning became one the most prominent branches of AI. Yet, some problems remained. Suppose the goal of a machine learning algorithm is to recognize a face in a picture, then the machine may use the presence of a nose as a feature -- i.e., piece of relevant information of the real-world to extract for that task. However, describing exactly what a nose is in terms of pixel composition can be difficult since there are countless different shapes, shadows can modify and obscure part of the nose, and the viewing angle can further change the shape. All these attributes are known as factors of variations, essentially constructs in the human mind that can be thought of as high-level abstractions that help us make sense of the rich variability of the observed data. Traditional machine learning methods encountered enormous difficulties in extracting these high-level abstract features from raw data \citep{nilsson_2009, goodfellow_2016}.

The `deep learning' approach to machine intelligence turned out to be a good solution to this problem. A DL system learns from experience and understands the world in terms of a hierarchy of concepts, with each concept defined through its relation to simpler concepts \citep{schmidhuber_2015, lecun_2015, goodfellow_2016}. The DL approach has two important advantages. First, as with simple machine learning algorithms, the machine collects knowledge from past experience, hence the human does not need to embed in the machine all the formal knowledge necessary to attain a given goal. Second, the level of complexity and abstraction of concepts is no longer a barrier, since the machine can reconstruct and aggregate them on top of each other. Returning to the previous example, a DL system can represent the concept of a nose by combining simpler concepts such as angles and contours that are then aggregated in terms of edges. This hierarchy of concepts makes the learning process a process that can be thought of as being structured into multiple layers, hence the term `deep'. 

The function mapping from a set of features to an output can be often very complicated. Deep learning breaks down the complex desired mapping into a series of simple nested mappings, each described by a different layer of the model. The variables that we observe are presented at the input or visible layer. Then a series of hidden layers extracts increasingly abstract features from the data. The term `hidden' represents the idea that there is no predetermined structure but it is the model itself that determines which concepts are useful to explain the relationships observed in the data. The general architecture of a DL system can therefore be thought of as a neural network because nodes in the input, hidden and output layers are vaguely similar to biological neurons, and the connections between these nodes can be thought of as somehow reflecting the connections between neurons \citep{hassabis_2017}. Today there is no consensus on how much depth a learning system requires to be considered as `deep'. However, there is consensus on the fact that DL involves a greater amount of learned functions or concepts compared to traditional machine learning methods. This allows DL to achieve great performance in an incredible variety of tasks.

\subsection{Trends in deep learning research}
\label{sec:trends_dl}

``\textit{Deep learning, as it is primarily used, is essentially a statistical technique for classifying patterns, based on sample data, using neural networks with multiple layers [...] Deep learning is a perfectly fine way of optimizing a complex system for representing a mapping between inputs and outputs, given a sufficiently large data set}'' \cite[p.3-15]{marcus_2018}. 
Although the term `deep learning' is recent, DL has a long and rich history dating back to the 1940s. The field of research has been re-branded many times, reflecting the influence of researchers who have contributed to its development over time and who came from different backgrounds. For the narrative of our study, we find it useful to understand why DL has only begun to spread in recent years.

In their in-depth review on the history of deep learning research, \cite{goodfellow_2016} identify three major waves of developments: (i) cybernetics in the 1940s--1960s has marked important developments in theories of biological learning and the training of simple models with a single neuron; (ii) connectionism in the 1980s--1990s has brought methodological advances that have allowed faster training of neural networks with a few hidden layers; and (iii) the recent wave that started around 2006 (and is still ongoing) during which the appellative `deep learning' was coined.

The first predecessors of DL were simple linear models aimed at emulating computational models of the biological brain. The neuroscience perspective was motivated by the idea that the creation of intelligent machines could be achieved by reverse engineering the computational principles, albeit greatly simplified, behind the biological brain and replicate some of its basic functionalities \citep{nilsson_2009, hassabis_2017}. These models were intended to provide in turn some insights to better understand the brain and the principle of human intelligence. The McCalloch-Pitts neuron is perhaps considered the first linear model of brain function that could classify two categories of input on the basis of a set of weights, although defined by human operators. A few years later, \cite{rosenblatt_1958} proposed the first model, known as perceptron, that could learn weights directly from examples of inputs without human intervention.

Neuroscience has inspired many principles that today form the backbone of DL architectures, such as artificial neural network or theories of mammalian visual system for computer vision. The intuition that many computational units become intelligent only via their interaction with each other is indeed regarded as the dawn of DL systems. However, the limited knowledge of the brain to use it as a guide, soon posed a barrier to further theoretical and practical developments of the field. Several critiques were levelled mainly against the excessive simplification of biological learning, and this approach to artificial intelligence lost its popularity. Although neuroscience is still regarded as an important source of inspiration for DL, a recent bibliometric analysis of the evolution of artificial intelligence research and its related fields from the 1950s to the present suggests that neuroscience is no longer the predominant guide for the field. Modern DL research predominantly refers to many other areas including mathematics, information theory and computer science \citep{frank_2019}.\footnote{There is still a tendency, often reinforced by the media, to perceive deep learning as an attempt to simulate the human brain. Our knowledge of the function of the human brain is yet very limited, and there is broad agreement in the scientific community that deep learning cannot be seen as an accurate model of how the brain actually works.}

In the 1980s a new wave of neural network research emerged in the context of cognitive science with a movement known as connectionism \citep{nilsson_2009}. During the early 1980s, models of symbolic reasoning (knowledge-based approach discussed in Section \ref{sec:machine_intell}) were slowly overtaken by models of cognition that could be anchored in neural implementation. Connectionists shared the idea that a large number of neuron-like processing units can achieve intelligent behaviour when they are intensely networked together, thereby emphasizing the role of hidden layers as a way to increase the complexity of interconnections between units. Great achievements were made during those years, including techniques and models that still play a fundamental role in modern deep learning. Examples include the concept of distributed representation aimed at capturing meaningful `semantic similarity' between data through concepts, the successful use of back-propagation algorithm to train deep neural networks which had previously been insoluble, and long-short term memory networks (LSTMs) for modelling sequences with long-term dependencies \citep{goodfellow_2016}. However, deep networks were too computationally expensive to empower real-world applications with the hardware available at the time, so research on neural networks began again to lose some popularity. The decline was further accentuated by the introduction of other machine learning techniques, in particular kernel machines, which could achieve similar performances in various applications with much lower computational requirements.

In spite of the difficult period, neural network research was kept alive. The Canadian Institute for Advanced Research (CIFAR) via its Neural Computation and Adaptive Perception (NCAP) research programme brought together some leading machine learning groups led by Geoffrey Hinton (University of Toronto), Yoshua Bengio (University of Montreal) and Yann LeCun (New York University). This programme paved the way for the last wave of DL research, which officially began in 2006, with a major breakthrough in the efficiency of neural network training. An important constraint to the training of deep neural networks was traditionally due to a problem known as vanishing gradient, which meant that weights in layers close to the input level were not updated in response to errors calculated on the training data set. Geoffrey Hinton and colleagues introduced a topology of network known as deep belief network that could be efficiently trained using a technique called greedy layer-wise pretraining \citep{hinton_2006}.\footnote{This technique is `layer-wise' as the model is trained one layer at a time, and `greedy' as the training process is divided into a succession of layer-wise training processes. The procedure acts as a shortcut leading to an aggregate of locally optimal solutions, which in turn results in a reasonably good global solution.} This strategy has finally enabled very deep neural networks to be successfully trained and to achieve cutting-edge performance. Since then, researchers have begun to popularize the term `deep learning' and to focus on the theoretical implications of depth in neural networks. Further innovations and important milestones in the field of DL have been achieved over the last decade, including convolutional neural networks (CNNs) and generative adversarial networks (GANs). Today we are still in this third wave of research and deep learning outperforms any other machine learning technique in almost any real-world application.

\subsection{Deep learning in science matters}

Scientific discovery can been seen as the process or product of successful scientific inquiry.\footnote{In the narrowest sense, the term discovery would refer to the alleged `eureka moment' of having new insights, although here we adopt its broadest sense -- i.e., we use the term discovery as a synonym for `successful scientific endeavour' as a whole.} Historically, the process of scientific inquiry has evolved through paradigms, seen as symbolic generalizations, metaphysical commitments, values and exemplars that are shared by a community of scientists and that guide the research of that community \citep{kuhn_1962}. 

For most of human history, scientists have been observing phenomena, postulating laws or principles to generalize the complexity of observations into simpler concepts. The laws of science can be viewed in fact as compressed, elegant mathematical representations that offer insights into the functioning of the universe. Originally there were only experimental and theoretical sciences. \cite{hey_2009} refer to empirical observation and logical (theory) formulation as the first and second scientific paradigm, respectively. Towards the middle of the last century, however, many problems turned out to be too complicated to be solved analytically and researchers had to start simulating. Science entered into a third paradigm, a paradigm characterized by the development of computational models and simulations to understand complex phenomena. Data and information have begun to grow and accumulate on an unprecedented scale, also benefiting from the advent of other technologies such as remote sensing and the Internet of Things. The search over an increasingly vast combinatorial search space have soon become prohibitive for humans. Common to all scientific paradigms is in fact the idea that scientists use existing bits of knowledge to produce new knowledge and this new knowledge becomes then part of the knowledge base from which subsequent discoveries are made. As the volume and the complexity of the knowledge landscape increase, human cognition becomes a major limitation to experiment and re-combine distinct elements of the knowledge stock, understand the landscape and push the knowledge frontier further \citep{fleming_2001, jones_2009}. We are moving towards a fourth scientific paradigm, a paradigm in which scientific exploration is grounded in data-intensive computing with a massive deployment of intelligent machines capable of finding representations, rules and patterns from an ever-increasing volume of structured and unstructured data \citep{hey_2009, king_2009}. Much of this paradigm shift can be attributed to AI systems enhanced by deep learning \citep{brynjolfsson_2014, agrawal_book_2018}. 

DL redefines and enriches the knowledge base by observing the real world through examples, thus affecting both the process of `search' and `discovery' \citep{agrawal_2018}. As for search, DL can support access to knowledge at a time where we are witnessing an explosion of data and information, predicting which pieces of knowledge are most relevant to the researcher. DL-based cross domain recommender systems, for instance, offer high-quality cross domain recommendation by exploiting numeric measurements, images, text and interactions in a unified joint framework. Transfer learning can further improve learning tasks in one domain by using knowledge transferred from other domains, in turn catching the generalizations and differences across different domains. DL is well suited for transfer learning as it learns high-level abstractions that disentangle the variation of different domains \citep{zhang_2019}. 

As for discovery, DL allows a better prediction of which pieces of knowledge can be combined to produce new knowledge, and the value of that knowledge. In other terms, DL allows the researcher to identify valuable combinations in a rugged landscape where knowledge interacts in highly complex ways. Humans can indeed consider a few hypotheses at a time whilst machines can generate and test an almost unlimited number of (more complex) hypotheses, explore unknown experimental landscapes and select which hypotheses are worthy of testing using a sort of economic rationality \citep{nilsson_2009, daugherty_2018}. There are very efficient forms of learning that allow intelligent machines to reduce the uncertainty associated with regions of experiment space that are sparsely populated with results. Deep active learning systems, for instance, dynamically pose queries during the training process with the aim at maximising the information gains in the search space. The machine proposes which regions to navigate on the basis of the amount of new knowledge that is likely to be obtained in a given region, whilst the researcher can further screen the regions according to priorities and insights.\footnote{Active learning is a successful example of the `human-in-the-loop' approach to balance exploration and exploitation in the presence of uncertainty. Human scientists and DL systems have indeed different strengths and weaknesses, and combining forces can create synergies and forms of human-machine collaboration that will eventually produce a better science than can be performed alone. Humans are characterized by creativity, improvisation, dexterity, judging, social and leadership abilities. On the other hand, machines are characterized by speed, accuracy, repetition, prediction capabilities, and scalability. Intelligent machines may augment human scientific reasoning, free up time, creativity and human capital, ``\textit{[l]etting people work more like humans and less like robots}'' \cite[p.20]{daugherty_2018}.} DL may therefore overcome the `knowledge burden' within a scientific domain, but also act as a cross-fertilizer for knowledge recombination across domains. All these properties have allowed DL to qualify as the nucleus of a General-Purpose Invention in the Method of Invention \citep{cockburn_2018, klinger_2020}.\footnote{Other advantages offered by DL as a research tool are worth mentioning. While human scientists get tired, a machine can work without interruption and preserve the same degree of performance on a given collection of tasks. Intelligent machines can also improve data sharing and scientific reproducibility since they can easily record experimental actions, metadata and procedures, and results at no (or very limited) additional cost.}

One important consequence to be gained from thinking of DL as a General-Purpose Invention in the Method of Invention is that its impact is not limited to its ability to reduce the costs of specific scientific activities, but also to enable a new approach to science itself, by altering the scientific paradigm in the domains where the new research tool is deployed \citep{hey_2009, king_2009, cockburn_2018}. Exploring the rise of DL as a research tool and the impact it can entail on scientific development represents the backbone of our study.

\section{Identifying deep learning research}
\label{sec:DLIdentification}

\subsection{Identification strategy}

We are interested in the development of DL in science across disciplines. Our empirical analysis of scientific publications rests on two databases: arXiv.org and Web of Science (WoS). In a first step, we use arXiv.org to develop an appropriate list of \textit{search terms} referring to DL through Natural Language Processing of scientific abstracts from publications in `Computer Science', `Mathematics' and `Statistics' subject areas. In a second step, these search terms are used to query the WoS database and extract our sample of DL papers across all scientific fields.\footnote{WoS is widely used for scientometric analyses, and seems also appropriate in our case. WoS lists all scientific papers published in a defined set of journals and conference proceedings. This shortlist approach necessarily introduces some heterogeneity in the coverage of the different scientific disciplines depending on the type of outlet preferred. Nevertheless, WoS coverage of journals and proceedings seems vast enough to capture science dynamics more generally. We were able to gather detailed information about each publication, including title, keywords, abstract, publication year, journal information, topical information, author and institutional affiliations, as well as cited references.} This second step is straightforward. How to create the list of search terms requires more in-depth discussion.

Reliance on a list of search terms for document retrieval is a common practice in research on emerging technologies or sciences, where no relevant structural information is available (e.g., an a priori appropriate classification). Unfortunately, extant studies do not provide us with an authoritative `ready-to-use' list of search terms. \cite{cockburn_2018} use a list of AI-related search terms to extract publications from WoS, but their definition of AI (also if restricted to what they characterized as `learning' approach) is too broad for our focus on DL. The same applies for \cite{vanroy_2020} who adopt a keyword-based approach to select AI patents. \cite{klinger_2020} explicitly deal with DL publications, but they do not define search terms. Instead, they identify DL papers through topic modeling of abstracts from arXiv.org. This approach essentially results in the relevance of the topic `deep learning' for each article. Hence, it allows them to tag papers as DL papers, but does not result (immediately) in search terms.

Our first step is therefore to create a list of DL search terms.\footnote{Other alternatives would be available. A first alternative could be to start with a set of `core' deep learning papers and include additional papers citing them. Two arguments, however, speak against such an approach. DL core papers should be defined in the first place. Secondly, papers that cite citing papers (the logical second step of this snowball sampling approach) may or may not apply deep learning as research tool, and that would distort the sample for our subsequent analysis by increasing the presence of false positives. The second alternative could be to query WoS simply using the term `deep learning'. Although simple, this approach would lead to the opposite problem, that is to increase the risk of false negatives. Further details about this issue can be found in Section \ref{sec:robcheck} when we discuss robustness checks.} There are by now various methods to create search terms and queries (see, e.g., \citealp{huang_2011, huang_2015}). In general, an additional search term in the list may be useful by adding correctly identified documents to the sample (here documents related to DL), but may also be harmful when adding incorrectly identified documents (here documents not related to DL). In order not to `forget' relevant terms, evolutionary lexical queries start with a set of core terms, retrieve the corresponding documents and then use them to identify further terms, possibly in a loop. Additional bibliometric information such as citations, journals or authors can also be used. However, over-expanding the search query quickly raises the risk of false positives. 

To find the right balance, studies often rely on experts with domain-specific knowledge to establish appropriate search terms. Due to the nature of emerging technologies, however, experts often lack a shared perspective. Thus, the delineation of the sub-domain and the identification of domain-specific terminology intertwine, eventually exacerbating the problem of validity and reliability.

We propose a data-driven approach to delineate the perimeter of the deep learning domain, and to identify recurrent terms in that domain. In a nutshell, our approach consists in training the word embedding model `Word2Vec' \citep{mikolov_2013a} with scientific abstracts from arXiv.org's documents in order to \textit{learn} DL-related terms.  

Roughly speaking, text embedding algorithms such as Word2Vec process text by vectorizing words -- i.e., they project words from a text corpus into a common vector space. The structure of the projected space reflects the semantics of the text, so that the semantically related words tend to cluster together in vector space. We build on this idea. Using a text embedding algorithm, we project terms from scientific abstracts into a vector space. In that space, we identify the word cluster that includes the term `deep learning' in order to obtain other terms that relate semantically and syntactically to the DL domain.\footnote{Word embedding algorithms have recently gained prominence in Natural Language Processing community \citep{li_2018}, but also in scientometrics and bibliometrics. On the one hand, text embedding may enrich the analysis of a given corpus. For example, citation analysis may be enriched through word embedding methods to identify why a citation has been given, the sentiment associated with the citation, and what exactly has been cited \citep{jha_2017}. On the other hand, text embedding is used for the identification of the corpus itself, such as for document retrieval. For example, \cite{dynomant_2019} propose and discuss various word embedding techniques to identify `similar' documents in PubMed database.} 


\subsection{Learning a list of deep learning search terms}

Text analysis is more a practice than a science. Many decisions have to be made, often iteratively depending on what works and what does not: from cleaning text input, hyper-parameter settings, to various post-processing steps. The following Section conveys the main ideas and discusses the choices we have made to produce the list of DL search terms. Appendix \ref{appendix:embedding} provides details.

Our training data consists of scientific abstracts from arXiv.org. Recall from Section \ref{sec:trends_dl} that DL blends statistics and informatics, but develops predominantly within computer sciences. Informatics is a fast-developing field in which conference proceedings are traditionally very important. More recently, however, the rapid dissemination of research is (better) achieved via open access journals and platforms. Of these, arXiv.org is the most prominent. It hosts not only many, but also very recent and highly cited research papers on AI in general and DL in particular. Therefore arXiv.org provides us with a rich corpus for the identification of DL-related terms. We downloaded a total of 197,439 abstracts of papers that fall in the subject areas `Computer Science', `Mathematics' and `Statistics', over the period 1990--2018. The three areas represent roughly 50\% of all arXiv.org documents in 2018, and only 10\% in the early 2000s.

We use the abstracts as input for the word embedding algorithm `Word2Vec' \citep{mikolov_2013a, mikolov_2013b}. The words of a vocabulary of size $V$ are positioned in a $D$-dimensional space, giving rise to $V \times D$ parameters to fit. Hence, each word is represented by a $D$-dimensional continuous vector -- i.e., the word representation. The Word2Vec algorithm fits word representations in such a way that the probability that two words are close together in the corpus increases with the vector product of their word representations. A very intuitive outcome is that words that tend to appear close to each other in the text will have similar vector representations, since word representations only produce a high vector product if their larger values are in the same components. A less intuitive, but more striking and useful outcome is that two words that do not co-occur together but co-occur with the same other words are also close in vector space. Both effects together produce clusters of syntactically and semantically related terms in the projected space.

This is particularly useful in our case. The term `deep learning' is certainly to be included in the list of search terms. Hence, we can identify in the vector space the cluster of terms including the term `deep learning'. Synonyms and other semantically related elements (e.g., `neural network') are likely to be identified. `Neural network\textit{s}', a syntactically related term, may also be found in the same cluster. As terms show up in similar text contexts, they appear in the same cluster in the projection. On the other hand, terms that are not closely related to DL but, say to informatics or other machine learning methods, remain excluded (e.g., `support vector machine'). These terms sometimes also appear in text contexts similar to DL, but even more often in other contexts. Therefore, their word representation will be different and so will their cluster. By looking at the terms in the cluster `deep learning' in the projected space, we make sure we do not miss relevant search terms. In addition, the boundaries of the word cluster provide an indication on how to delineate the DL domain. 

Clustering of syntax-related words is convenient as it reduces the needs for preprocessing. Stemming or lemmatization is no longer necessary because variants of the same word stem are clustered by design. In contrast, preprocessing $n$-grams that refer to idiomatic phrases is essential. Many technical terms are idiomatic phrases that consist out of multiple words. For example, the term `neural network' refers to one specific concept although it consists of two words. The Word2Vec algorithm produces for each word exactly one vector (embedding) and that vector does not vary with the context. Thus, by default, the word `neural' will be one vector and the word `network' another vector. This algorithmic feature is inconsistent with the fact that many words take on different meanings depending on the context. The word `neural', for instance, has a different meaning in brain research (probably referring to neural activity) than AI (probably referring to computational neural architectures). As proposed in \cite{mikolov_2013b}, we take this issue into account during preprocessing by pasting together subsequent words (bi-grams) into single tokens whenever these subsequent words co-occur frequently in our full corpus, as explained in the Appendix \ref{appendix:embedding}. For instance, `neural' and `networks' are pasted into one token `neural\_networks'.\footnote{\cite{camacho_2018} discuss other (often more elaborated) approaches that are explicitly designed to handle this issue. However, pasting $n$-grams during preprocessing, as proposed in \cite{mikolov_2013b} is very simple, intuitive, and turned out to be satisfactory.} Finally, the use of acronyms is a standard practice in the AI community. Several terms in our clusters were acronyms (e.g., ANN). We replaced the most frequent acronyms with their appropriate full names.\footnote{Acronyms are also quite convenient in the learning process as they allow to capture concepts without inflating the size of the vocabulary. For example, without any preprocessing of the corpus, a concept such as `deep convolutional neural network' can be represented by a single token (DCNN). The list that we have developed manually to convert the acronyms to full names is reported in the Appendix \ref{appendix:embedding}.}

The data is then used to train the Word2Vec model in its Skip-Gram with Negative Sampling (SGNS) version, as discussed in \cite{mikolov_2013b}. Fitting the model involves various parameter settings that are described in Appendix \ref{appendix:embedding}. The main outcome of the model is one vector representation for each term in the vocabulary. We identify those terms that appear in the same cluster as `deep\_learning'.\footnote{The results were obtained with $k$-mean clustering. The optimal number of clusters via the `gap statistic'. We also tried different clustering methods and the results were very robust in this respect; apparently because the estimated vector space is clearly structured.}

\begin{table}[t!]
 \centering  
 \caption{Deep learning search terms from word embedding}
	\begin{threeparttable}
  \label{tab:DLTerms} 
\begin{tabular}{@{\extracolsep{5pt}}lrlr}  
\toprule 
$n$-gram & Count & $n$-gram & Count  \\ 
\hline \\[-1.8ex] 
neural network & 402,996 & long short term memory & 3,122 \\ 
neural networks & 173,470 & hidden layers & 2,080 \\ 
artificial neural & 100,749 & restricted boltzmann & 1,635 \\ 
artificial neural network &  99,794 & auto encoder & 1,444 \\ 
deep learning &  24,104 & generative adversarial & 1,242 \\ 
convolutional neural &  20,742 & encoder decoder & 1,198 \\ 
convolutional neural network &  20,595 & adversarial network & 1,192 \\ 
recurrent neural &  14,355 & generative adversarial network & 1,085 \\ 
recurrent neural network &  13,965 & fully convolutional network &  688 \\ 
deep neural &   9,418 & convolutional layers &  568 \\ 
multilayer perceptron &   9,352 & variational autoencoder &  216 \\ 
deep neural network &   9,181 & adversarial attacks &  197 \\ 
hidden layer &   7,810 & adversarial examples &   92 \\ 
deep convolutional &   4,263 & variational autoencoders &   75 \\ 
deep convolutional neural network &   3,384 & adversarial perturbations &   24 \\ 
\bottomrule
\end{tabular}
\begin{tablenotes}
 \footnotesize
 \item {\it Notes:} The count refers to how many times a given term occurs in the Web of Science corpus, as discussed in Section \ref{sec:DiffusionInScience}. Note that a document may include several terms.
\end{tablenotes}
 \end{threeparttable}
\end{table}

The resulting list of potential search terms includes individual words (uni-grams) but also technical terms consisting of multiple words. We decided to retain only those terms consisting of multiple words -- i.e., to remove all uni-grams from the list of search terms -- in order to remain conservative and include only terms that relate unambiguously to DL. Moreover, we retained only the 30 most frequent $n$-grams after having dropped terms that are too generic (e.g., `short\_term' or `supervised\_learning').\footnote{Since there is no established canon, our approach was a fairly exploratory trail-and-error process, aiming to balance false positives and false negatives. Considering 30 terms and removing too general terms, we were able to retrieve 260,459 DL documents from Web of Science (our sample for the analysis). To get an idea of the sensitivity, if limit the list to 30 terms but include generic terms, the query results into 639,317. If we limit the list to 20 terms, we identify 616,349 documents; if we increase the list up to 50 terms, we get 685,631 documents. And 711,905 documents when considering 80 terms. Therefore, as these numbers suggest, we preferred to adopt a conservative strategy and limit the presence of false positive in the sample.} A manual check of various random extractions from WoS confirmed that this choice greatly reduces the presence of false positives in the final sample. The exception being the term `neural\_network' which may in fact refer to a biological neuronal network. We decided to keep that term, however, because the confusion between biological neural networks and artificial neural networks seems to be confined to the field of neuroscience. This issue is therefore marginal and does not affect descriptive statistics across scientific fields, nor any subsequent results on the role of the DL in health sciences (see also robustness checks in Section \ref{sec:robcheck}). The final list of search terms is shown in Table \ref{tab:DLTerms}. A more complete list of terms for all clusters identified through word embedding can be found in the Appendix \ref{appendix:embedding}.\footnote{It is fair to remark that a neural network architecture may not necessarily be deep, although what is meant by depth is still a matter of contention (Section \ref{sec:machine_intell}). This implies that our sample may also include `shallow' networks with only one hidden layer; a potential problem that also characterizes previous research on deep learning \citep{cockburn_2018, klinger_2020}. Aware of this, we made sure that the diffusion patterns presented in Section \ref{sec:DiffusionInScience} are robust with respect to a much more restrictive definition of deep learning -- that is, restricting the sample to documents that contain only the terms with the prefix `deep' (e.g. `deep\_learning', `deep\_neural'). At the same time, a strict separation between deep learning and neural network makes no sense. The two are closely linked because one relies on the other to function. Put simply, without neural networks, there would be no deep learning. We omit any further reference to this issue in what follows.}

\section{Diffusion of deep learning in science}
\label{sec:DiffusionInScience}

This Section documents the diffusion of DL in science across geographies and scientific areas. Our sample includes all publications in the WoS Core Collection that were published between 1990 and 2018, and have at least one of the search terms (Table \ref{tab:DLTerms}) in their title, keywords or abstract. In total, we identify 260,459 DL documents (144,095 articles; 39,925 conference proceedings; 76,439 others).\footnote{To get an idea of sensitivity with respect to the most influential terms, if we consider only the term `deep learning' the query provides us with 15,085 documents. Adding `neural network' to the query pushes the number up to 219,782 documents.}  

\subsection{Deep learning is a global phenomenon}

In this subsection, we provide some insights into the spatial diffusion of DL on a global scale. Spatial diffusion is not the focus of the paper at hand. However, the fact that DL spreads globally and that countries show patterns of specialization in DL research supports the idea that DL is a general and relevant method in science.\footnote{Our spatial observations also open up research questions that go beyond the scope of this paper. Particularly interesting seems to be the issue of competition and cooperation between national scientific systems, as well as alignment and differentiation in DL research which, in turn, can shape the trajectory of the technology. We leave these issues for future research.}

\begin{figure}[h!]
     \caption{Global diffusion of deep learning in science across countries}
   \begin{center}
	  \includegraphics[scale=0.65]{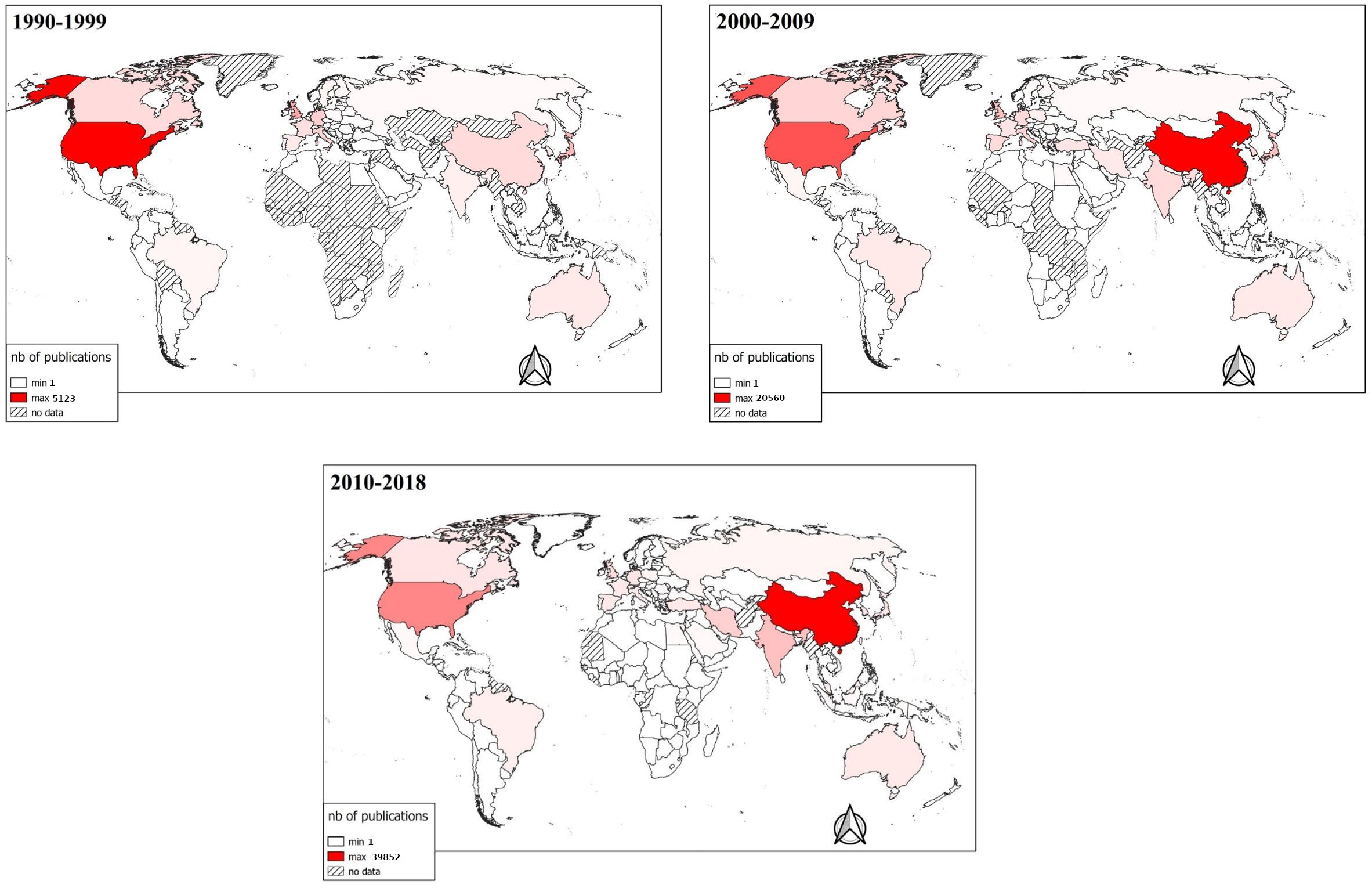}
   \end{center}
  \label{fig:geo1}
\footnotesize
\textit{Notes:} The intensity of colour reflects the country's relative number of DL publications in a given period, with no observed DL publication activity in hatched countries [WoS sample]. 
\end{figure}

Figure \ref{fig:geo1} shows deep learning science dynamics at the country level. A complete table with numbers is provided in the Appendix \ref{appendix:country}. Each DL document is attributed to a given country when at least one author's affiliation is in that country. The upper left panel shows the pattern for the first period, 1990--1999. During that period, most of the documents (about 5,000) were published by scientists in the United States. Publishing activity is relatively low in absolute numbers in the European countries, Australia and China, and negligible or non-existent in most other countries. In the following decade, 2000--2009, China becomes the most prolific country with about 20,000 DL documents. The US ranks second with around 14,000 articles, whereas European countries and Australia grow sufficiently to preserve their relative strength. Interestingly, in an increasing number of countries DL research activity starts. These trends are reinforced in the third, last period, 2010--2018. Compared to the previous decade, China has doubled its DL research output, thus widening the gap with the US and, to a lesser extent, with the EU.\footnote{Two remarks are noteworthy here. First, a document with multiple affiliations in different countries is counted multiple times. We verified that weighted paper counts yield essentially the same patterns. Second, considering the EU as one single player would rank the EU first in the first and second period, and second in the third period.} 

\begin{figure}[h!]   
    \caption{Specialization patterns of world regions in deep learning research} 
  \begin{center}
	  \includegraphics[scale=0.55]{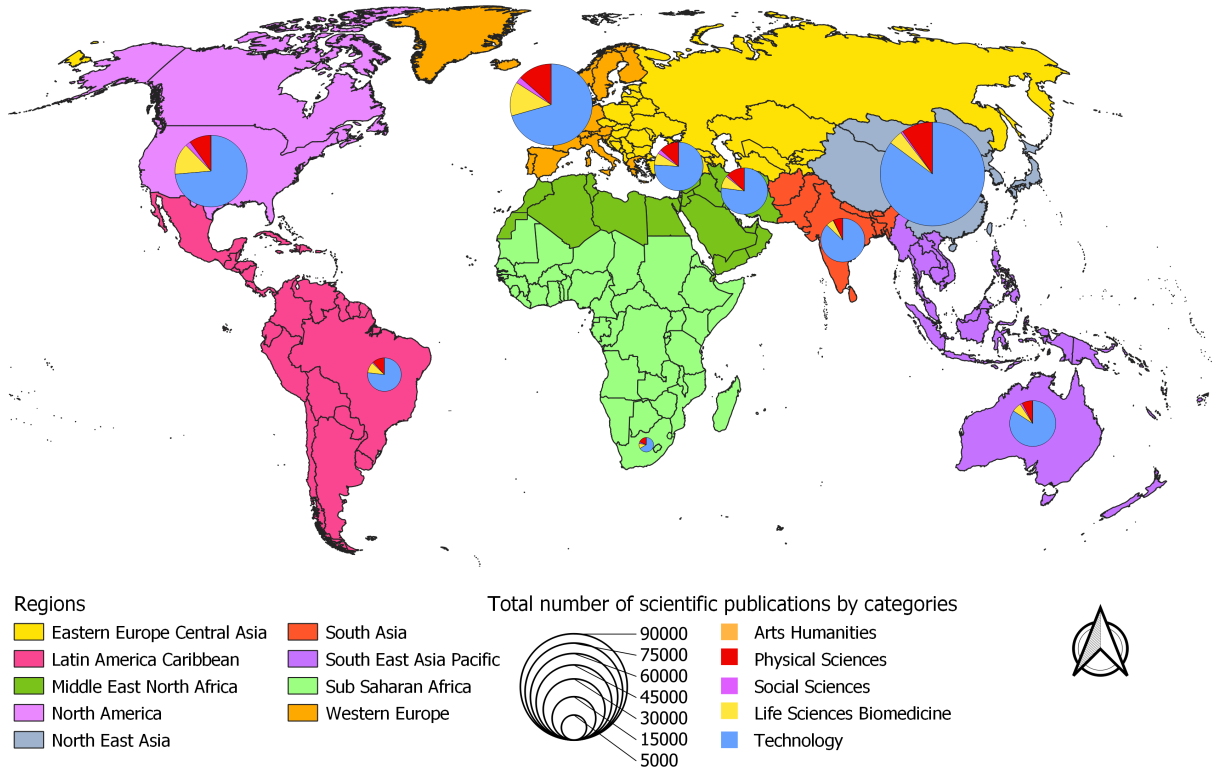}
  \end{center}
  \label{fig:geo2}
\footnotesize
\textit{Notes:} Scientific publications cumulated from 1990 to 2018. The pie charts reflect the shares of WoS research areas [WoS sample].
\end{figure}

Figure \ref{fig:geo2} documents regional specialization in DL research. Throughout, the scientific area (defined according to the WoS research areas) `Technology' takes high shares, accounting for around 70\% to more than 80\% of all publications. Yet, there is substantial variation across geographies. In Asia and Eastern Europe, DL activity centers in particular on `Technology' and `Physical Sciences'. In Western Europe and North America a larger proportion of DL research takes place also in `Life Sciences \& Biomedicine'. 

Taken together, we can conclude that DL diffuses rapidly on a global scale. A high volatility in the rankings has characterized the early stages of development of DL research, with some countries rapidly climbing up the ranking while others lagging behind.  The main players are China, the United States and Europe. Also note that DL research activity is now present virtually everywhere. DL as a research tool seems to find applications in a variety of domains, and world regions show heterogeneous patterns of `specialization' in different scientific areas. These trends are in line with previous evidence \citep{cockburn_2018, oecd_2019, wipo_2019, klinger_2020, vanroy_2020} and consistent with the diffusion process of pervasive technologies.

\subsection{A general method of invention?}

Conceptually, a General Method of Invention (GMI) blends the concepts of Method of Invention (MI) \citep{griliches_1957} and General Purpose Technology (GPT) \citep{bresnahan_1995}. The seminal study of Griliches discussed double-cross hybridization as a MI -- i.e., a way to breed new corn varieties for specific local environments. Double-cross hybridization however is not very `general' because it only applies to corn breeding. On the other hand, there are GPTs. These are technologies that originate in one sector and are usefully applied in many other sectors of the economy. Taking a dynamic perspective, innovation complementarities create positive feedback loops whereby innovations in the originating and application sectors reinforce each other. Combining these ideas, we refer to GMI as a MI that is applicable across many domains. Similar to GPTs, development of the GMI and complementary developments in application domains are mutually reinforcing.\footnote{Instead of GMI, \cite{cockburn_2018} use the term `General-Purpose Invention in the Method of Invention'. We note that a technology that qualifies as a GMI could be used for other purposes than invention (or more generally knowledge creation) and thus also qualifies as a GPT. For DL this is probably the case but is not the focus of this study.} \\

As discussed in Section \ref{sec:DeepLearning}, the originating domain of DL is predominately computer science. Looking at the scientists involved and the influential publications that have pushed forward DL methods over time leaves little room for doubt. On that ground, it seems appropriate to follow \cite{cockburn_2018}, and assume that DL publications in all areas apart from computer science represent \textit{applications} of DL methods to address field-specific research questions -- i.e., adoption of DL as a research tool. However, when it comes to science, one may doubt whether such a strict separation between originating and application domain is actually useful and tenable. The history of scientific instruments, for instance, suggests that the development of instruments often coincide with their scientific use \citep{rosenberg_1992}. In the case of DL, the connection between the `instrument' and the scientific application is given by definition because DL derives predictions from specific examples. This could flip the role of the application domains: Do these domains just supply data for DL instead of requiring DL for their science? We turn to that question below, after examining some dynamics of DL activity across domains.

\begin{figure}[t!]
  \caption{Trends of deep learning publication activity in scientific areas}
  \begin{center}
	  \includegraphics[scale=0.56]{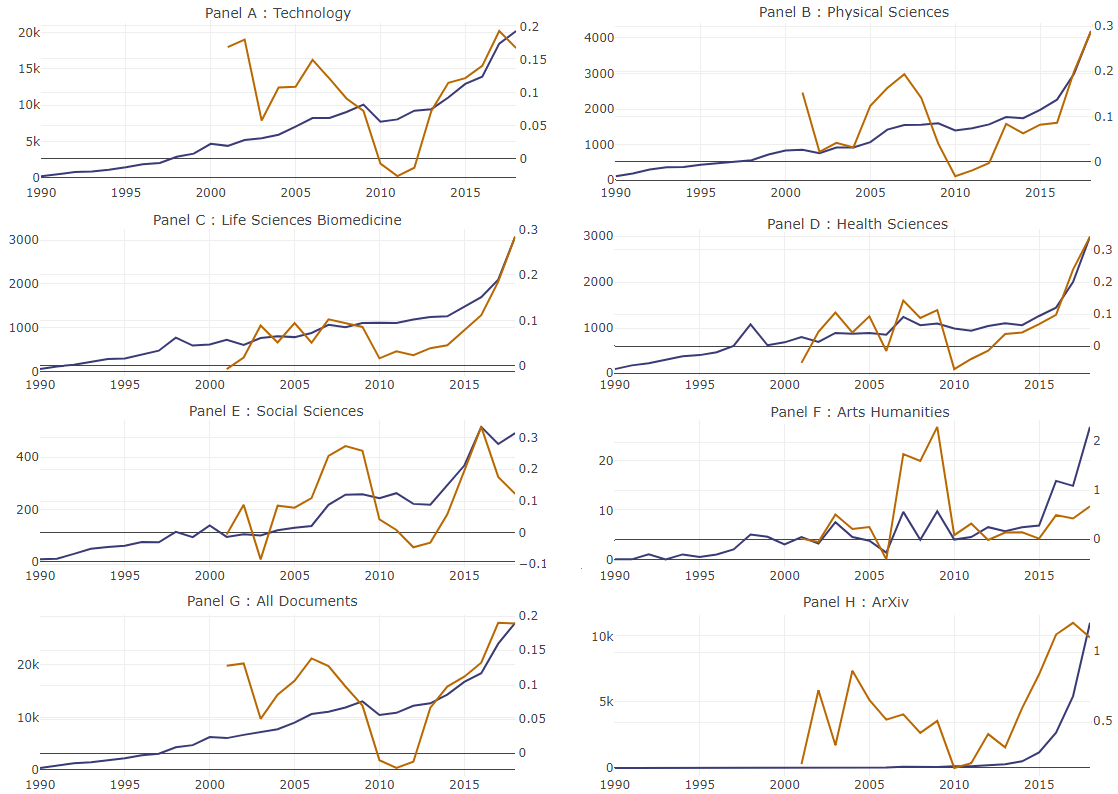}
  \end{center}
  \label{fig:time_trend}
 \footnotesize
\textit{Notes:} These plots show time trends in publication activity related to deep learning. The blue curve corresponds to the number of publications in a given scientific area. The orange curve corresponds to growth rates. Growth rates are calculated as three-year moving averages and omitted before 2001. Scientific areas correspond to WoS research areas. Health Sciences (Panel D) are defined by the set of WoS categories reported in the Appendix \ref{appendix:health}. ArXiv (Panel H) refers to deep learning research published on arXiv.org, based on the sample discussed in Section \ref{sec:DLIdentification}. 
\end{figure}

Figure \ref{fig:time_trend} shows time trends of DL publications in our WoS sample by scientific area (Panels A, B, C, E, F). Panel D refers to `Health Sciences', defined as a subset of `Life Sciences \& Biomedicine'; health science is at the core of our analysis in the next Section. Cross-classified papers are included in each relevant panel. Panel G, `All Documents', simply combines all papers from the WoS sample. Panel H, `arXiv', provides complementary insights on our arXiv.org sample (discussed in Section \ref{sec:DLIdentification} and Appendix \ref{appendix:embedding}). Looking at Figure \ref{fig:time_trend} as a whole, we note a rapid growth in DL research activity in all scientific areas. Yet, the volume of DL papers (blue line) is highly different across areas. `Technology' (Panel A) dominates all others, which is at least partly explained by the fact that it includes `Computer Science', the main originating field. With about five times fewer papers, `Physical Sciences' (Panel B) comes second, closely followed by `Life Sciences and Biomedicine' (Panel C). `Health Sciences' (Panel D) parallels `Life Sciences' (Panel C) because both documents sets are highly overlapping. Publication counts in `Social Sciences' (Panel E) are relatively low, and negligible for `Arts and Humanities' (Panel F). Panel G combines all WoS documents into one picture. In that panel, the (three-year average) growth rates (orange line) show a high growth of 10\% in DL publication activity around 2005, a decline around 2010, and a subsequent recovery with steady growth rates reaching 20\% at the end of the observation period. This growth pattern is close in form and magnitude to the one observed for `Technology' (Panel A), trivially because that is the dominating area; however, also the other areas exhibit very similar growth patterns. 

The publication activity on arXiv.org (Panel H) follows essentially the same dynamics. Growth rates mimic the same shape over time but are about five times higher than growth rates in WoS panels. The comparatively higher growth rates may result from the fact that open platforms are increasingly popular as an efficient and fast way of communication between researchers, particularly in machine learning and computer science communities \citep{sutton_2017}. The arXiv.org dataset corroborates the finding on the WoS dataset: strong growth of DL research in two waves across the sciences.

The overall number of DL-related documents varies over sub-disciplines within scientific areas (not displayed). The general trend in `Technology' is mainly driven by `Computer Science' (103,729 documents), `Engineering' (95,638) and `Automation \& Control Systems' (24,721). In the case of `Physical Sciences', we find `Physics' (7,239), `Mathematics' (5,123) and `Chemistry' (3,702). And for `Life Science \& Biomedicine' we see the preponderant role of `Environmental Sciences \& Ecology' (2,632), `Neurosciences \& Neurology' (2,032), and `Biochemistry \& Molecular Biology' (1,728).

DL publication activity increases not only in absolute numbers but also relative to the overall number of papers in scientific areas; albeit from a low level. In 2018, for example, DL documents account for 2.6\% of all papers in the category `Technology', 1.02\% in Physical sciences, and 0.3\% in `Life Sciences and Biomedicine'. Thus, DL publications still account for only a tiny fraction of the whole research volume, in particular in application domains. Yet, recent growth rates of shares are remarkable. DL has the highest growth rates in the `Life Sciences \& Biomedicine' with 47.3\% from 2017 to 2018. `Physical Sciences' comes second with a DL growth rate of 42\%, and `Technology' shows roughly 18\%.

\begin{table}[t!]\footnotesize
 \centering 
 \caption{Influential deep learning publications} 
    \scalebox{0.96}{
	\begin{threeparttable}
  \label{tab:most_cit_DL} 
\begin{tabular}{@{\extracolsep{5pt}} lccc} 
\toprule
Title $|$ Journal & Cluster & \# Citations  & Share [\%] \\
\midrule \\[-1.8ex] 
Multilayer feedforward networks are universal approximators $|$ NN             & 1 & 5,904 & 0.14 \\
Neural networks and physical systems with emergent ... $|$ PNAS                & 1 & 4,658 & 0.11 \\
Learning representations by back-propagating errors $|$ Nature                 & 1 & 4,645 & 0.11 \\
Learning internal representations by error propagation $|$ MIT Press           & 1 & 3,921 & 0.09 \\
Approximation by superpositions of a sigmoidal function $|$ MCSS               & 1 & 3,657 & 0.09 \\
Training feedforward networks with the Marquardt algorithm $|$ IEEE TNNLS      & 1 & 3,128 & 0.07 \\
ANFIS: adaptive-network-based fuzzy inference system $|$ IEEE SMC              & 1 & 2,909 & 0.07 \\
Identification and control of dynamical systems using ... $|$ IEEE TNNLS       & 1 & 2,551 & 0.06 \\
Cellular neural networks: theory $|$ IEEE CAS                                  & 1 & 2,267 & 0.05 \\
\addlinespace
ImageNet classification with deep convolutional neural networks $|$ NeurIPS    & 2 & 7,177 & 0.17 \\
Gradient-based learning applied to document recognition $|$ IEEE Proceedings   & 2 & 3,590 & 0.09 \\
Deep learning $|$ Nature                                                       & 2 & 3,542 & 0.08 \\
Long short-term memory $|$ NC                                                  & 2 & 3,074 & 0.07 \\
A fast learning algorithm for deep belief nets $|$ NC                          & 2 & 2,710 & 0.06 \\
Reducing the dimensionality of data with neural networks $|$ Science           & 2 & 2,621 & 0.06 \\
Very deep convolutional networks for large-scale image recognition $|$ arXiv   & 2 & 2,582 & 0.06 \\
Particle swarm optimization $|$ IEEE Proceedings ICNN                          & 2 & 2,568 & 0.06 \\
Deep residual learning for image recognition $|$ IEEE Proceedings CVPR         & 2 & 2,160 & 0.05 \\
\bottomrule
\end{tabular}
\begin{tablenotes}
 \footnotesize
 \item {\it Notes:} This table reports the references (title and journal) of the most cited articles from the WoS publication sample over the period 2000--2018. From a total of 4,190,306 references (1,618,836 unique) cited by the documents in our sample, we selected the five most used references for each year. This gives us 18 time series that were clustered. Clustering is obtained via $k$-medoid and dynamic time wrapping. References within clusters ranked by total number of citations.
\end{tablenotes}
 \end{threeparttable}
 }
\end{table}

The growth pattern of DL research with a first boom, subsequent decline, and a second (bursting) boom, reminds the double-boom-cycle that has been observed before for emergent technologies \citep{schmoch_2007}. The narrative goes like this: a new, emerging technology seems at first to offer a high potential. High expectations trigger high development efforts -- the first boom. However, during these early development activities, actors learn about the difficulties to put the principle into practice. Most fail and stop their innovation activities, which puts an end to the first boom. Some continue and, as time goes by, may overcome important practical hurdles and demonstrate real benefits in praxis -- starting a second boom.

\begin{figure}[htbp!]
\caption{Trends in annual citations of influential deep learning publications}
  \begin{center}
	  \includegraphics[scale=0.60]{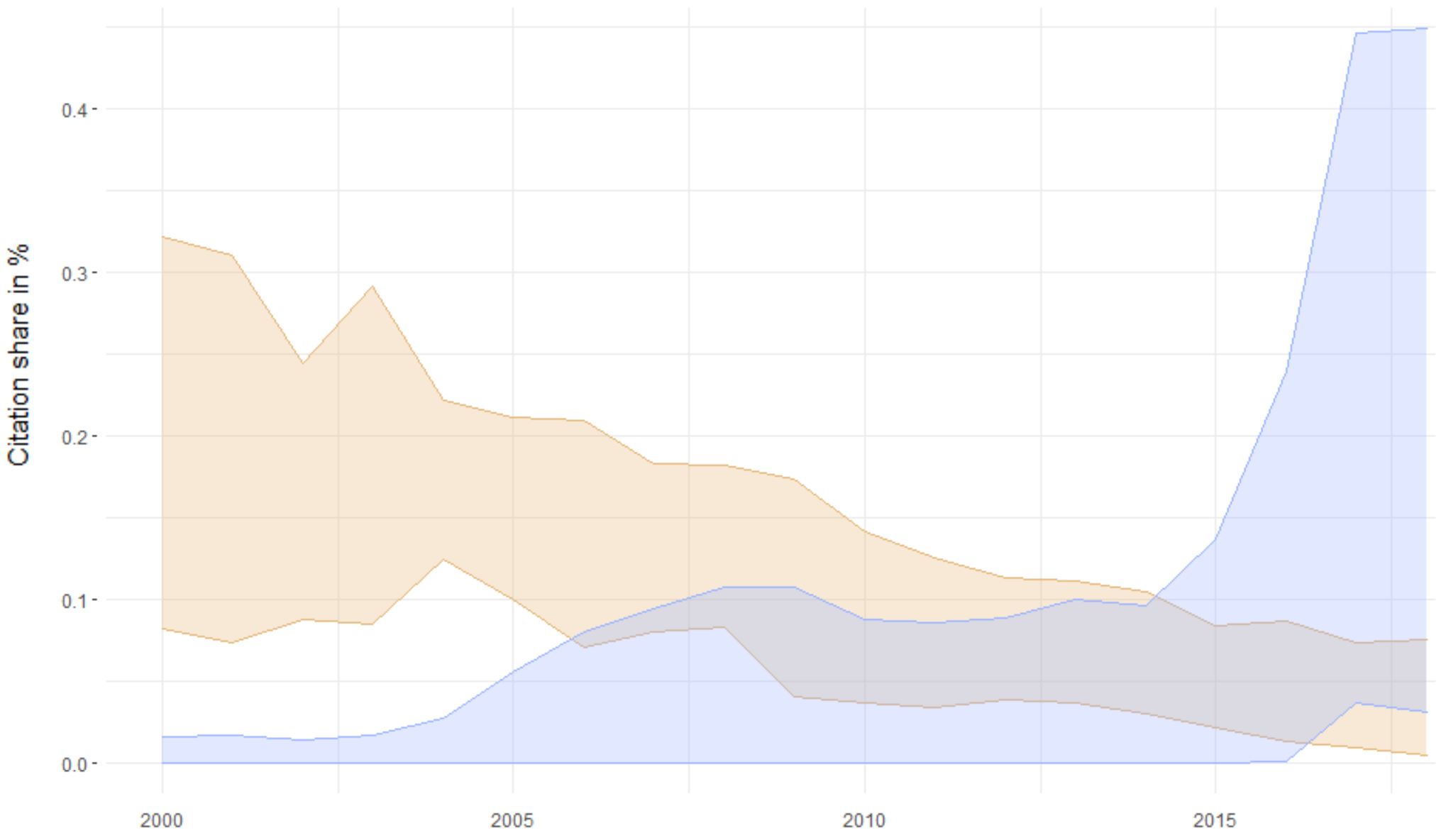}
  \end{center}
    \label{fig:clusters}
\footnotesize
\textit{Notes:} This plot shows the annual share of all citations in the Web of Science sample for the two clusters of most cited deep learning articles. Shaded areas display time series intervals defined by minimum and maximum citation shares. The red profile mostly represents `theoretical' contributions while the blue profile represents `applications'. Due to the limited number of articles that can be cited in the initial period, we clustered the time series from 2000.   
\end{figure}

An in-depth analysis of citation dynamics suggests that the double-boom-cycle story holds for DL. We consider the top five cited references in each year of the observation period (i.e., documents with the highest annual shares of all cited references in our DL publications). This gives us a list of 18 unique articles and their corresponding citations counts, as shown in Table \ref{tab:most_cit_DL}. Using dynamic time warping (DTW) to measure dissimilarity between time series, we cluster these temporal sequences by mean of $k$-medoid \citep{berndt_1994}. As shown in Figure \ref{fig:clusters}, we obtain two clusters. In the first period, most cited articles in our sample are theoretical contributions, including the possibility of using multilayer feedforward networks as universal function approximators, training algorithms (backprop), and parallel computing theories (cellular NN). In the second period, the most influential articles are no longer theoretical contributions, but rather articles that have shown how to put theoretical principles into practice -- with tremendous success in various AI competitions. These contributions include inventions that have brought enormous performance gains on real-world tasks, particularly for image and text analysis (deep convolutional neural networks and LSTM, as discussed in Section \ref{sec:trends_dl}).

Observed dynamics are consistent with the idea of positive feedback between DL development -- within `Technology' -- and DL applications -- mostly in `Physical Sciences' and `Life Sciences and Biomedicine'. However, one might question the extent to which different scientific domains are inclined to incorporate deep learning methods as a practice in their disciplinary research. In other words, is there indeed a diffusion of DL \textit{into applications} or rather a cross-disciplinary effect of DL? 

We investigate this question by considering the cross-classification of publications in our sample. Each document is labelled by WoS as belonging to at least one subject category according to the journal in which it was published. In most of the cases a document falls into more than one scientific category. The extent to which publications in a given scientific area are cross-classified as computer science contributions may therefore proxy cross-disciplinarity with respect to computer science. For each broad scientific area and year, we calculate the fraction of deep learning documents that are (also) labeled as `Computer Science'.\footnote{We define `Computer Science' the set of the following Web of Science subcategories: `Computer Science, Artificial Intelligence'; `Computer Science, Cybernetics'; `Computer Science, Hardware \& Architecture'; `Computer Science, Information Systems'; `Computer Science, Interdisciplinary Applications'; `Computer Science, Software Engineering'; `Computer Science, Theory \& Methods'.} 

\begin{figure}[h!]
  \caption{Deep learning publications cross-classified as `Computer Science'}
  \begin{center}
    \includegraphics[scale=0.65]{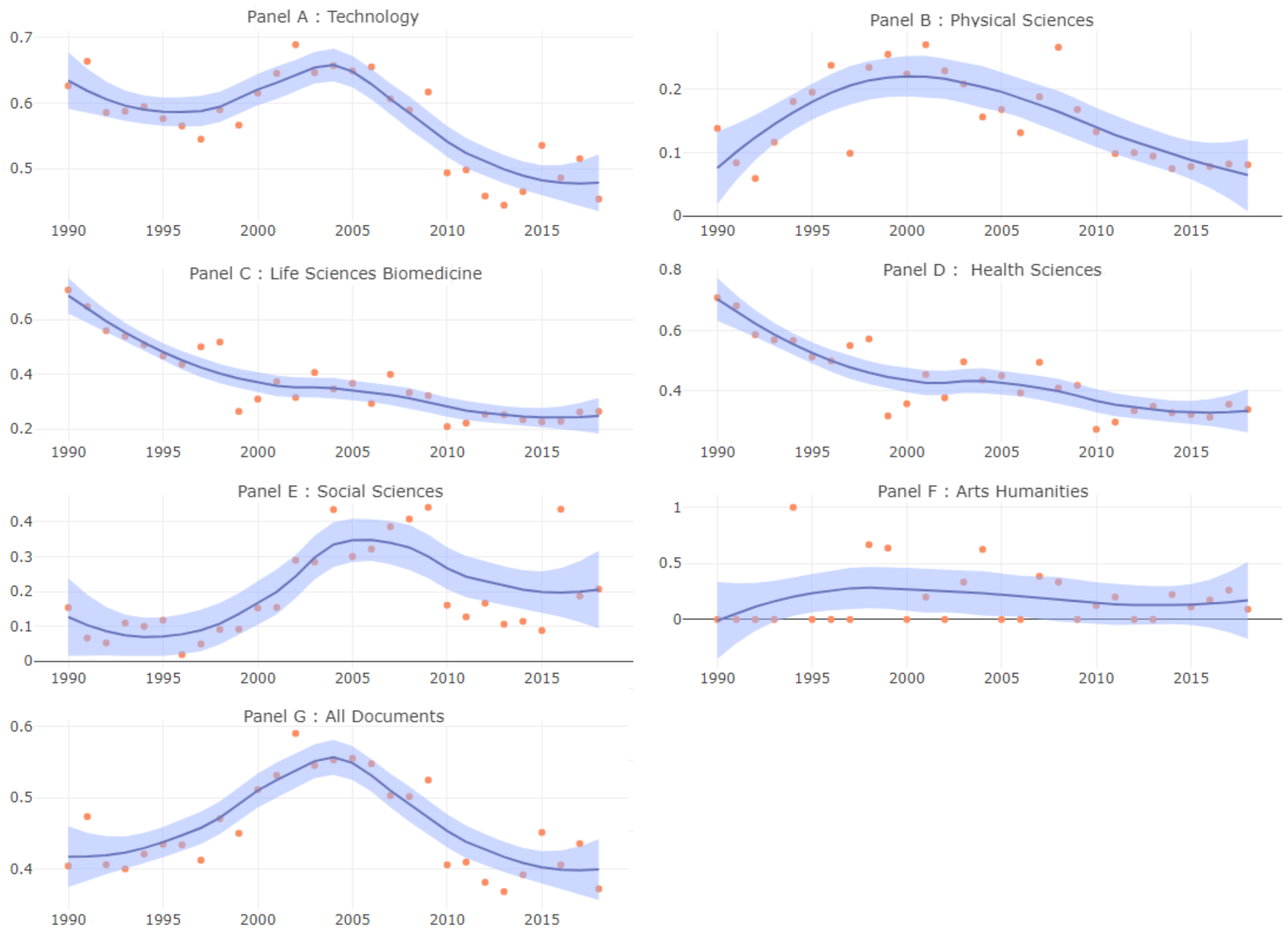}
  \end{center}
  \label{fig:CSApplication}
\footnotesize
\textit{Notes:} These plots show the fraction of deep learning documents cross-classified as `Computer Science'. Red dots represent the share of cross-classified papers in each year. The blue curve corresponds to a simple local regression, with the surrounding shaded area representing the 95\% confidence interval around the mean.   
  \end{figure}

Figure \ref{fig:CSApplication} provides results. Turn now to the first, upper-left panel `Technology' (Panel A). Each point of the plot represents the average number of `Technology' DL documents cross-classified as `Computer Science' in a given year. For example, in 1990 about 60\% of `Technology' DL publications fell (also) into the `Computer Science' category (the first red dot). The trend (blue line) follows a flat U-shape approaching around 70\% in 2005, before decreasing to less than 50\% by the end of the observation period. In 2018, over 50\% of DL documents in the area `Technology' are no longer labeled as computer science contributions. The upper-right panel `Physical Sciences' (Panel B) shows an inverse U-shape, with an increase in cross-classified computer science documents of up to 20\% in 2000, before falling down again to 10\% by the end of the period. The pattern in `Life Sciences \& Biomedicine' (Panel C) is different in that there is no increase of computer science cross-classification. Instead, there is a very high share of 70\% at the beginning of the period which continuously decreases to about 20\%, with significant drops around 2000 and again in 2010. The `Health Sciences' (Panel D) experience the same pattern. `Social Sciences' (Panel E) increase their share of computer science documents to 40\% around 2010, followed by a sharp downturn. Finally, for `Arts \& Humanities' the fraction of computer science documents is very noisy, and so we do not observe any particular tendency. 

To interpret these patterns, recall that the areas `Technology', `Physical Sciences', and `Life Sciences \& Biomedicine' exhibit the strongest DL publication activity in terms of absolute and relative numbers, as well as growth rates. For these three areas, DL activity takes off around 2010 (Figure \ref{fig:time_trend}). At the same time the fraction of publications that are cross-classified as `Computer Science' decreases, as shown in Figure \ref{fig:CSApplication}. These dynamics indicate that DL diffuses indeed \textit{from} computer science, the originating discipline, \textit{into} other application-oriented scientific disciplines. The high share of cross-disciplinary research involving computer science in the early periods could be explained by the fact that the transfer of DL and its adoption to the field of application requires close interaction between researchers from both domains. Computer scientists need to learn what can be done in practice and scientists in the application sector need to understand the potential of the (new) technology for their research.

In sum, the evidence presented so far points to a simple statement: DL meets the conditions of a General Method of Invention. The technology originates predominantly from computer science and is increasingly integrated as a research tool into many other scientific fields. Hence, it is a method of invention that is generally applicable in various domains. Moreover, the (joint) dynamics across sciences are consistent with mutual reinforcement between originating and application domains, a core idea of GPT. This opens the question of how DL affects scientific development in its application domains. We turn to that question in the next Section.

\section{Scientific impact of deep learning in health sciences}
\label{sec:DLHealth}

This Section deals with the impact of DL specifically in health sciences.\footnote{We delineate the `Health Sciences' by 83 Web of Science subject categories within `Life Sciences \& Biomedicine' research area. The complete list of included categories can be found in Appendix \ref{appendix:health}.} We focus on health sciences because it is among the scientific domains where the adoption of DL is more widespread and dynamic, as shown in Section \ref{sec:DiffusionInScience}. Furthermore, AI in general and DL in particular have already led to a variety of innovations in the health realm -- improving healthcare systems, supporting clinicians in surgery, monitoring patient diseases. DL research demonstrated high societal impact in the short run \citep{miotto_2018}. Investigating the impact of DL in scientific domains other than health sciences would certainly be interesting, but is beyond our limits. At this early stage of research on DL in science, contextualizing the empirical analysis is essential. Doing so for several scientific domains would not only exceed the page limit of an article but also our expertise in those domains. Furthermore, the empirical analysis itself is highly demanding in terms of data requirements and computational burden. We keep that manageable by focusing on one scientific area.

The next subsection \ref{sec:dp_health_rev} illustrates how DL research has advanced several areas within the health sciences. Subsection \ref{sec:novelty_impact} develops the conceptual framework for the empirical analysis. Subsection \ref{sec:empirics} discusses the data, methods, and results of the analyses.

\subsection{Deep learning in health sciences}
\label{sec:dp_health_rev}

In recent years, several areas within health sciences have seen a shift from systems with hand-crafted features (i.e., systems completely designed by humans) to intelligent machines that learn the features from data. As discussed in Section \ref{sec:machine_intell}, this approach is foundational to the DL principle. Provided that the architecture is optimally weighted, DL leads to an effective high-level abstraction of raw data, thus increasing perceptive and predictive capabilities \citep{lecun_2015, schmidhuber_2015}. DL has enabled the development of multiple data-driven solutions in health informatics and biomedical research, making it possible to automatically generate features and reduce the amount of human intervention in the process \citep{ravi_2017}. Below we provide a cursory review of the main areas in which DL finds applications, highlighting the benefits that the technology can bring in the process of knowledge search, experimentation, knowledge production and patient health care.

One domain in which DL has gained considerable success in recent years is translational bioinformatics, understood as the study of biological processes at the molecular level by means of biomedical and genomic data and informatics methodologies \citep{leung_2015, ravi_2017}. The application of DL in translational bioinformatics has turned out to be particularly relevant for genomic research. This area of research aims to determine how variations in the DNA of individuals can affect the risk of different diseases and find causal explanations in order to design targeted therapies. The details of the mechanisms at work in the cell are hidden. What we can observe is the outcome of many layers of biophysical processes and interactions, most of which are not fully understood. 

Progress in biotechnology has contributed to reducing the costs of genome sequencing and shifted the research focus on prognostic, diagnostic and treatment of diseases through gene and protein analysis. Modern biology allows also high-throughput measurement of many cell variables, including gene expression, splicing, and proteins binding to nucleic acids, all of which can be treated as training targets for predictive models \citep{marx_2013}. Modern deep learning systems can accurately interpret the text of the genome just as the machinery inside the cell does, making it possible to explore the effects of genetic variations and potential therapies quickly, cheaply and more accurately than can be achieved using `standard' laboratory experiments \citep{leung_2015}.

Most of the DL applications have been deployed for the accurate prediction of splicing patterns and gene variations, which is a key to providing early diagnosis of various diseases and disorders such as cystic fibrosis, Parkinsonism, spinal muscular atrophy, myotonic dystrophy, amyotrophic lateral sclerosis, premature aging, and dozens of cancers. An exhausting review of the literature on the advent of intelligent machines in genetic research suggests that computational methods will not be able to completely replace laboratory and clinical diagnosis, but should significantly reduce the time needed for these methods of analysis by reducing the search space for the hypotheses that need to be validated \citep{leung_2015, angermueller_2016}. 

Genomic medicine is not the only area where DL can bring benefits. The ability to abstract large, complex and unstructured data makes deep learning also an effective solution for the prediction of protein-protein and protein-compound interactions (CPI). CPI are crucial for drug discovery, the identification of new compounds and toxic substances, and for advances in pharmacogenomics. DL allows a richer representation of possible interactions beyond the genetic and molecular structural information encoded in large datasets, thus paving the way for data-driven discoveries. For example, a team of machine learning researchers took advantage of a number of different Quantitative Structure -- Activity Relationships (QSAR) datasets from Merck's drug discovery effort and significantly improved the state-of-the-art drug discovery pipeline, despite having no prior knowledge about the biochemical properties of training features \citep{ma_2015}. The purely data-driven approach is not the only one possible when intelligent machines are available as research tools. Instead of examining the parameters of the model and coming up with an interpretation, researchers can also ask the system for presumable relationships between inputs and outputs that cannot be `checked by eye'.

Medical imaging is another domain in which the advent of deep learning, and especially of CNNs, has played a central role. As soon as it was possible to collect and catalogue medical images, computer systems were used to assist researchers and doctors in the analysis of these images. Indeed, whereas diagnosis based on the interpretation of images runs the risk of being highly subjective, computer-aided diagnosis (CAD) can result in a more objective assessment of the underlying disease processes. However, CADs have long suffered from some limitations mainly due to differences in shape and intensity of abnormal tissues (tumours or lesions) and variations in imaging protocol. Non-isotropic resolution in Magnetic Resonance Imaging (MRI), for instance, has been a major challenge to manage with traditional machine learning methods. Recent advances in deep CNNs have found fertile ground in the medical imaging research community due to their outstanding performance in various computer vision tasks, challenging the accuracy of experts in some tasks \citep{litjens_2017, shen_2017}.

Image or exam classification (e.g., disease present or not) was one of the first areas in which deep learning made a decisive contribution to medical image analysis. Yet, the current range of applications is much wider and includes organ, region and landmark localization, object or lesion detection, organ and substructure segmentation, and lesion segmentation.\footnote{The detection task typically consists in identifying and localizing small lesions in the full image space. It represents an important pre-processing step in the clinical workflow for therapy planning and intervention. The segmentation task consists instead in identifying the set of voxels that make up the contour or the interior of the objects of interest. The segmentation of organs and other substructures enables quantitative analysis of clinical parameters related to volume and shape (e.g., brain analysis).} Both detection and segmentation tasks play a key role in the diagnosis of tumours and other diseases, and represent one of the most labour-intensive activities for doctors since accurate classification requires both local analysis on lesion appearance and global contextual analysis on lesion location \citep{litjens_2017}. DL methods have also found applications in other medical imaging tasks, such as content-based image retrieval (CBIR) and combining image data with text reports (Rav\`i et al., 2017). Comprehensive reviews of the literature on the impact of DL in the field of medical imaging highlight that deep learning algorithms empower machines for automatic discovery of object features and automatic exploration of features hierarchy and interaction, once again supporting and facilitating the work of scientists and clinicians \citep{litjens_2017, shen_2017}.\footnote{Despite incredible progress, the potential of deep learning in image analysis has not been fully unfolded. One of the main limitations seems to be the shortage of large labeled datasets. Labeling data is a time-consuming and expensive activity that requires some expertise. There is an ongoing debate about the possibility of crowdsourcing (even by non-experts) as a viable alternative for creating low-cost, truth-based medical imaging datasets \citep{litjens_2017}.}

The development of intelligent devices and cloud computing has allowed the generation and collection of an incredibly high volume of health data from various sources in real-time. Wearable, implantable and ambient sensors, as well as the data they provide, enable continuous monitoring of health and well-being \citep{marx_2013, raghupathi_2014}. The adoption of DL has increased the benefits of pervasive sensing in a wide range of health applications such as measure of food calorie intake, energy expenditure, activity recognition, sign language interpretation and the detection of anomalous events in vital signs (e.g., blood pressure and respiration rate). Most applications use DL algorithms to achieve greater efficiency and performance for real-time processing on low power devices. These devices are beneficial for science because they increase the understanding of diseases by enabling a more thorough and systematic analysis of the patient's condition.

Given its versatility, deep learning has also proven to be efficient in handling multimodal unstructured information by combining several neural network architectural components on big data infrastructures stored in hospitals, cloud providers and research organizations. An example of this type of data are the Electronic Health Records (EHR). EHR provides an extremely rich source of patient information that includes history details such as diagnoses, diagnostic exams, medications and treatment plans, immunization records, allergies, radiological images, laboratory and test results. DL permits an efficient navigation, extraction and analysis of these data, hence providing valuable information on disease management and the discovery of new patterns (e.g., long-term time dependencies between clinical events and disease diagnosis and treatment) that result in completely new hypotheses and research questions \citep{rajkomar_2018}.

Public health has also witnessed an upsurge in deep learning applications. The latter involve epidemic surveillance, modelling lifestyle diseases (e.g., smoking and obesity) with relation to geographical areas, monitoring and predicting air quality, contamination of food and water supplies, and many more \citep{miotto_2018, ravi_2017}. Traditional machine learning methods can accurately model several phenomena but have the limited ability to incorporate real-time information. On the contrary, current systems in public health studies are based on online deep learning and can build hierarchical models and encode information sequentially as new training datasets become available \citep{zhang_2018}.

Finally, the neuroscience and cognitive sciences communities have begun to argue that the brain is, at least in part, an optimization machine \citep{marblestone_2016, hassabis_2017}. As such, optimization algorithms in DL such as backpropagation may have analogies in biological brains, although, from a practical point of view, the design of an AI system does not require any adherence to biological plausibility. It is not yet well known which processes in the brain emerge from the optimization of cost functions, which emerge from other forms of self-organization, which are pre-structured through genetics and which are based on the interplay of all these mechanisms. Yet, progress in understanding the architecture and behaviour of artificial neural networks could play a vital role for understanding biological brains. For example, recent work on improving the performance of deep CNNs has yielded new insights into the nature of neural representations in high-level visual areas. Further, some properties of the LSTM architecture offered key ideas that motivated the development of working memory models \citep{hassabis_2017}. Hence, by focusing on the computational and algorithmic levels when building intelligent machines \textit{in silico}, researchers can obtain transferable insights into the general mechanisms of brain function, in turn opening new prospects for neuroscience. As \cite{hassabis_2017} put it: ``\textit{[D]istilling intelligence into an algorithmic construct and comparing it to the human brain might yield insights into some of the deepest and the most enduring mysteries of the mind, such as the nature of creativity, dreams, and perhaps one day, even consciousness''}[p.255].\footnote{As discussed in Section \ref{sec:trends_dl}, the architectural and algorithmic constraints underlying deep learning were originally inspired by neuroscience. The role played by this discipline has gradually diminished and many of the major developments in artificial intelligence have been driven by insights into the mathematics of efficient optimization. Notwithstanding this, neuroscience may still inform machine learning in many respects. \cite{lake_2017}, for instance, have recently suggested some directions for neuroscience-inspired AI research. These studies emphasize the ways through which neural networks could be modified and improved, especially by trying to incorporate more powerful, human-like cognitive abilities such as attention, imagination and planning, efficient and transfer learning, and intuitive understanding of the physical world. In short, exchange of ideas between AI and neuroscience may create a virtuous circle advancing the objectives of both fields.}

\subsection{Novelty and impact in science}
\label{sec:novelty_impact}

The idea of a scientific contribution commonly includes two aspects: \textit{scientific novelty} and \textit{impact}. Different terms for essentially the same idea have been employed in earlier research on science. Back then, debates were focused on originality, discovery or breakthrough and their contribution to scientific progress \citep{price_1986, merton_1957, bourdieu_1975}. \cite{kuhn_1962} coined the term `novelty' to describe a more radical contribution that does not advance incrementally the `normal science' in place, but instead breaks the current paradigm. In the more recent literature, the term novelty partly lost this radical connotation, but still carries the idea of a higher degree of originality. The concept of `re-combinatorial novelty' adds to this the idea that new knowledge comes from the recombination of previously generated bits of knowledge \citep{weitzman_1998, fleming_2001, uzzi_2013, wang_2017}. 

The main argument is that we realize only a small part of the potential of useful recombinations in the knowledge space because humans face inherently cognitive limits. In other words, scientific development depends not only on the nature of the object of investigation, but also on the nature of the investigator herself. Deep learning may change the way science develops because it affects research behaviour and helps to connect ideas from seemingly different scientific domains \citep{agrawal_2018, cockburn_2018, furman_2020}. However, even if DL has this alleged re-combinatorial potential, one may doubt whether it is actually used in that way. The way in which DL is employed in science as a research tool depends not only on the properties of the technology, but also on the scientific complex in which it is introduced \citep{callon_1994}. This implies that the same technology may be used either to further deepen established research trajectories or to explore new avenues, leading to lower and higher re-combinatorial novelty respectively. The prevailing effect most likely depends on the stage of a scientific domain's life-cycle and the degree of integration of the tool into existing scientific practices.\footnote{A case study on scientific creativity at a large research infrastructure (European Synchrotron) confirms this conjecture \ \citep{ferrando_2019}. The synchrotron offers favourable conditions for creative research, but this potential depends heavily on the discipline, its degree of maturity and the structure of user communities.} The extent to which DL is associated with `re-combinatorial novelty' in health sciences is therefore an empirical question that we address in the analysis.

The second aspect of our study concerns the scientific impact of deep learning research. Impact is related to but different from novelty. If research provides novelty, that novelty must be taken up by the scientific community in order to unfold impact. On the other hand, research may well have impact on subsequent research for reasons other than (re-combinatorial) novelty, especially by providing new insights within established knowledge structures.

Yet, impact is in general not independent of novelty. The reasons can be found in the process of knowledge production itself as well as in the social dimension. Throughout the knowledge creation process, a higher level of novelty is likely to increase the risk of delays and failures \citep{azoulay_2011}. On the one hand, the complexity and ruggedness of knowledge landscapes makes navigation difficult. On the other hand, more novelty sometimes necessitates more complex and risky collaborative social structures \citep{fleming_2007, foster_2015}. Both arguments suggest that research carrying high novelty may be subject to considerable variations in `quality' \citep{fleming_2001, wang_2017}. This alone would translate into an increased variance of the impact. Social arguments further reinforce this variation. Although originality is praised as fundamental for scientific advancement, high novelty may encounter high resistance in the society because it can cause structural changes in roles and norms \citep{merton_1957, bourdieu_1975}. These structural changes are particularly relevant in the context of AI where the intersection of ethical and legal considerations shapes the future of both individuals and society as a whole.\footnote{Taking a broader perspective, we note that new digital technologies have a profound effect on almost every aspect of people's lives. Think of algorithmic decisions which can be used for recruitment, credit authorization, policy decision-making, and many other purposes. These decisions are often seen as opaque, unregulated, and amplifying discrimination \citep{oneil_2016}.}

Most of the empirical evidence on the nexus between re-combinatorial novelty and impact comes from bibliometric studies on scientific publications and patents. Re-combinatorial novelty is commonly measured by the combination of different elements figuring in the document. Citations, keywords, and scientific/technological classes are the most common elements to proxy ideas, concepts, and domains, respectively. Incremental science advances mostly within domains through the addition of concepts and their relations. Compared to novel keyword combinations, novel re-combinations of classes are therefore closer to novelty in the Kuhnian sense. 

Most studies build on citations to measure novelty, citations being a way of systematically tracking individual ideas through scientific documents. \cite{small_1973} suggests that the co-occurrence of document pairs in the list of references is indeed a good measure of ideas' association. Citations are therefore useful to trace the flow and connection of ideas within and across domains (or scientific categories).

\cite{uzzi_2013} lift this idea at the journal level. They contend that different academic journals tend to reflect different scientific domains. In light of this, the co-occurrence of journal references signals domain combinations, and hence can be used to proxy re-combinatorial novelty. More precisely, \cite{uzzi_2013} count how often each journal pair is referenced across papers within a given year. Frequently observed journal pairs indicate conventional domain combinations, while rarely observed journal pairs indicate novel (or `atypical') domain combinations. They find that papers incorporating many conventional and some atypical combinations are particularly likely to be among the top 5\% cited documents (high impact). In addition, the right balance between high conventionality and some degree of atypical combinations is more likely to be observed in team work than in individual work. Finally, impact propensity is higher for teams than individuals, given any combination pattern. One interpretation of these findings is that a good research strategy is to balance conventionality and novelty, and that teams may be instrumental in implementing that strategy. \cite{lee_2015}, based on a very similar methodology complemented by a survey, investigates more closely the role of scientific teams in the novelty-impact nexus. Their findings suggest that medium-sized teams, covering a variety of fields, are particularly successful in producing novel combinations. 

\cite{wang_2017} shift from measuring `atypical' combinations to `novel' combinations. Atypical combinations are journal combinations that are \textit{rarely} observed in the \textit{same} year. Novel combinations are journal combinations that have \textit{never} been observed in \textit{prior} years. Therefore, the difference lies in the degree (rare vs. never) and in the reference set (same year vs. previous years). Also note that first-time-ever combinations of referenced journals may have different degrees of difficulty; the latter takes into account the knowledge distance between the newly combined journals based on their co-cited journal profiles (i.e., their `common friends', using authors' terminology). The findings confirm that novelty in research is characterized by a `high risk/high gain' profile, delayed recognition, and publication biases. 

For the main analysis, we operationalize the concept of `re-combinatorial novelty' as the first appearance of a knowledge combination, closely following \cite{wang_2017}. Novelty \textit{\`{a} la} Wang et al. complies more with the idea of DL as a general method of invention (i.e., a method for creating something new and not only unusual), as discussed in the Section \ref{sec:DiffusionInScience}. In Section \ref{sec:robcheck}, we also implement the method \textit{\`{a} la} Uzzi et al., to test the robustness of our results and further examine the degree of conventionality of DL research.

\subsection{Empirical analysis}
\label{sec:empirics}

\subsubsection{Data and methods}

We measure scientific knowledge creation on scientific papers published in peer-reviewed journals and conference proceedings in health sciences.\footnote{Henceforth, the term `journal' refers interchangeably to both peer-reviewed scientific journals and conference proceedings. We restrict the focus on journals that are not cross-classified as `Computer Science' journals. The sample of DL publications used in this Section is therefore a subset of the sample used in Section \ref{sec:DiffusionInScience}. This subsample is further complemented with additional non-DL publications, as discussed hereafter.} The principle idea is to compare publications that apply DL with those that do not apply DL, while controlling for a set of confounding factors. Comparison is made in terms of re-combinatorial novelty and subsequent use of the paper measured by the number of citations received. 

\paragraph{Sample.} Sample selection is at the journal level. We aim to include all the articles for the whole observation period published in those health journals where research involving deep learning has been the most prominent. This provides us with a relatively coherent knowledge base against which we may judge on novelty and impact. Some sampling is necessary because we do not have the entire universe of health science papers at our disposal. Yet, the focus on journals with the bulk of DL research activity is advantageous in at least two ways. First, we keep a relatively large number of DL articles in the sample for a given number of journal downloads. Second, our final sample contains DL and non-DL publications that are more likely to be comparable: on the one hand, DL is currently undergoing a diffusion process, and therefore its current level of adoption is unevenly distributed across research (sub)domains; on the other hand, re-combinatorial novelty and subsequent use vary depending on these domains \citep{wang_2017}. It seems plausible that `early DL adopters' come from domains that are generally characterized by high re-combinatorial novelty. We somehow limit this potential bias by selecting those journals in which we indeed observe deep learning research.\footnote{To see the argument more clearly, imagine we would sample from all domains, while DL research would only appear in specific domains that are particularly apt to novel re-combinations within their `normal science' mode. This would lead to the conclusion that DL is likely to provide new combinations. But that would be mainly due to a `domain'--effect rather than a `DL'--effect. Sampling on domains where DL is present limits the influence of the `domain'--effect on the results, and thereby helps to capture the `DL'--effect we care about.} 

In total, we count 26,461 DL health papers in about 5,000 health journals and proceedings. Roughly 45\% (11,520) of these documents are published in the top 100 most frequent health journals in the sample. The list of journals is reported in the Appendix \ref{appendix:health}. Following our sampling strategy, we downloaded the entirety of these journals for the period 1990--2018. Combining DL and non-DL publications, our final sample contains 1,081,223 articles over the period 1990--2018. A general characteristic of the sample is that the number of observations increases over time. We start in 1990 with 14,317 sampled papers and gradually increase to 62,342 sampled papers in 2018. Although we sample all the journals over the whole observation period, our sample includes less journals at the beginning and more at the end of the period. This is caused by the positive trend in DL adoption, as documented in Section \ref{sec:DiffusionInScience}. Since most of DL papers are published recently, we are likely to sample `younger' journals. 
This aspect will be taken into account in the analysis.

\paragraph{Variables.} All variables for the analysis are measured on the sampled papers. Our main explanatory variable is a binary indicator of DL content in the paper (\textit{DL}): 1 if the paper is classified as DL, 0 otherwise. Our main dependent variables are (various measures of) re-combinatorial novelty and scientific impact based on citation counts. In the following we describe the construction of our main response variables and other covariates.

Re-combinatorial novelty is measured on referenced journals. The basic idea consists in examining for each paper whether it makes first-time-ever combinations of referenced journals -- i.e., its list of references contains journal pairs that have never appeared jointly in any prior list of references. In order to exclude journal pairs that simply formed once by happenstance, we further impose the condition that journal pairs be observed again within the next three years. A paper with at least one journal pair in the reference list that is both novel and re-used, is judged as providing some novelty. This way we construct a binary indicator of novelty, henceforth defined as \textit{Novelty Dummy}. 

One further consideration is that a novel journal pair may span more or less distant domains. This subtlety is captured through the co-citation profiles of the two journals forming a novel pair. The idea is that if both journals are often (rarely) cited with the same third journal(s), they are likely to span less (more) distant domains. This can be used to construct a weighted (continuous) measure of novelty, in the following \textit{Novelty}. Calculations of the binary and weighted novelty measures follow \cite{wang_2017}, therefore we provide only a concise formal description of the procedure. Other mathematical details and parameter settings are reported in the Appendix \ref{appendix:novelty}.

Each paper includes a set of references to journals, say $\{J_1, \ldots, J_i, J_j, \ldots, J_J \}$. \textit{Novelty Dummy} assumes a value equal to 1 if at least one journal pair $(J_i, J_j)$ has never been observed before and is re-used five times within the next three years, 0 otherwise.\footnote{Five times is motivated by the fact that by adopting the threshold used by \cite{wang_2017} (3 times), the amount of novel documents was too high, being about 50\%. Using a more restrictive threshold, we lowered the share of novel papers to roughly 30\%. The patterns discussed in the following are not sensitive to changes in the threshold.}

Although the pair $(J_i, J_j)$ is novel, the two journals may be related through co-occurrence with other journals in prior reference lists. This in turn leads to a weighted version of the novelty measure. We construct a journal co-occurrence matrix $C$ of dimension $J \times J$. Its $i$th column, $C_{i}$, provides counts of how often journal $J_i$ appeared together with any other journal in a journal reference set from the prior three years. The cosine similarity between $C_{i}$ and $C_{j}$, say $\textit{COS}_{i,j}$, measures the `closeness' of journals $J_i$ and $J_j$, and it ranges between zero and one. 
One if both journals share exactly the same profile in their co-cited journals; zero if they do not have any co-cited journal in common. Therefore, $(1-\textit{COS}_{i,j})$ measures the co-citation distance of the journal pair. There may be more than one novel journal pairs in a paper's reference list. The weighted novelty measure of the paper is obtained by summing $(1-\textit{COS}_{i,j})$ over all novel journal pairs $(J_i, J_j)$ in that paper. In a final step, we take the log transform (we apply $log(x+1)$). In sum, \textit{Novelty} score increases with the number of novel journal pairs in its references and their co-citation distances.

We differentiate from \cite{wang_2017} in two important aspects. Firstly, we judge novelty and co-citation distance only on journal pairs that are observed in the reference lists of our sampled papers. Thus, we do not measure novelty per se but with respect to a knowledge base covered by the sampled (health) journals. The interpretation is that this adds an addressee -- i.e., novel for whom? The addressee here is the health sciences community.

Secondly, we calculate different measures of novelty by considering different sets of journals in the references. In this way, we try to capture the \textit{source} of novelty: where does the novelty come from? ICT, health, or other domains? Let us try to clarify this argument. Although all the articles in the sample are published in health science outlets, they can reference journals in various domains. For instance, a health science paper that involves deep learning as research tool is likely to cite computer science journals where DL methods were published.\footnote{A visual inspection of a random sample of articles in our sample confirms that this is the case.} This translates into a re-combinatorial novelty `simply' due to the adoption of the method, but does not necessarily reflect the re-combinatorial potential of DL to connect and re-combine knowledge pieces in complex knowledge landscapes. Recall that the general objective of our analysis is to investigate whether DL is mainly used to cope with knowledge explosion within a given domain or to facilitate cross-fertilization of knowledge across domains. In other words, we aim to measure whether the adoption of DL fosters novel re-combinations \textit{within} health sciences and/or \textit{between} health sciences and disciplines other than computer sciences. Therefore we calculate novelty not only on any journal pairs, indicated by `All Sciences', but also limit to journal pairs where: (i) no referenced journal is classified as a computer science journal, indicated by `No CS'; (ii) both referenced journals are uniquely classified as health sciences, indicated by `Only HS'. By way of example, the combination of `Biology \& Biochemistry' and `Computer Science' journals would be regarded as an `All Sciences' combination; `Engineering' and `Molecular Biology \& Genetics' as a `No CS' combination; and `Neuroscience \& Behavior' and `Psychiatry/Psychology' would be considered an intra-domain `Only HS' combination. 

Combining the above three re-combinatorial possibilities with the option to calculate the novelty either as a binary indicator or continuous score, we obtain six different novelty measures, namely: \textit{Novelty Dummy (All Sciences)}, \textit{Novelty Dummy (No CS)}, \textit{Novelty Dummy (Only HS)}, \textit{Novelty (All Sciences)}, \textit{Novelty (No CS)}, and \textit{Novelty (Only HS)}.

Impact is measured by the number of citations (\textit{\# Citations}) received by a paper from its year of publication up to 2019, the time of data extraction. Furthermore, we code dummy indicators for `big hits' contributions -- i.e., highly cited papers. Whether a paper is among the top 5\% or 10\% cited papers (\textit{Top 5\% Cited} and \textit{Top 10\% Cited}) is calculated with reference to other papers published in the same year and falling in the same WoS subject category.\footnote{A paper may fall into several WoS subject categories (two-character field WC). We consider all papers having at least one subject category in common with the focal paper to obtain the citation distribution and identify those papers that receive an exceptionally large number of citations.}

We consider a set of control variables to capture various characteristics of a focal paper. We control for the number of references made by a paper (\textit{\# References}) which might mechanically increase the likelihood of having new combinations and affect the number of received citations. In prior research, the number of authors has proven to be positively associated with both novelty and impact, hence we control for that (\textit{\# Authors}). The adoption of deep learning can indeed have an ambiguous effect on team size. Size may increase as new members are needed to manage the technology (at least in the early stages), but the technology may also automatize some tasks, thereby generating a replacement effect in the scientific workforce.\footnote{Although beyond the scope of our study, it interesting to note that the distribution of the number of authors of non-DL papers dominates the distribution of papers using DL. At the same time, we see that single-author contributions seem to have a lower probability of appearing within the DL category. We leave open the question as to whether DL acts as labour-saving or labour-augmenting technology in the scientific realm.} International collaborations may also be a source of novelty and impact, and may be instrumental to the adoption of DL. We proxy international collaboration by a dummy (\textit{International Collab.}) taking a value of 1 if there are at least two different countries in the authors' affiliations, 0 otherwise. For the same reason, we construct a dummy for private collaborations (\textit{Private Collab.}) taking a value of 1 if the paper has at least one non-university affiliation in the list. We consider the journal impact factor (\textit{JIF}) since high impact journals may be biased against novelty, on the one hand, but increase visibility and hence citations, on the other hand. We additionally control for the journal age (\textit{Journal Age}). Finally, we include a dummy indicating whether the paper provides a review or survey of existing literature (\textit{Survey}). A survey may in fact cover separate streams of research without really connecting them.\footnote{ \textit{Private Collab.} takes value of 1 if we detect in the authors' affiliation at least one of the acronyms present in the Wikipedia page: `List of legal entity types by country'. We use the SCImago Journal Rank to get the impact factor (\textit{JIF}) for each journal in each year. \textit{Journal Age} is calculated as the time elapsed from the date of the journal's creation and the publication year. \textit{Survey} takes value of 1 if we detect in the title of the paper the terms `Survey', `Overview' or `Review'.}

\paragraph{Descriptive statistics.} DL health papers represent about 1.3\% of our sample. Table \ref{tab:Descriptives} provides basic statistics to compare papers involving DL (first column) with papers not involving DL (second column). 

First consider novelty. The first row in Table \ref{tab:Descriptives} shows the \textit{Novelty Dummy (All Sciences)} based on journal references to all sciences. We see that DL documents exhibit novelty more often than others (36\% vs. 30\%). The second row, \textit{Novelty Dummy (No CS)}, excludes journal references to computer sciences. That reduces the novelty gap. Novelty of DL papers decreases by four percentage points, while the novelty of non-DL papers remains unchanged. The third row, \textit{Novelty Dummy (Only HS)}, restricts further to journal references to health sciences only. This results in a significant drop of novelty for both DL and non-DL papers. However, the reduction is larger for DL papers than for non-DL papers. This flips the pattern: only 21\% of DL papers provide novel recombinations within health sciences, which is less than 23\% of non-DL papers.  
 
Now consider the continuous novelty measures (rows four to six). In general, novelty distributions for DL and non-DL publications are highly skewed and heavily overlapping in such a way that no distribution dominates the other. The mean for \textit{Novelty (All Sciences)} of DL papers is somewhat higher than the mean of non-DL papers (0.81 vs. 0.74). However, excluding computer science journals and restricting further to health science journals, again leads to a drop in novelty, which is even stronger on DL papers. The standard deviations of novelty for DL papers are relatively small, from 60 to 70\% of the others. Taken together, these statistics point to two main patterns: DL papers offer as much or more novelty than non-DL papers through new recombinations including computer science journals; once journal references to computer sciences are excluded, DL papers bring on average less novelty than others.

\begin{table}[t!]\footnotesize
\centering  
\caption{Descriptive statistics of the variables}
  \begin{threeparttable}
  \label{tab:Descriptives} 
  \begin{tabular}{@{\extracolsep{5pt}} lccc} 
  \toprule
  & DL Papers & Non-DL Papers & Total \\  
  \midrule
  \addlinespace
  \textit{\underline{Re-combinatorial Novelty}} & & \\
  Novelty Dummy (All Sciences) & 36.43 & 30.32 & 30.40\\
  Novelty Dummy (No CS)    & 32.39 & 29.55 & 29.59\\
  Novelty Dummy (Only HS)     & 20.96 & 23.52 & 23.49\\
  \addlinespace
  Novelty (All Sciences) & 0/0.81 (2.39) & 0/0.74 (3.10) & 0/0.74 (3.09) \\ 
                                                    
  Novelty (No CS)     & 0/0.65 (2.12) & 0/0.71 (3.07) & 0/0.71 (3.06) \\ 
                                                    
  Novelty (Only HS)   & 0/0.37 (1.62) & 0/0.50 (2.40) & 0/0.50 (2.39) \\ 
  \addlinespace
  \textit{\underline{Scientific Impact}} & & \\
  Top 5\% Cited & 8.33 & 5.77 & 5.80 \\
  Top 10\% Cited & 15.68 & 11.38 & 11.43 \\
  \# Citations (Raw Count)  & 17/38.34 (114.43) & 18/35.48 (82.67) & 18/35.51 (83.15) \\ 
  Citations (Yearly Normalized) & 2.06/4.06 (8.16) & 2.08/3.75 (8.02) & 2.08/3.75 (8.02) \\ 
  \addlinespace
  \textit{\underline{Controls}} & & \\
  \# References & 40/45.92 (29.59) & 33/37.12 (25.66) & 33/37.23 (25.73) \\
  \# Authors & 4/4.07 (2.37) & 4/4.90 (3.50) & 4/4.89 (3.49) \\
  International Collab. & 26.21 & 23.02 & 23.06 \\
  Private Collab. & 6.80 & 7.09 & 7.09 \\
  JIF & 1.39/2.12 (2.06) & 1.73/2.42 (2.13) & 1.73/2.41 (2.13) \\
  Journal Age & 22/28.57 (26.07) & 33/38.47 (29.08) & 32/38.35 (29.06) \\
  Survey   & 0.72 & 0.78 & 0.77 \\
  
  \addlinespace
  \midrule
  Time Period & [2001 -- 2015] & [2001 -- 2015] & [2001 -- 2015] \\
  \# Scientific Fields & 46 & 48 & 48 \\
  \# Journals & 92 & 92 & 92 \\
  \# Papers & 4,560(1.28\%) & 351,477(98.72\%) & 356,037(100\%) \\
  \bottomrule
  \end{tabular}
  \begin{tablenotes}
  \footnotesize
  \item {\it Notes:} Binary indicators in [\%], for continuous measures [median/mean (s.d.)]. The statistics refer to the period used for the econometric analysis.
  \end{tablenotes}
 \end{threeparttable}
\end{table}

The second aspect of interest is the impact of publications that we measure by the number of citations. There are clear differences in the citation profiles of DL and non-DL papers. We do not see differences in terms of the median number of citations, but we do see a slightly larger mean for DL contributions (38 vs. 35). Interestingly, DL papers are characterized by a higher dispersion in citation performance (114 vs. 82). The most cited papers (top 5\% and 10\%) are more likely to be contributions involving DL. 

As far as controls are concerned, we see that DL papers have on average (i) more references, (ii) slightly smaller team size, (iii) more international collaborations, (iv) slightly less private collaborations, (v) and are published in younger and lower ranking journals.

\paragraph{Estimation methods.} We model three different types of outcomes: (i) binary indicators of novelty and impact, (ii) positive continuous measures of novelty, and (iii) positive discrete measures of impact (number of received citations). Each type of outcome requires a specific econometric setting.

All binary indicators are modeled with a Probit. Our continuous novelty measure is censored at zero, hence we use a Tobit model. The Tobit model introduces a latent (unobserved) outcome variable. Unlike the observed outcome, the latent outcome is unrestricted and may thus be modeled with a simple linear model with normally distributed error term. When the latent variable exceeds a certain threshold, in our case zero, the realization of the latent variable is assumed to equal the observed outcome. If the latent variable realizes below zero, the observation is assumed to be zero. The distributional assumption on the error term makes it possible to parameterize the model through Maximum Likelihood.

Citations are count data for which the Poisson and Negative Binomial models are natural candidates. In the Poisson model, expectation and variance necessarily coincide for each draw. The sum of Poisson distributed realizations is itself Poisson distributed, and this implies equality of mean and variance in the sample. In the Negative Binomial model the two moments may differ. Over-dispersion and conditional mean of the outcome variable much lower than its variance are the most common arguments to favor the Negative Binomial on Poisson model. Both empirical arguments hold in our case. Furthermore, prior empirical findings suggest that novelty may affect not only the expectation but also the variance of received citations \citep{fleming_2001, wang_2017}. Therefore, we use Negative Binomial to model mean and dispersion separately, each with a linear predictor incorporating our main left-hand side variables and controls. 

In all estimations, we include the control variables discussed above and a set of dummies to control for scientific fields and cohort effects. We proxy scientific fields through WoS categories (field WC). A paper may fall into several categories, hence we code dummy variables taking value of 1 for each category. Throughout, robust standard errors clustered at the journal-level are obtained via bootstrapping all journals.\footnote{Bootstrapping is essential in this context because there are group effects induced by the fact that some journals are more inclined to accept articles with a certain degree of novelty. Bootstrapping allows us to calculate the variance of estimates taking into account clustering at the journal level. Note that non-bootstrapped robust standard errors are much smaller, and all the DL effects discussed hereafter would be much stronger in terms of statistical significance. To get an idea of the difference, bootstrapped standard errors of DL coefficient estimates are always at least twice as large as the non-bootstrapped ones.}

\subsubsection{Results}

We begin with the results of how DL correlates with novelty, as shown in Table \ref{tab:NoveltyReg}. Columns 1--3 refer to Tobit regressions of our continuous measures of novelty, \textit{Novelty}. Columns 4--6 report Probit estimates of the binary novelty indicators, \textit{Novelty Dummy}. Recall that both measures are calculated on three different sets of journal references: all journal references (all sciences), journal references excluding computer sciences (no CS), and journal references only to health science journals (only HS).\footnote{The tables presented in this Section provide the main results, i.e. the results on the six outcome measures that always include the entire set of control variables and fixed effects. We have verified that the DL coefficient estimates are robust to various configurations of controls.} 

First, consider the Tobit regressions and focus on the first row that reports DL coefficient estimates for different novelty measures. When considering re-combinatorial novelty across all sciences (Column 1), the estimated DL coefficient is positive but insignificant. The exclusion of computer science references (Column 2) turns the coefficient into negative but keeps it insignificant. Restricting further to health sciences only (Column 3) magnifies the negative coefficient and turns it significantly below a 1\% significance level. We observe exactly the same pattern when considering the results of the Probit regression of the novelty dummy. How much does DL adoption change our expectations on associated re-combinatorial novelty within health sciences? We see that adopting DL decreases by 18.6\% the degree of novelty. In addition, the marginal effects for Probit (model of Column 6) tell us that, for the median observation, DL decreases by 0.031 the probability for an article to be novel (0.037 for the average observation).

In sum, controlling for confounding factors, deep learning is not significantly correlated with novel recombinations across the entire knowledge landscape, nor with novel recombinations involving anything but computer sciences. Deep learning is significantly and negatively correlated with novel recombinations within the health sciences though. These findings suggest that deep learning tends to be adopted within a `balancing strategy' in which the risk of DL adoption is counterbalanced by keeping the health science research landscape stable. Another way to interpret our result is that deep learning works mainly as a research tool for deepening existing knowledge structures rather than exploring/establishing new connections between more distant domains. This evidence is consistent with the idea of extending science while maintaining the advantages of conventional domain-level thinking \citep{uzzi_2013, wagner_2019}. Both explanations are likely to carry some truth. The first emphasizes agency in complex environments. The second points to the properties of DL as `scientific instrument'.

\begin{table}[h!]\footnotesize
\centering  
\caption{Novelty profile of deep learning publications}
\label{tab:NoveltyReg}
\vspace*{1em}
\begin{threeparttable}
\begin{tabular}{@{\extracolsep{5pt}}lcccccc} 
\toprule
& \multicolumn{3}{c}{\textit{Tobit: Novelty}}
& \multicolumn{3}{c}{\textit{Probit: Novelty Dummy}} \\ 
\cline{2-4}\cline{5-7} 

\\[-1.8ex] & \multicolumn{1}{c}{All Sciences}  &\multicolumn{1}{c}{No CS} &\multicolumn{1}{c}{Only HS} & \multicolumn{1}{c}{All Sciences}  &\multicolumn{1}{c}{No CS}&\multicolumn{1}{c}{Only HS} \\ 
\\[-1.8ex] & (1) & (2) & (3)& (4) & (5) & (6)\\ 
\midrule 
DL & 0.044  & -0.031  & -0.186$^{***}$ & 0.053  & -0.008  & -0.150$^{***}$ \\
   & (0.038) & (0.034) & (0.040) & (0.037) & (0.033) & (0.037) \\
   &    &    &    &    &    &    \\
\# References (log) & 1.046$^{***}$ & 1.050$^{***}$ & 1.029$^{***}$ & 0.878$^{***}$ & 0.879$^{***}$ & 0.843$^{***}$ \\
   & (0.033) & (0.033) & (0.033) & (0.026) & (0.026) & (0.023) \\
   &    &    &    &    &    &    \\
\# Authors (log) & 0.177$^{***}$ & 0.184$^{***}$ & 0.227$^{***}$ & 0.184$^{***}$ & 0.189$^{***}$ & 0.223$^{***}$ \\
   & (0.021) & (0.022) & (0.024) & (0.020) & (0.020) & (0.022) \\
   &    &    &    &    &    &    \\
International Collab. & -0.053$^{***}$ & -0.058$^{***}$ & -0.084$^{***}$ & -0.050$^{***}$ & -0.054$^{***}$ & -0.076$^{***}$ \\
   & (0.010) & (0.010) & (0.010) & (0.009) & (0.010) & (0.009) \\
   &    &    &    &    &    &    \\
Private Collab. & -0.004  & -0.004  & -0.027$^{*}$ & -0.007  & -0.008  & -0.026$^{**}$ \\
   & (0.012) & (0.012) & (0.014) & (0.012) & (0.013) & (0.013) \\
   &    &    &    &    &    &    \\
JIF & -0.026  & -0.024  & -0.017  & -0.025  & -0.024  & -0.017  \\
   & (0.019) & (0.019) & (0.021) & (0.017) & (0.017) & (0.018) \\
   &    &    &    &    &    &    \\
Journal Age (log) & -0.098  & -0.082  & -0.044  & -0.074  & -0.061  & -0.030  \\
   & (0.099) & (0.100) & (0.108) & (0.090) & (0.090) & (0.095) \\
   &    &    &    &    &    &    \\
Survey & 0.225$^{***}$ & 0.216$^{***}$ & 0.181$^{***}$ & 0.206$^{***}$ & 0.199$^{***}$ & 0.163$^{***}$ \\
   & (0.049) & (0.047) & (0.050) & (0.049) & (0.047) & (0.046) \\
   &    &    &    &    &    &    \\
\midrule
Log Likelihood & 
-263,098  &  -258,255  &  -221,241  &  -180,701  &  -178,639  &  -161,710 \\
$\chi^2$ [Null Model] &  96,074$^{***}$  &  94,950$^{***}$  &  77,374.6$^{***}$  &  75,936$^{***}$  &  75,187$^{***}$  &  64,730$^{***}$ \\
$\chi^2$ [w/o DL Model]  &  4.90$^{*}$  &  2.20   &  60.90$^{***}$  &  6.70$^{**}$  &  0.10   &  44.60$^{***}$ \\
\# Obs   &  356,037  &  356,037  &  356,037  &  356,037  &  356,037  &  356,037 \\
\bottomrule 
\end{tabular} 
\begin{tablenotes}
\footnotesize
\item {\it Notes:} This table reports coefficients of the effect of deep learning (\textit{DL}, dummy) on re-combinatorial novelty built by considering different knowledge landscapes. Bootstrapped (500 replications) standard errors clustered at the journal-level in parentheses: ***, ** and * indicate significance at the 1\%, 5\% and 10\% level, respectively. The effect of DL on the positive continuous novelty measure is estimated using a Tobit regression (Columns 1--3). The effect on the novelty dummy is estimated using a Probit (Columns 4--6). Each novelty measure is calculated on three different sets of journal references: `All Sciences' -- All cited journals, `No CS' -- All cited journals except for computer science journals, and `Only HS' -- Only citations to health science journals. Constant term, scientific field (WoS subject category) and time fixed effects are incorporated in all model specifications. Likelihood-ratio test are used to compare the goodness of fit of two statistical models: (i) null model against complete model; (ii) model without the $DL$ variable against the complete model.
\end{tablenotes}
\end{threeparttable}
\end{table}

\begin{table}[h!]\footnotesize
 \centering
  \caption{Impact profile of deep learning publications}
  \vspace*{1em}
    \scalebox{0.88}{
    \label{tab:ImpactReg}
  \vspace*{1em}
	\begin{threeparttable}
\begin{tabular}{@{\extracolsep{5pt}}llccc}
\toprule
\\[-1.8ex] & & \multicolumn{1}{c}{NegBin: \# Citations}&\multicolumn{1}{c}{Probit: Top 5\% Cited} &\multicolumn{1}{c}{Probit: Top 10\% Cited}\\
\\[-1.8ex]&  & (1) & (2) & (3)\\
\midrule

\textit{Panel A: Mean}
  & DL & 0.101$^{**}$ & 0.147$^{***}$ & 0.155$^{***}$ \\
   &   & (0.040) & (0.041) & (0.043) \\
   &    &    &    &    \\
   & Novelty (All Sciences) & 0.153$^{***}$ & 0.200$^{***}$ & 0.191$^{***}$ \\
   &   & (0.023) & (0.016) & (0.015) \\
   &    &    &    &    \\
   & \# References (log) & 0.491$^{***}$ & 0.429$^{***}$ & 0.477$^{***}$ \\
   &   & (0.064) & (0.075) & (0.062) \\
   &    &    &    &    \\
   & \# Authors (log) & 0.237$^{***}$ & 0.166$^{***}$ & 0.194$^{***}$ \\
   &   & (0.026) & (0.039) & (0.036) \\
   &    &    &    &    \\
   & International Collab. & 0.064$^{***}$ & 0.083$^{***}$ & 0.085$^{***}$ \\
   &   & (0.013) & (0.014) & (0.013) \\
   &    &    &    &    \\
   & Private Collab. & -0.029$^{*}$ & -0.027  & -0.034$^{**}$ \\
   &   & (0.015) & (0.018) & (0.015) \\
   &    &    &    &    \\
   & JIF & 0.205$^{***}$ & 0.167$^{***}$ & 0.179$^{***}$ \\
   &   & (0.022) & (0.017) & (0.018) \\
   &    &    &    &    \\
   & Journal Age (log) & 0.050  & -0.066  & -0.048  \\
   &   & (0.036) & (0.086) & (0.079) \\
   &    &    &    &    \\
   & Survey & 0.541$^{***}$ & 0.667$^{***}$ & 0.627$^{***}$ \\
   &   & (0.060) & (0.054) & (0.049) \\
   &    &    &    &    \\
\midrule
                                                      \textit{Panel B: Dispersion}
   & DL & 0.136$^{***}$ &   &   \\
   &   & (0.051) &  &  \\
   &    &    &    &    \\
   & Novelty (All Sciences) & 0.093$^{***}$ &   &   \\
   &   & (0.017) &  &  \\
   &    &    &    &    \\
   & \# References (log) & -0.496$^{***}$ &   &   \\
   &   & (0.038) &  &  \\
   &    &    &    &    \\
   & \# Authors (log) & -0.213$^{***}$ &   &   \\
   &   & (0.044) &  &  \\
   &    &    &    &    \\
   & JIF & 0.040  &   &   \\
   &   & (0.031) &  &  \\
   &    &    &    &    \\
   & Journal Age (log) & -0.118$^{***}$ &   &   \\
   &   & (0.029) &  &  \\
   &    &    &    &    \\
\midrule
Log Likelihood & &
-1,519,720  &  -69,222  &  -110,788 \\
$\chi^2$ [Null Model] & & 318,463$^{***}$  &  19,317$^{***}$  &  31,564$^{***}$ \\
$\chi^2$ [w/o DL Model]  & & 8.70$^{***}$   &  24.80$^{***}$  &  40.00$^{***}$ \\
\# Obs   & & 356,037  &  356,037  &  356,037 \\
\bottomrule
\end{tabular}

\begin{tablenotes}
 \footnotesize
 \item {\it Notes:} This table reports coefficients of the effect of deep learning (\textit{DL}, dummy) on scientific impact proxied by the number of received citations (Column 1) and `big hits' (Columns 2 and 3). Bootstrapped (500 replications) standard errors clustered at the journal-level in parentheses: ***, ** and * indicate significance at the 1\%, 5\% and 10\% level, respectively. The effect of DL on the citation count is estimated using a Negative Binomial regression. Estimates for the expectation and variance are reported in Panel A and B, respectively. The effects on the binary indicators is estimated using a Probit. Constant term, scientific field (WoS subject category) and time fixed effects are incorporated in all model specifications. Likelihood-ratio test are used to compare the goodness of fit of two statistical models: (i) null model against complete model; (ii) model without the $DL$ variable against the complete model.
\end{tablenotes}
 \end{threeparttable}
 }
\end{table}

Estimates of control variables echo previous research. Larger teams are associated with more novelty \citep{fleming_2007, lee_2015}. International collaborations are negatively associated with novelty \citep{wagner_2019}. Mechanically, the chances of providing a new combination of journal references increase with the number of references \citep{wang_2017}. Literature reviews also tend to draw from a wider range of sources leading to novel combinations of references. We find a negative effect of private collaboration. On the other hand, the journal age and its impact factor  seem to play no role for novelty. 

We now turn to how DL correlates with impact, as shown in Table \ref{tab:ImpactReg}. Column 1 presents the results of the Negative Binomial regression of citation counts. Here, the mean and dispersion parameters may vary with various right-hand side factors.\footnote{We excluded dummy variables other than $DL$ to model the dispersion of citations because these variables caused problems of convergence of maximum likelihood estimator. In modeling the dispersion, we also tried simpler specifications by progressively incorporating a few variables at a time. The results are in line with those presented in this document.} We find that DL adoption positively and significantly affects the number of citations received, both in terms of expectation and variance. The same applies to novelty. These two effects are akin to the `high-risk/high-gain' profile that characterizes the adoption of emerging technologies and novel research \citep{wang_2017}.\footnote{In the main analysis at hand we enter the continuous novelty measure on all sciences, but we have verified that the results are robust to the various novelty measures discussed in this paper. We also find, not shown here, that novelty and DL act rather independently of each other. Estimates hardly change when one or the other variable is removed from the model, implying that DL and novelty add to each other when it comes to impact. This is perhaps not too surprising if we consider that the adoption of DL is often a specific type of novelty per se.} Compared with non-DL papers, ceteris paribus, DL papers receive on average 10.32\% more citations. The citation count expectation increases in median by 6.01 citations if DL is used.  The dispersion of the citation distribution is 19.57\% higher for DL papers than non-DL papers. \\

As for the controls, the number of authors is positively related to impact \citep{lee_2015}; a new insight here is that it also reduces impact variation. International collaborations also increase citation expectations \citep{glanzel_2001}. Publishing in a high impact factor journal further increases the average number of citations, being this through increased visibility, signaling or selection effects. Moreover, surveys and other papers with many references tend to attract more citations. Finally, we find a negative effect, albeit not particularly significant, between private collaborations and scientific impact.

The Probit regressions used to model the probability that a paper falls in the right-tail (top 5\% or 10\%) of the year--field citation distribution confirm the above results. The marginal effects suggest that, in median, deep learning papers have a 0.019 higher probability to fall in the 10\% of the most cited documents (0.027 in mean), and a 0.009 higher probability to fall in the top 5\% (0.014 in mean). 

In sum, the econometric analysis shows that publications involving deep learning methods exhibit citation patterns consistent with the `high-risk/high-gain' profile associated with breakthrough research. DL seems to play a positive effect on citation count, on average; however, DL papers have a higher variance in citation performance than do other health counterparts, confirming their risky profile. As such, research involving deep learning has a high potential for major impact, on the one hand, but also carries a higher uncertainty of having impact, on the other. As discussed in Section \ref{sec:novelty_impact}, uncertainty (which could in turn affect success and scientific quality) may find several explanations, including the challenge of integrating the scientific instrument into existing scientific practices, the capacity to extract the true potential from the instrument and not adopt it simply because `everybody does' and, to a lesser extent, possible social resistance especially in sensitive domains as some areas of health sciences may be. We do not find any effect of deep learning on the degree re-combinatorial novelty on the entire knowledge landscape, and a systematic negative effect on the novelty intra-field. These results suggest that deep learning acts mainly as a research tool to support and deepen already formalized and well-defined research in the health sciences community.

\subsubsection{Robustness analysis}
\label{sec:robcheck}

Our results are robust across a wide range of additional tests. Here below, we discuss the main exercises that were carried out. Tables and further material can be found in the Appendix \ref{appendix:robcheck}. 

First of all, we excluded all articles that fall into the WoS `Neurosciences' category. As pointed out throughout the document, this domain can be potentially problematic in that some terms (neural network in primis) may not necessarily refer to artificial intelligence but rather to human intelligence and biological brain. The sample drops by about 30\% and the number of DL articles almost halves. However, the main results are consistent when replicating the estimates on the reduced sample. 

Second, we excluded all articles that contain exclusively the terms `neural\_network' and `neural\_networks' in their title, keywords or abstract.
Keep in mind that an article may still contain a term such as `artificial\_neural\_network' or `convolutional\_neural\_network' which should now refer unambiguously to artificial intelligence. In this case, neuroscience papers may be part of the sample but we make sure that they explicitly deals with deep learning. This restriction is severe insofar as the number of DL articles decreases by more than 70\%. Yet our results are robust to this constraint. The only pattern we lose is the significant effect of DL on the dispersion of citations received. A plausible explanation is due to the smaller sample size and the consequent drop in the variance of the number of citations, as suggested by the descriptive statistics reported in the Appendix \ref{appendix:robcheck}.

The third exercise consists of a different econometric approach. Instead of regression analysis we compared each DL papers with `twin non-DL papers'. More precisely, the empirical strategy considers the adoption of DL as a `treatment', hence we employ exact matching and 1:1 nearest neighbor matching on propensity scores (PSM) to select an appropriate control group of untreated papers. Exact matching is performed considering Web of Science categories, publication year, and journal (i.e., we compare a DL article in terms of novelty and impact with a article belonging to the same domain(s), published in the same year and on the same journal). As for PSM, we obtain the propensity scores associated with the binary treatment via the estimation of the Probit model (or selection equation) containing the original set of variables. Although we refrain from giving any causal interpretation, the average treatment effects (ATT) for the selected variables bring further support to our results.

\cite{fontana_2020} show that different novelty indicators are often inconsistent with each other as they return different sets of novel contributions. Thus, in the forth exercise, we implemented the indicator proposed in \cite{uzzi_2013} to define the `atypical' (novelty/conventionality) quadrant. For the period 2001--2015, we computed pairwise combinations of cited references of each article. We then compared the observed frequency with the frequency from a randomized citation network that preserves the realistic aspects of the data and its network structure (i.e., same number of references to prior work, the same number of citations from subsequent papers, and the same distribution of these citations over time). We obtained a distribution of $z$-scores for each document and calculated two statistics: the median $z$-score and the 10th percentile $z$-score of that article. High conventionality is coded as 1 if the median $z$-score value is above or equal to the overall median, 0 otherwise. High novelty is coded as 1 if the 10th percentile $z$-score value is below the median value of the overall 10th percentiles. We combined these two binary variables into a nominal variable with four categories, namely: high-conventionality/high-novelty (HC--HN); high-conventionality/low-novelty (HC--LN); low-conventionality/high-novelty (LC--HN); and low-conventionality/low-novelty (LC--LN).\footnote{Although we replicated the \cite{uzzi_2013} method, we differ in two respects. Firstly, we simulated 50 random citation networks (instead of 10) to calculate the $z$-scores. This substantially increases the computational burden, but reduces the likelihood that an article makes combinations that do not exist in the simulated networks; to notice that a combination that does not exist in the simulation has no standard deviation and therefore its score $z$-score cannot be defined. Secondly, we consider the high novelty threshold as dependent on the citation network instead of using a default value (0 in the original paper). This allows us to set the share of high-novel papers to 50\%. Considering a threshold equal to 0, we would get roughly 70\%. The discrepancy between our shares and those in Uzzi et al. is simply due to the size of the sample on which the indicators are built (1 million vs. 18 million articles).} 

The four categories are employed in a Multinomial Logistic regressions. Category LC--HN acts as the reference category for all models. This category represents the group of papers characterized by the most atypical combination profiles. Within the health sciences knowledge landscape, we find that deep learning articles are more likely to draw on highly conventional mixtures of knowledge. Ceteris paribus, the estimates suggest that even when DL injects some highly (field-specific) unusual combinations, it does so primarily in an exceptionally conventional knowledge space.

\section{Discussion and conclusion}
\label{sec:concl}

We developed this paper with the aim of studying the diffusion and impact of the DL in the scientific system. We proceeded in three steps. In the first part, we developed a list of DL search terms through Natural Language Processing of scientific abstracts from arXiv.org. We used a word embedding algorithm that projects words into a mathematical space that reflects the semantic structure of the abstracts. Words cluster in that mathematical space by `topic', which allows us to define the DL perimeter. The second part of this paper documents the diffusion of DL research in science. Our sample includes Web of Science documents across all sciences identified through our DL search terms. Starting from a low level in the early 2000s, DL research activity is growing exponentially in almost all sciences and worldwide. Asia and Eastern Europe are focusing their research mainly on DL methodologies, while Western Europe and North America seem to have a `relative advantage' in application areas such as life sciences and biomedicine. Scientific co-classification of articles suggests that the diffusion of DL into application domains began with an interdisciplinary effort involving the computer sciences, breaking its way into `pure' field-specific research within the various application domains. The third part deals with the consequences of DL adoption on scientific development, making the case for the health sciences. We find that the adoption of DL is negatively correlated with re-combinatorial novelty. On the other hand, it is positively correlated with the expectation and dispersion of citations received, increasing the likelihood for a contribution to become an influential `big hit'. 

Conceptually we considered scientific development as a re-combinatorial process in which existing pieces of knowledge are recombined to create new knowledge. This continues perpetually in a dynamic knowledge landscape. A traditional image of science is one in which the knowledge landscape is made up of islands, (sub)-disciplines or scientific fields, where most recombination takes place. The islands reflect the structure of nature but also the necessity for a scientific mind to organize the complexity of the world. The scientist is the sailor. The aim of the sailor is to navigate through the islands, figure out their structure, and explore the surrounding landscape. So she can stay in the `comfort zone' by further deepening her knowledge of one (or neighboring) island(s). Or she can sail to more distant islands and connect new areas of the landscape. Both actions enrich the knowledge space, one digging into a well-formalized knowledge structures, the other reshaping and rearranging the landscape. Our findings suggest that, at least as it is used today, deep learning -- the boat or the compass to keep the analogy -- seems to be more in line with the first option. Yet the possibility of discovering new valuable things about the already known islands is not obvious. This may result from the fact that the sailor needs to know how to use the new equipment but also from the fact that the crew -- the society and/or the rest of the scientific community -- have confidence and trust in it.

Our findings lead us to take a moderate position in the recent debate on how DL affects the development of knowledge. A passing fad in science? We don't think so. DL does not (yet?) work as an autopilot to navigate a sea of knowledge, but stands as an extremely powerful and versatile research tool that impacts knowledge creation in measurable ways.

With the results of our study in mind, let us conclude with some considerations for policy and management. The evidence that deep learning owns the characteristics of a general purpose invention in the method of invention implies that its socio-economic impact in terms of innovation, growth, productivity gains and societal well-being will be pervasive. Impact pathways when it comes to AI are very diverse and often indirect. The new research paradigm creates huge opportunities but also comes with some challenges. 

First, intelligent machines as input in the research production process question the organization and management of science. Deep learning may trigger short-term substitution towards capital and away from highly skilled labour in the knowledge production process. Whether or not such a substitution effect is occurring is questionable and clearly requires further empirical investigation. At the same time, it is undeniable that the arrival of deep learning as a research tool jeopardizes a wide range of research tasks, either by reducing the cost of performing these tasks or by surpassing the performance of human scientists in performing them. Some tasks within the occupation may be suitable for automation while others may not, and the overall effects on employment in science are very complex. Research-oriented organizations need to better understand the collection of tasks that make up the job of their scientists, how to coordinate these tasks, the strengths and weaknesses of humans (H) and machines (M), to finally unleash the benefits from H + M cooperation.

Although we cannot rule out the possibility that deep learning has the potential to ease researchers in predicting which combinations of distant knowledge in the landscape will produce useful new discoveries, our findings suggest that the technology is not used in that way. Deep learning seems to be adopted in a scientific field by maintaining the existing knowledge structure relatively stable. Therefore, the full potential of deep learning (and its future development) could be achieved by further spanning the boundaries between scientific areas. Bringing together expertise and knowledge from various domains could help to see blind spots and opportunities in the knowledge landscape. The concepts of `knowledge communities' and `communities of practice' seem particularly apt in this context. Although communities often self-organize and self-sustain, they may also benefit from policy endorsement. It seems crucial to us to develop institutions and a policy environment that is conductive to enhancing dialogue and cross-fertilization between communities. This can be achieved, for instance, through the reinforcement of horizontal (intra-field) but also vertical (inter-field) knowledge management. Digital platforms and knowledge hubs should be complemented by physical `collaborative spaces' where the tacit knowledge of different communities can be transferred face-to-face, documented and made accessible for later use. Another standard instrument is research funding, which should not target individual areas but mainly research `priorities' (i.e., fighting a given disease) involving different communities framing research questions together. 

However, promoting international collaboration between communities may pose some challenges in terms of governance and data ownership. Data is a polymorphous category, so the standards, principles and rules governing the various types of data are not homogeneous across countries. No data, no AI. Bad data, bad AI. This opens the question of how data should be generated/used in compliance with different regulations, and also how the value of the data should be distributed (see the discussion in \citealp{savona_2019}).


The diffusion of deep learning, also as a research tool, can be self-sustaining if and only if there is social acceptance -- i.e., if the crew trusts the captain and the equipment. Several DL applications are innovations that can change almost every aspect of our daily lives. These social innovations can have negative and unintended consequences in terms of security, privacy and social equity \citep{oneil_2016, agrawal_book_2018, furman_2019}. People will no longer tolerate being excluded from the debate. Here again the scientific and policy community have a key role to play. Both parties may improve the channelling of scientific evidence into the public debate and fight the dangers of fake news, particularly common when it comes to AI, robotics and automation. On the one hand, policy can promote communication by setting the right, often intrinsic, incentives to encourage as many scientists as possible to engage with various segments of the public. However, communicating science to non-scientific audiences might be difficult since it requires a different approach from communicating science to scientific audiences. Scientists need to understand how to detach the layers of scientific complexity of their research in order to deliver some clear messages to the public. These messages should include both potential impacts and ethical issues. `Listening mechanisms' can also be used to bring citizens' knowledge, expectations, and imaginaries about intelligent machines. There are a variety of forms ranging from in-depth interviews, material deliberations, to citizen science. In the context of AI, we believe that citizen science may probably bring the greatest benefits to both the public and the scientific system. The non-professional involvement of volunteers in the scientific process, either in the data collection phase or in other stages of research, offers great opportunities for the public to become familiar with the technology but also great opportunities for researchers to improve their results \citep{bonney_2014}. To get a sense of impact, the research participation by more than 300,000 volunteers over 1 year could provide nearly 33 million classifications of subcellular localization patterns \citep{sullivan_2018}. But accountable institutional mechanisms are a precondition for guaranteeing trust between scientists and the public, thus ensuring continuity in their relationship. For instance, the results and the process used for arriving at these results should be open to scrutiny. Policy should promote feedback activities to get back to citizens and explain how their inputs were used for research aims, reconcile conflicting values and objectives, and put in place collective intelligence mechanisms to help citizens develop a systemic understanding of the future implications of technological progress and make better consensus decision-making; very much in line with the notion of `decisions 2.0' \citep{bonabeau_2009}. Finally, we fully embrace the conception of `boundary organisations' specifically designed to deal with socio-economic transformations in the digital age. These organisations would sit at the intersection of scientific and political spheres and allow scientists and policy-makers to maintain a constant dialogue with each other. 

Although the AI revolution is the subject under scrutiny, ironically that revolution offers the tools that can bring the greatest potential for a radical transformation in the interaction between the public, the scientific community and the policy environment. These interactions, if conducted fruitfully enough, will give a boost to the human attempt to better understand the largest mystery we know: the origin and function of the world and our place in it. That's science! \\

\footnotesize{
\noindent \textbf{Acknowledgement}. Earlier versions of this paper were presented at the INNOVA MEASURE III Expert Workshop `Brainstorming in Ispra'; 7th European Conference on Corporate R\&D and Innovation, Seville (Spain); EMAEE 2019 `Economics, Governance and Management of AI, Robots and Digital Transformations', Brighton (UK); OLKC 2019 `The Human Side of Innovation -- Understanding the Role of Interpersonal Relations in an Increasingly Digitised Workplace'; CAGE-NESTA Workshop on `Data Science for the Economics of Science, Technology and Innovation', London (UK); and the Workshop Series on `The Economics and Management of AI Technologies', Copenhagen (Denmark) and Strasbourg (France). The authors thank seminar participants and in particular Giacomo Damioli, Mirko Draca, Daniel S. Hain, Bj\"{o}rn Jindra, Bertrand Koebel, Roman Jurowetzki, Patrick Llerena, Juan Mateos-Garcia, Simone Vannuccini, Daniel Vertesy. Stefano Bianchini gratefully acknowledges the financial support of the CNRS through the MITI interdisciplinary programs [reference: Artificial intelligence in the science system (ARISE)] and the French National Research Agency [reference: DInnAMICS -ANR-18-CE26-0017-01].}


\bibliography{biblio}

\newpage

\appendix

\section{From word embeddings to search terms}
\label{appendix:embedding}

This Appendix complements Section \ref{sec:DLIdentification} by adding technical details on the learning of search terms for data retrieval. Source data and codes are fully accessible upon request.

\subsection{Preparation of the training data}

The training data is a bulk download of article abstracts from arXiv.org via its API provided by the R package `aRxiv' \citep{ram_2019}. We obtained in total 197,439 documents submitted between 1990 (2 documents) and 2019 (16,533 documents at the time of downloading -- July; 35,807 in 2018) in the `Computer Science', `Mathematics' and `Statistics' sections of arXiv.org. Preprocessing entails removing all abstracts with less than 15 words; a pre-defined set of stop words; all words occurring less than 5 times in the corpus. In addition, we paste unigrams into bi-grams depending on the frequency of co-occurence. Following \cite{mikolov_2013a, mikolov_2013b} a bi-gram is created when the score of the two words, $w_i$ and $w_j$, pass a given threshold. The score is calculated as follows: $score(w_i,w_j) = \frac{count(w_i,w_j) - \delta}{count(w_i) \cdot count(w_j)}$, where $\delta$ is used as a discounting coefficient and prevents too many bi-grams consisting of very infrequent words to be formed. We choose a threshold of 50 to increase the number of bi-grams generated (default is 100). 

After preprocessing, the training data includes 14,458,777 words from a vocabulary of size 87,990. This leads to a weight matrix of dimension  45,050,880 (87,990 $\times$ 512 dimensions).

\subsection{Estimation of word representations}

We estimate word representations with the continuous Skip-Gram model introduced in \cite{mikolov_2013a, mikolov_2013b}. The Skip-Gram model is one specific variant of a set of word embedding algorithms that have become popular under the label of Word2Vec. 

We use negative sampling. In `old school' parlance, this is essentially a Logit model. The binary dependent variable indicates whether or not two terms are close in the text corpus, at distance $c$. For each observed neighboring term pair (success), one adds $k$ `negative samples' (failures). The scalar product of word representations enters the model as the linear predictor. Sequential processing is achieved through stochastic gradient descent. 

The results presented along the main text have been obtained with the following parameter settings. The main free parameter is the dimensionality of the dense word representation, which we set to 512 dimensions.\footnote{We tried several dimensions to represent dense representation: 256, 300, 512 and 1,024. Our choice was guided by the results of the $k$-mean clustering; we opted for the dimension for which the DL cluster was best defined.} We define a context window (distance $c$) of 7 words from both sides around the target. For each observed neighboring term pair, we draw $k=15$ negative examples. A negative example is obtained by replacing one word of the observed neighboring terms by another word from the vocabulary that is drawn randomly with probability proportional to its frequency -- i.e., $P(w_i) = \frac{f(w_i)^{3/4}}{\Sigma_{j=0}^n (f(w_j)^{3/4})}$, which is close to draws uniformly at random. Further, we make use of sub-sampling by specifying 30 epochs, so that the whole dataset is passed 30 times through the network.



\subsection{Word clustering}

Estimation results -- i.e., estimated word embeddings -- serve as input to a cluster analysis. Term clusters are identified with the $k$-means clustering procedure. We used the gap statistics to determine the optimal number of clusters. The most frequent $n$-grams for the 22 identified clusters are reported in Tables \ref{tab:embedding1} and \ref{tab:embedding2}.

\begin{landscape}
\begin{table}[h]\footnotesize
\renewcommand{\thetable}{\arabic{table}a}
\caption{Word embedding obtained via Word2Vec [arXiv.org sample]}
 \centering  
    \scalebox{0.6}{
	\begin{threeparttable}
  \label{tab:embedding1} 
\begin{tabular}{@{\extracolsep{1pt}} lllllllllll} 
\toprule
Cluster 1 (Deep   Learning)      & Cluster 2                   & Cluster 3                 & Cluster 4                 & Cluster 5                     & Cluster 6                  & Cluster 7                & Cluster 8                & Cluster 9                  & Cluster 10              & Cluster 11             \\
\midrule
\addlinespace
\textit{\underline{n-grams}}                        \\
adversarial\_examples            & alternating\_direction      & audio\_visual             & adjacency\_matrix         & action\_spaces                & asymptotic\_properties     & alzheimer's\_disease     & alice\_bob               & achieves\_art              & algorithm\_computes     & 2d\_3d                 \\
artificial\_neural               & batch\_size                 & automatic\_speech         & analysis\_pca             & actor\_critic                 & bayesian\_inference        & association\_studies     & base\_station            & benchmark\_datasets        & approximation\_ratio    & 3d\_reconstruction     \\
\st{attention\_mechanism}             & belief\_propagation         & bag\_words                & community\_detection      & agent\_learns                 & central\_limit             & brain\_activity          & brute\_force             & compared\_art              & asymptotically\_optimal & 3d\_shape              \\
convolutional\_neural            & boundary\_conditions        & character\_level          & component\_analysis       & atari\_games                  & computationally\_efficient & brain\_regions           & coding\_scheme           & compared\_traditional      & bipartite\_graph        & action\_recognition    \\
convolutional\_neural\_network   & combinatorial\_optimization & collaborative\_filtering  & compressed\_sensing       & contextual\_bandit            & confidence\_intervals      & brain\_tumor             & cognitive\_radio         & computational\_efficiency  & bipartite\_graphs       & autonomous\_driving    \\
\st{data\_augmentation}               & computational\_cost         & context\_aware            & compressive\_sensing      & contextual\_bandits           & covariance\_matrix         & breast\_cancer           & communication\_channels  & conduct\_experiments       & bounded\_degree         & bounding\_box          \\
deep\_convolutional              & constrained\_optimization   & contextual\_information   & covariance\_matrices      & control\_policies             & cross\_validation          & causal\_relationships    & communication\_protocols & current\_art               & competitive\_ratio      & bounding\_boxes        \\
deep\_learning                   & convergence\_rate           & cross\_lingual            & dimension\_reduction      & decision\_processes           & density\_estimation        & clinical\_practice       & elliptic\_curve          & demonstrate\_effectiveness & connected\_components   & computer\_vision       \\
deep\_neural                     & convergence\_rates          & domain\_specific          & dimensionality\_reduction & deep\_q\_learning             & expectation\_maximization  & clinical\_trial          & encoding\_decoding       & demonstrate\_efficacy      & constant\_factor        & facial\_expression     \\
deep\_neural\_network            & convex\_optimization        & emotion\_recognition      & distance\_metric          & deep\_reinforcement           & gaussian\_process          & clinical\_trials         & encryption\_scheme       & demonstrate\_superiority   & convex\_hull            & features\_extracted    \\
\st{domain\_adaptation}               & differential\_equations     & fake\_news                & euclidean\_distance       & deep\_reinforcement\_learning & gaussian\_processes        & computed\_tomography     & error\_correcting        & effectiveness\_proposed    & directed\_graphs        & ground\_truth          \\
encoder\_decoder                 & divide\_conquer             & feature\_engineering      & euclidean\_space          & expected\_reward              & hidden\_markov             & computer\_aided          & error\_correction        & empirical\_evaluation      & edge\_coloring          & human\_pose            \\
\st{feature\_extraction}              & dynamic\_programming        & image\_captioning         & feature\_selection        & experience\_replay            & hypothesis\_testing        & computer\_assisted       & fusion\_center           & error\_rate                & epsilon\_0              & image\_denoising       \\
\st{feature\_maps}                    & dynamical\_systems          & information\_retrieval    & fourier\_transform        & exploration\_exploitation     & importance\_sampling       & cross\_sectional         & hash\_function           & error\_rates               & fixed\_parameter        & image\_patches         \\
\st{feed\_forward}                    & easy\_implement             & knowledge\_base           & gaussian\_mixture         & game\_playing                 & joint\_distribution        & ct\_images               & hash\_functions          & evaluation\_metrics        & induced\_subgraph       & image\_registration    \\
\st{fine\_tuning}                     & empirical\_risk             & latent\_dirichlet         & gaussian\_noise           & heuristic\_search             & latent\_variable           & disease\_progression     & hoc\_networks            & experimental\_evaluation   & log\_log                & image\_restoration     \\
generative\_adversarial          & evolutionary\_algorithms    & link\_prediction          & graph\_laplacian          & imitation\_learning           & latent\_variables          & dna\_sequences           & homomorphic\_encryption  & experimental\_results      & lower\_bound            & image\_retrieval       \\
generative\_adversarial\_network & faster\_convergence         & low\_resource             & hilbert\_space            & imperfect\_information        & linear\_regression         & electronic\_health       & information\_leakage     & experiments\_conducted     & lower\_bounds           & image\_super           \\
\st{hand\_crafted}                    & finite\_element             & machine\_translation      & ill\_posed                & infinite\_horizon             & log\_likelihood            & false\_negative          & information\_theoretic   & experiments\_synthetic     & lower\_upper            & instance\_segmentation \\
\st{latent\_space}                    & forward\_backward           & manually\_annotated       & linear\_combination       & intelligent\_agents           & logistic\_regression       & functional\_magnetic     & key\_exchange            & extensive\_experimental    & maximum\_degree         & low\_resolution        \\
long\_short\_term\_memory        & generalization\_error       & named\_entity             & linear\_combinations      & inverse\_reinforcement        & markov\_chain              & gene\_expression         & leader\_election         & extensive\_experiments     & minimum\_degree         & multi\_modal           \\
\st{loss\_function}                   & gradient\_descent           & natural\_language         & low\_dimensional          & markov\_decision              & markov\_chains             & genome\_wide             & min\_entropy             & f1\_score                  & np\_complete            & object\_detection      \\
\st{loss\_functions}                  & linear\_programming         & news\_articles            & low\_rank                 & markov\_decision\_process     & maximum\_likelihood        & gold\_standard           & multi\_hop               & https\_github.com          & np\_hard                & object\_recognition    \\
\st{multi\_label}                     & local\_minima               & nlp\_tasks                & matrix\_completion        & motion\_planning              & monte\_carlo               & heart\_rate              & multi\_party             & magnitude\_faster          & omega\_log              & object\_tracking       \\
\st{multi\_layer}                     & min\_max                    & processing\_nlp           & matrix\_factorization     & multi\_agent                  & numerical\_examples        & human\_brain             & noise\_ratio             & outperform\_art            & perfect\_matching       & optical\_flow          \\
networks\_cnns                  & mini\_batch                 & question\_answering       & means\_clustering         & multi\_armed                  & parameter\_estimation      & imaging\_mri             & physical\_layer          & outperforms\_art           & planar\_graph           & person\_identification \\
neural\_network                  & mixed\_integer              & relation\_extraction      & nearest\_neighbor         & partially\_observable         & posterior\_distribution    & low\_dose                & pseudo\_random           & outperforms\_existing      & planar\_graphs          & pixel\_level           \\
neural\_networks                 & numerical\_experiments      & semantic\_similarity      & nearest\_neighbors        & policy\_gradient              & probability\_density       & lung\_cancer             & public\_key              & pascal\_voc                & polynomial\_time        & pose\_estimation       \\
\st{pre\_trained}                     & objective\_function         & sentence\_level           & nuclear\_norm             & policy\_iteration             & probability\_distributions & magnetic\_resonance      & quantum\_computers       & random\_forest             & regular\_graphs         & post\_processing       \\
recurrent\_neural                & optimal\_transport          & sentiment\_analysis       & positive\_definite        & processes\_mdps               & random\_fields             & medical\_diagnosis       & rate\_distortion         & significant\_improvement   & running\_time           & remote\_sensing        \\
recurrent\_neural\_network       & partial\_differential       & sentiment\_classification & principal\_component      & regret\_bound                 & random\_variable           & medical\_image           & ratio\_snr               & significant\_improvements  & shortest\_path          & semantic\_segmentation \\
\st{semi\_supervised}                 & particle\_swarm             & sequence\_sequence        & principal\_components     & regret\_bounds                & random\_variables          & medical\_images          & secret\_key              & significantly\_improve     & spanning\_tree          & spatial\_resolution    \\
\st{short\_term}                      & primal\_dual                & similarity\_measures      & reproducing\_kernel       & reinforcement\_learning       & rate\_convergence          & medical\_imaging         & secret\_sharing          & significantly\_improved    & spanning\_trees         & spatial\_temporal      \\
\st{shot\_learning}                   & sample\_complexity          & speech\_recognition       & signal\_noise             & reinforcement\_learning       & sample\_size               & presence\_absence        & secure\_communication    & significantly\_improves    & sqrt\_log               & spatio\_temporal       \\
\st{source\_target}                   & statistical\_physics        & text\_mining              & signal\_processing        & reward\_function              & simulation\_studies        & resonance\_imaging       & security\_protocols      & significantly\_outperforms & undirected\_graph       & style\_transfer        \\
\st{spiking\_neural}                  & stochastic\_gradient        & topic\_modeling           & sparse\_coding            & sequential\_decision          & smoothing                  & risk\_factors            & sensor\_networks         & simulated\_real            & upper\_bound            & super\_resolution      \\
\st{supervised\_learning}             & strongly\_convex            & web\_pages                & spectral\_clustering      & temporal\_difference          & squared\_error             & rna\_seq                 & sensor\_nodes            & source\_code               & upper\_bounds           & video\_frames          \\
\st{target\_domain}                   & theoretical\_guarantees     & word\_embedding           & subspace\_clustering      & thompson\_sampling            & time\_series               & sensitivity\_specificity & source\_destination      & superior\_performance      & upper\_lower            & video\_sequences       \\
variational\_autoencoder         & variance\_reduction         & word\_embeddings          & support\_vector           & time\_horizon                 & time\_varying              & skin\_lesion             & wireless\_communication  & synthetic\_real            & vertex\_cover           &                        \\
\st{weakly\_supervised}               & variational\_inference      & word\_vectors             & total\_variation          & upper\_confidence             & variable\_selection        & white\_matter            & wireless\_sensor         & times\_faster              & vertices\_edges         &                        \\
\addlinespace
\textit{\underline{Unigrams}}                        \\
architecture                     & approximation               & annotation                & clustering                & action                        & asymptotic                 & biological               & adversary                & accuracy                   & bound                   & 3d                     \\
architectures                    & approximations              & annotations               & coefficients              & actions                       & bayesian                   & brain                    & bit                      & achieve                    & bounded                 & color                  \\
classification                   & convergence                 & corpus                    & decomposition             & actor                         & conditional                & cancer                   & bits                     & achieved                   & bounds                  & depth                  \\
classifier                       & convex                      & document                  & dictionary                & agent                         & density                    & clinical                 & channel                  & achieves                   & conjecture              & detection              \\
deep                             & equations                   & documents                 & inverse                   & agent's                       & dependence                 & diagnosis                & channels                 & art                        & connected               & image                  \\
feature                          & formulation                 & embeddings                & kernel                    & arm                           & distributions              & disease                  & codes                    & benchmark                  & constant                & images                 \\
features                         & gradient                    & entities                  & kernels                   & arms                          & estimated                  & gene                     & coding                   & compare                    & delta                   & motion                 \\
layer                            & greedy                      & extraction                & manifold                  & bandit                        & estimates                  & genes                    & communication            & compared                   & edge                    & object                 \\
learn                            & heuristic                   & linguistic                & matrices                  & demonstrations                & estimating                 & genetic                  & decoding                 & comparison                 & edges                   & objects                \\
learned                          & iteration                   & retrieval                 & matrix                    & exploration                   & estimation                 & imaging                  & exchange                 & dataset                    & epsilon                 & recognition            \\
learning                         & iterative                   & semantic                  & nonlinear                 & horizon                       & estimator                  & longitudinal             & message                  & datasets                   & graph                   & reconstruction         \\
loss                             & minimization                & sentence                  & projection                & learner                       & estimators                 & medical                  & messages                 & demonstrate                & graphs                  & regions                \\
network                          & numerical                   & sentences                 & rank                      & mdps                          & gaussian                   & molecular                & protocol                 & evaluate                   & log                     & resolution             \\
networks                         & optimization                & speech                    & regularization            & planner                       & likelihood                 & patient                  & protocols                & evaluated                  & maximum                 & scene                  \\
neural                           & quadratic                   & style                     & signal                    & planning                      & multivariate               & patients                 & quantum                  & experiments                & minimum                 & segmentation           \\
representations                  & smooth                      & text                      & sparse                    & policies                      & parametric                 & protein                  & routing                  & performance                & polynomial              & shape                  \\
tasks                            & solve                       & topic                     & sparsity                  & policy                        & regression                 & screening                & scheme                   & proposed                   & prove                   & spatial                \\
train                            & solving                     & translation               & spectral                  & regret                        & statistics                 & subjects                 & schemes                  & results                    & trees                   & tracking               \\
trained                          & stochastic                  & word                      & vector                    & reward                        & tests                      & survival                 & secure                   & robustness                 & vertex                  & video                  \\
training                         & variational                 & words                     & vectors                   & rewards                       & variance                   & treatment                & transmission             & test                       & vertices                & visual                   \\ 

\bottomrule
\end{tabular}
\begin{tablenotes}
\footnotesize
 \item {\it Notes:} This table reports the most frequent $n$-grams per cluster, sorted by alphabetical order. Cluster 1 concerns deep learning. The terms excluded from the list used to retrieve data from WoS are crossed out.
\end{tablenotes}
 \end{threeparttable}
 }
\end{table}
\end{landscape}

\begin{landscape}
\begin{table}[t!]\footnotesize
  \addtocounter{table}{-1}
  \renewcommand{\thetable}{\arabic{table}b}
\caption{Word embedding obtained via Word2Vec [arXiv.org sample]}
 \centering  
    \scalebox{0.6}{
	\begin{threeparttable}
  \label{tab:embedding2} 
\begin{tabular}{@{\extracolsep{1pt}} lllllllllll} 
\toprule
Cluster 12            & Cluster 13               & Cluster 14                    & Cluster 15                 & Cluster 16            & Cluster 17                 & Cluster 18        & Cluster 19             & Cluster 20   & Cluster 21 & Cluster 22         \\
\midrule
\addlinespace
\textit{\underline{n-grams}}                        \\
access\_control       & answer\_programming      & algebraic\_geometry           & computational\_resources   & cooperative\_game     & computer\_science          & machine\_learning & autonomous\_vehicles   &              &            & 0\_leq             \\
cloud\_computing      & association\_rules       & binary\_trees                 & computationally\_intensive & decision\_maker       & decision\_makers           & real\_world       & numerical\_simulations &              &            & a\_i               \\
cloud\_services       & background\_knowledge    & cayley\_graphs                & computing\_resources       & differential\_privacy & decision\_support          &                   &                        &              &            & a\_n               \\
cyber\_physical       & boolean\_function        & combinatorial\_interpretation & energy\_consumption        & expected\_utility     & health\_care               &                   &                        &              &            & alpha\_0           \\
cyber\_security       & boolean\_functions       & combinatorial\_objects        & energy\_efficiency         & game\_players         & public\_health             &                   &                        &              &            & alpha\_alpha       \\
internet\_iot         & cellular\_automata       & dyck\_paths                   & external\_memory           & game\_theoretic       & search\_engine             &                   &                        &              &            & alpha\_beta        \\
intrusion\_detection  & constraint\_satisfaction & equivalence\_classes          & fault\_tolerance           & incentive\_compatible & search\_engines            &                   &                        &              &            & c\_n               \\
mobile\_devices       & dempster\_shafer         & explicit\_formula             & fault\_tolerant            & nash\_equilibria      & social\_media              &                   &                        &              &            & delta\_0           \\
mobile\_phone         & expressive\_power        & explicit\_formulas            & graphics\_processing       & nash\_equilibrium     & social\_sciences           &                   &                        &              &            & delta\_delta       \\
operating\_system     & finite\_automata         & expressed\_terms              & hardware\_software         & pareto\_optimal       & software\_development      &                   &                        &              &            & delta\_geq         \\
peer\_peer            & fuzzy\_logic             & generating\_function          & linear\_algebra            & price\_anarchy        & software\_engineering      &                   &                        &              &            & epsilon\_epsilon   \\
privacy\_concerns     & knowledge\_bases         & generating\_functions         & load\_balancing            & pure\_nash            & statistically\_significant &                   &                        &              &            & frac\_log          \\
privacy\_preserving   & kolmogorov\_complexity   & hopf\_algebra                 & low\_latency               & resource\_allocation  &                            &                   &                        &              &            & g\_n               \\
quality\_service      & logic\_programming       & infinite\_family              & massively\_parallel        & social\_choice        &                            &                   &                        &              &            & ge\_2              \\
resource\_constrained & object\_oriented         & lie\_algebra                  & matrix\_multiplication     & social\_welfare       &                            &                   &                        &              &            & ge\_3              \\
resource\_management  & pattern\_matching        & partition\_function           & memory\_footprint          & stable\_matching      &                            &                   &                        &              &            & geq\_0             \\
safety\_critical      & programming\_languages   & schur\_functions              & memory\_requirements       &                       &                            &                   &                        &              &            & geq\_3             \\
security\_privacy     & pspace\_complete         & simplicial\_complex           & message\_passing           &                       &                            &                   &                        &              &            & geq\_4             \\
service\_providers    & quantum\_mechanics       & simplicial\_complexes         & multi\_core                &                       &                            &                   &                        &              &            & k\_n               \\
smart\_contracts      & relational\_database     & strongly\_regular             & power\_consumption         &                       &                            &                   &                        &              &            & lceil\_frac        \\
smart\_grid           & relational\_databases    & symmetric\_functions          & programming\_language      &                       &                            &                   &                        &              &            & le\_le             \\
supply\_chain         & semantic\_web            & tensor\_product               & shared\_memory             &                       &                            &                   &                        &              &            & lfloor\_frac       \\
user\_interface       & temporal\_logic          & tutte\_polynomial             &                            &                       &                            &                   &                        &              &            & log\_2             \\
virtual\_machines     & turing\_machine          & vector\_spaces                &                            &                       &                            &                   &                        &              &            & m\_n               \\
web\_service          & turing\_machines         &                               &                            &                       &                            &                   &                        &              &            & mathbb\_mathbb     \\
web\_services         &                          &                               &                            &                       &                            &                   &                        &              &            & mathbb\_times      \\
                      &                          &                               &                            &                       &                            &                   &                        &              &            & mathbf\_mathbf     \\
                      &                          &                               &                            &                       &                            &                   &                        &              &            & mathcal\_mathcal   \\
                      &                          &                               &                            &                       &                            &                   &                        &              &            & positive\_integer  \\
                      &                          &                               &                            &                       &                            &                   &                        &              &            & positive\_integers \\
                      &                          &                               &                            &                       &                            &                   &                        &              &            & rightarrow\_infty  \\
                      &                          &                               &                            &                       &                            &                   &                        &              &            & s\_n               \\
                      &                          &                               &                            &                       &                            &                   &                        &              &            & subset\_mathbb     \\
                      &                          &                               &                            &                       &                            &                   &                        &              &            & sum\_\_0           \\
                      &                          &                               &                            &                       &                            &                   &                        &              &            & varepsilon\_0      \\
                      &                          &                               &                            &                       &                            &                   &                        &              &            & x\_1\_x\_2         \\
                      &                          &                               &                            &                       &                            &                   &                        &              &            & x\_1\_x\_n         \\
                      &                          &                               &                            &                       &                            &                   &                        &              &            & x\_i               \\
                      &                          &                               &                            &                       &                            &                   &                        &              &            & x\_n               \\
\addlinespace
\textit{\underline{Unigrams}}  		\\
access                & answer                   & algebra                       & code                       & agents                & activities                 & analysis          & activity               & algorithm    & 5          & 0,1                \\
attack                & causal                   & algebraic                     & computation                & allocation            & ai                         & applications      & control                & algorithms   & 6          & alpha              \\
attacks               & definition               & combinatorial                 & computations               & costs                 & authors                    & approaches        & dynamics               & based        & 7          & beta               \\
cloud                 & formal                   & enumeration                   & distributed                & decisions             & collected                  & complex           & energy                 & class        & 8          & cdot               \\
devices               & language                 & families                      & execution                  & demand                & communities                & data              & environment            & distribution & 9          & ell                \\
internet              & languages                & formula                       & hardware                   & equilibrium           & community                  & design            & event                  & efficient    & 10         & frac               \\
management            & logic                    & formulas                      & implement                  & game                  & individuals                & existing          & events                 & function     & 12         & gamma              \\
mobile                & notion                   & integers                      & implementation             & games                 & papers                     & framework         & flow                   & functions    & 15         & geq                \\
platform              & operators                & lattice                       & implementations            & items                 & people                     & information       & location               & introduce    & 16         & lambda             \\
privacy               & probabilistic            & odd                           & implemented                & market                & project                    & knowledge         & locations              & linear       & 20         & le                 \\
resources             & program                  & partitions                    & machines                   & mechanism             & public                     & level             & measurements           & multiple     & 30         & leq                \\
security              & programs                 & permutation                   & memory                     & mechanisms            & research                   & methods           & monitoring             & optimal      & 50         & mathbb             \\
service               & proof                    & permutations                  & operations                 & outcome               & researchers                & process           & physical               & parameters   & 100        & mathcal            \\
services              & proofs                   & polynomials                   & parallel                   & outcomes              & science                    & provide           & sensor                 & properties   & 2010       & max                \\
sharing               & queries                  & polytope                      & processing                 & player                & scientific                 & real              & sensors                & random       & 2012       & mu                 \\
software              & query                    & prime                         & run                        & players               & social                     & scale             & signals                & simple       & 2014       & pi                 \\
technology            & reasoning                & rational                      & scalable                   & private               & students                   & system            & simulated              & size         & 2015       & sigma              \\
user                  & relations                & symmetry                      & scheduling                 & resource              & survey                     & systems           & simulation             & space        & 2016       & sqrt               \\
users                 & rules                    & theorem                       & speed                      & strategies            & topics                     & task              & simulations            & structure    & 2017       & theta              \\
web                   & semantics                & theorems                      & storage                    & utility               & world                      & techniques        & traffic                & time         & million    & varepsilon       \\ 

\bottomrule
\end{tabular}
\begin{tablenotes}
\footnotesize
 \item {\it Notes:} This table reports the most frequent $n$-grams per cluster, sorted by alphabetical order. Cluster 1 concerns deep learning. The terms excluded from the list used to retrieve data from WoS are crossed out.
\end{tablenotes}
 \end{threeparttable}
 }
\end{table}
\end{landscape}

\subsubsection{Acronyms and full names}

\begin{footnotesize}
\begin{longtable}{ll}
\caption{List of acronyms replaced by full name} \\
\toprule
Acronym & Full name  \\
\hline
\endfirsthead 
\caption{List of acronyms replaced by full name -- continued}\\
\midrule
\endhead
\multicolumn{2}{r}{Continued on next page}\\
\endfoot
\hline
\multicolumn{2}{r}{End of table.} \\
\endlastfoot
\addlinespace
ann & artificial\_neural\_network\\
anns & artificial\_neural\_networks\\
\addlinespace
blstm & bidirectional\_long\_short\_term\_memory\\
bns & bayesian\_networks\\
bpn & bidirectional\_pyramid\_networks\\
\addlinespace
cav & computer\_aided\_verification\\
cnn & convolutional\_neural\_network\\
cnns & convolutional\_neural\_networks\\
crf & conditional\_random\_fields\\
ctc & connectionist\_temporal\_classification\\
\addlinespace
dan & deep\_alignment\_network\\
dbm & deep\_boltzmann\_machine\\
dbms & database\_management\_systems\\
dbn & deep\_belief\_network\\
dcn & dynamic\_coattention\_network\\
dcnn & deep\_convolutional\_neural\_network\\
dcnns & deep\_convolutional\_neural\_networks\\
dl & deep\_learning\\
dek & deep\_embedding\_kernel\\
dnn & deep\_neural\_network\\
dnns & deep\_neural\_networks\\
dqn & deep\_q\_network\\
dqns & deep\_q\_networks\\
drcn & deeply\_recursive\_convolutional\_network\\
drl & deep\_reinforcement\_learning\\
\addlinespace
elm & extreme\_learning\_machine\\
\addlinespace
fcn & fully\_convolutional\_network\\
fhmms & factorial\_hidden\_markov\_model\\
\addlinespace
ga & genetic\_algorithm\\
gan & generative\_adversarial\_network\\
gans & generative\_adversarial\_networks\\
gcns & graph\_convolutional\_networks\\
grnn & general\_regression\_neural\_network\\
grus & gated\_recurrent\_units\\
gsn & generative\_stochastic\_network\\
gssl & graph\_based\_semi\_supervised\_learning\\
\addlinespace
knn & k\_nearest\_neighbors\\
\addlinespace
lmnn & large\_margin\_nearest\_neighbor\\
lstm & long\_short\_term\_memory\\
lstms & long\_short\_term\_memory\\
\addlinespace
mdp & markov\_decision\_process\\
ml & machine\_learning\\
mlp & multilayer\_perceptron\\
mtl & multi\_task\_learning\\
\addlinespace
nn & neural\_network\\
nns & neural\_networks\\
\addlinespace
pmvge & probabilistic\_multi\_view\_graph\_embedding\\
pnn & probabilistic\_neural\_network\\
pso & particle\_swarm\_optimization\\
psrnns & predictive\_state\_recurrent\_neural\_networks\\
\addlinespace
rbf & radial\_basis\_function\\
rbfn & radial\_basis\_function\_network\\
rbms & restricted\_boltzmann\_machines\\
rgp & recurrent\_gaussian\_process\\
rl & reinforcement\_learning\\
rlns & regularization\_learning\_networks\\
rmbs & restricted\_boltzmann\_networks\\
rnn & recurrent\_neural\_network\\
rnns & recurrent\_neural\_networks\\
\addlinespace
smffnn & supervised\_multilayers\_feed\_forward\_neural\_network\\
snn & spiking\_neural\_network\\
snns & spiking\_neural\_networks\\
ssrbm & spike\_slab\_restricted\_boltzmann\_machine\\
\addlinespace
svm & support\_vector\_machine\\
vae & variational\_autoencoder\\
vaes & variational\_autoencoders\\
\addlinespace
wae & wasserstein\_autoencoder\\
\addlinespace
zsl & zero\_shot\_learning\\

\bottomrule
\end{longtable}
\end{footnotesize}

\newpage

\section{Diffusion of deep learning in science: the sample}
\label{appendix:country}

This Appendix complements Section \ref{sec:DiffusionInScience} with details on the sample used for the analysis on the diffusion of deep learning in science.

\begin{table}[h!]\footnotesize
 \centering  
  \caption{Deep learning documents broken down by period and WoS research areas}
  \label{tab:SamplePapers}
  \vspace*{1em}
  \scalebox{0.75}{
	\begin{threeparttable}
\begin{tabular}{@{\extracolsep{5pt}}rrrrrrrr} 
\toprule
Year & All documents & Technology & Physical Sciences & Life Sciences \& Biomedicine & Health Sciences & Social Sciences & Art \& Humanities \\
& (1) & (2) & (3) & (4) & (5) & (6) & (7) \\
\midrule
1990 & 381 [0.15] & 193.33 [50.74] & 117.17 [30.75] & 61.83 [16.23] & 106 [27.82] & 8.67 [2.27] & 0 [0] \\
1991 & 836 [0.32] & 507.52 [60.71] & 193.73 [23.17] & 123.58 [14.78] & 188 [22.49] & 11.17 [1.34] & 0 [0] \\
1992 & 1,256 [0.48] & 759.17 [60.44] & 306.92 [24.44] & 159.25 [12.68] & 234 [18.63] & 29.67 [2.36] & 1 [0.08] \\
1993 & 1,477 [0.57] & 841.05 [56.94] & 365.37 [24.74] & 221.42 [14.99] & 315 [21.33] & 49.17 [3.33] & 0 [0] \\
1994 & 1,798 [0.69] & 1,074.10 [59.74] & 373.60 [20.78] & 293.53 [16.33] & 385 [21.41] & 55.77 [3.10] & 1 [0.06] \\
1995 & 2,220 [0.85] & 1,415.85 [63.78] & 436.92 [19.68] & 306.23 [13.79] & 412 [18.56] & 60.50 [2.73] & 0.50 [0.02] \\
\addlinespace
1996 & 2,791 [1.07] & 1,843.03 [66.03] & 478.30 [17.14] & 393.23 [14.09] & 475 [17.02] & 75.43 [2.70] & 1 [0.04] \\
1997 & 3,090 [1.19] & 2,002.12 [64.79] & 530.50 [17.17] & 481.38 [15.58] & 613 [19.84] & 74 [2.39] & 2 [0.06] \\
1998 & 4,330 [1.66] & 2,865.13 [66.17] & 566.70 [13.09] & 779.25 [18.00] & 1,083 [25.01] & 113.92 [2.63] & 5 [0.12] \\
1999 & 4,725 [1.81] & 3,302.34 [69.89] & 725.92 [15.36] & 598.61 [12.67] & 627 [13.27] & 93.58 [1.98] & 4.55 [0.10] \\
2000 & 6,259 [2.40] & 4,661.05 [74.47] & 835.07 [13.34] & 621.55 [9.93] & 691 [11.04] & 138.33 [2.21] & 3 [0.05] \\
\addlinespace
2001 & 6,062 [2.33] & 4,376.40 [72.19] & 859.52 [14.18] & 726.90 [11.99] & 806 [13.3] & 94.68 [1.56] & 4.50 [0.07] \\
2002 & 6,676 [2.56] & 5,191.35 [77.76] & 762.67 [11.42] & 614.42 [9.20] & 700 [10.49] & 104.37 [1.56] & 3.20 [0.05] \\
2003 & 7,230 [2.78] & 5,430.80 [75.11] & 923.27 [12.77] & 768.00 [10.62] & 897 [12.41] & 100.43 [1.39] & 7.50 [0.10] \\
2004 & 7,765 [2.98] & 5,907.90 [76.08] & 921.27 [11.86] & 811.58 [10.45] & 879 [11.32] & 119.75 [1.54] & 4.50 [0.06] \\
2005 & 9,023 [3.46] & 7,026.45 [77.87] & 1,072.60 [11.89] & 790.40 [8.76] & 896 [9.93] & 129.80 [1.44] & 3.75 [0.04] \\
\addlinespace
2006 & 10,654 [4.09] & 8,206.27 [77.03] & 1,424.57 [13.37] & 885.2 [8.31] & 859 [8.06] & 136.60 [1.28] & 1.36 [0.01] \\
2007 & 11,086 [4.26] & 8,234.85 [74.28] & 1,551.97 [14.00] & 1,072.22 [9.67] & 1,246 [11.24] & 217.38 [1.96] & 9.58 [0.09] \\
2008 & 11,891 [4.57] & 9,053.63 [76.14] & 1,562.03 [13.14] & 1,015.39 [8.54] & 1,067 [8.97] & 256 [2.15] & 3.95 [0.03] \\
2009 & 13,049 [5.01] & 10,066.02 [77.14] & 1,601.43 [12.27] & 1,113.80 [8.54] & 1,102 [8.45] & 258.05 [1.98] & 9.70 [0.07] \\
2010 & 10,467 [4.02] & 7,702.98 [73.59] & 1,399.65 [13.37] & 1,117.50 [10.68] & 992 [9.48] & 242.87 [2.32] & 4 [0.04] \\
\addlinespace
2011 & 10,872 [4.17] & 8,033.20 [73.89] & 1,462.38 [13.45] & 1,110.13 [10.21] & 943 [8.67] & 261.78 [2.41] & 4.50 [0.04] \\
2012 & 12,227 [4.69] & 9,238.63 [75.56] & 1,571.02 [12.85] & 1,189.90 [9.73] & 1,047 [8.56] & 220.95 [1.81] & 6.50 [0.05] \\
2013 & 12,691 [4.87] & 9,439.40 [74.38] & 1,779.75 [14.02] & 1,248.40 [9.84] & 1,106 [8.71] & 217.78 [1.72] & 5.67 [0.04] \\
2014 & 14,355 [5.51] & 11,044.90 [76.94] & 1,747.07 [12.17] & 1,263.52 [8.80] & 1,067 [7.43] & 293.02 [2.04] & 6.50 [0.05] \\
2015 & 16,764 [6.44] & 12,934.47 [77.16] & 1,978.12 [11.80] & 1,476.93 [8.81] & 1,267 [7.56] & 367.65 [2.19] & 6.83 [0.04] \\
\addlinespace
2016 & 18,425 [7.07] & 13,927.08 [75.59] & 2,265.67 [12.30] & 1,700.87 [9.23] & 1,449 [7.86] & 515.55 [2.80] & 15.83 [0.09] \\
2017 & 24,046 [9.23] & 18,488.38 [76.89] & 2,993.48 [12.45] & 2,099.37 [8.73] & 2,008 [8.35] & 449.93 [1.87] & 14.83 [0.06] \\
2018 & 28,013 [10.76] & 20,223.48 [72.19] & 4,192.73 [14.97] & 3,078.15 [10.99] & 3,001 [10.71] & 491.90 [1.76] & 26.73 [0.10] \\
\bottomrule
\end{tabular}
\begin{tablenotes}
 \footnotesize
 \item {\it Notes:} Number of deep learning documents (Column 1). Weighted count for all other columns. For `All Documents' the shares [\%] are calculated on the basis of the entire DL sample. For all other columns the share refers to the period. For example, the number of documents published in 2018 represents 10.76\% of all DL documents; of the 28,013 documents, 72.19\% belong to `Technology', 14.97\% to `Physical Sciences', and so on.
\end{tablenotes}
 \end{threeparttable}}
\end{table}

\begin{table}[h!]\footnotesize
 \centering  
  \caption{Deep learning publication activity broken down by country and period}
  \label{tab:SamplePapers}
  \vspace*{1em}
    \scalebox{0.88}{
	\begin{threeparttable}
\begin{tabular}{@{\extracolsep{5pt}}lrlrlr} 
\toprule
 \multicolumn{2}{c}{\textit{1990-1999}}
 & \multicolumn{2}{c}{\textit{2000-2009}} & \multicolumn{2}{c}{\textit{2010-2019}} \\ 
\cline{1-2}\cline{3-4}\cline{5-6}
\addlinespace
Country &  \# Documents & Country &  \#  Documents & Country & \#  Documents\\
\midrule
$[$UE$]$ & 6,205 &  $[$UE$]$ & 24,047 & China & 39,852\\
USA & 5,123 & China & 20,560 & $[$UE$]$ & 28,358\\
United Kingdom & 1,695 & USA & 13,665 & USA & 17,320\\
Japan & 1,368 & United Kingdom & 5,151 & India & 10,349\\
Germany & 991 & Japan & 5,076 & Iran (Islamic Republic of) & 7,008\\
Italy & 851 & Taiwan & 3,611 & United Kingdom & 4,917\\
China & 764 & Italy & 3,269 & Japan & 4,471\\
Canada & 721 & India & 3,225 & Taiwan & 4,027\\
France & 704 & Canada & 3,204 & Korea (Republic of) & 3,902\\
Spain & 474 & Spain & 2,898 & Turkey & 3,895\\
Taiwan & 456 & Korea (Republic of) & 2,872 & Spain & 3,438\\
Australia & 436 & Germany & 2,802 & Canada & 3,377\\
Korea (Republic of) & 427 & Turkey & 2,228 & Germany & 3,271\\
India & 354 & France & 2,183 & Italy & 3,063\\
Netherlands & 286 & Iran (Islamic Republic of) & 1,964 & Australia & 2,925\\
Brazil & 224 & Brazil & 1,874 & Malaysia & 2,631\\
\bottomrule
\end{tabular}
\begin{tablenotes}
 \footnotesize
 \item {\it Notes:} Top 15 countries for each period. [EU] represents EU28 as in 2018.
\end{tablenotes}
 \end{threeparttable}}
\end{table}

\newpage

\section{Deep learning in health sciences: data construction and sample details}
\label{appendix:health}

This Appendix provides additional statistics on the empirical analysis of Section \ref{sec:DLHealth}. The perimeter of the domain `health sciences' has been delineated using the WoS subject categories reported in Table \ref{tab:wos_cat}. Health sciences can be viewed as a subset of the broader WoS research area `Life Science \& Biomedicine'.

\begin{table}[h]\footnotesize
    \caption{WoS subject categories defining `health sciences'} 
 \centering  
    \scalebox{0.90}{
	\begin{threeparttable}
  \label{tab:wos_cat} 
\begin{tabular}{@{\extracolsep{5pt}} lrlr} 
\toprule 
Category & Count [Share] & Category & Count [Share] \\ 
\hline \\[-1.8ex] 
Neurosciences & 6,683 [2.56] &	Geriatrics \& Gerontology & 75 [0.03]\\
Biology & 6,084 [2.33] &	Anatomy \& Morphology & 73 [0.03]\\
Mathematical \& Computational Biology & 3,386 [1.30] &	Orthopedics & 73 [0.03]\\
Radiology, Nuclear Medicine \& Medical Imaging & 2,678 [1.03] &	Transplantation & 63 [0.02]\\
Medical Informatics & 2,218 [0.85] &	Dentistry, Oral Surgery \& Medicine & 62 [0.02]\\
Psychology & 1,932 [0.74] &	Virology & 61 [0.02]\\
Microbiology & 1,908 [0.73] &	Hematology & 59 [0.02]\\
Biochemistry \& Molecular Biology & 1,852 [0.71] &	Nursing & 50 [0.02]\\
Biotechnology \& Applied Microbiology & 1,727 [0.66] &	Reproductive Biology & 41 [0.02]\\
Pharmacology \& Pharmacy & 1,221 [0.47] &	Integrative \& Complementary Medicine & 36 [0.01]\\
Biophysics & 863 [0.33] &	Emergency Medicine & 35 [0.01]\\
Psychiatry & 733 [0.28] &	Rheumatology & 24 [0.01]\\
Cell Biology & 582 [0.22] &	Tropical Medicine & 21 [0.01]\\
Health Care Sciences \& Services & 549 [0.21] &	Mycology & 18 [0.01]\\
Oncology & 548 [0.21] &	Allergy & 7 [0]\\
Surgery & 465 [0.18] &	Medical Ethics & 4 [0]\\
Genetics \& Heredity & 443 [0.17] &	Psychology, Experimental & 0 [0]\\
Physiology & 431 [0.17] &	Psychology, Applied & 0 [0]\\
Behavioral Sciences & 419 [0.16] &	Psychology, Multidisciplinary & 0 [0]\\
Toxicology & 396 [0.15] &	Psychology, Biological & 0 [0]\\
Public, Environmental \& Occupational Health & 384 [0.15] &	Neuroimaging & 0 [0]\\
Endocrinology \& Metabolism & 283 [0.11] &	Engineering, Biomedical & 0 [0]\\
Pathology & 249 [0.10] &	Biochemical Research Methods & 0 [0]\\
Medical Laboratory Technology & 241 [0.09] &	Clinical Neurology & 0 [0]\\
Ophthalmology & 237 [0.09] &	Psychology, Developmental & 0 [0]\\
Urology \& Nephrology & 235 [0.09] &	Cardiac \& Cardiovascular Systems & 0 [0]\\
Rehabilitation & 230 [0.09] &	Psychology, Social & 0 [0]\\
Gastroenterology \& Hepatology & 184 [0.07] &	Critical Care Medicine & 0 [0]\\
Immunology & 174 [0.07] &	Medicine, Research \& Experimental & 0 [0]\\
Obstetrics \& Gynecology & 161 [0.06] &	Psychology, Mathematical & 0 [0]\\
Respiratory System & 129 [0.05] &	Chemistry, Medicinal & 0 [0]\\
Evolutionary Biology & 116 [0.04] &	Medicine, Legal & 0 [0]\\
Developmental Biology & 112 [0.04] &	Medicine, General \& Internal & 0 [0]\\
Anesthesiology & 111 [0.04] &	Peripheral Vascular Disease & 0 [0]\\
Pediatrics & 108 [0.04] &	Psychology, Clinical & 0 [0]\\
Nutrition \& Dietetics & 99 [0.04] &	Health Policy \& Services & 0 [0]\\
Otorhinolaryngology & 96 [0.04] &	Psychology, Educational & 0 [0]\\
Infectious Diseases & 82 [0.03] &	Social Sciences, Biomedical & 0 [0]\\
Audiology \& Speech-Language Pathology & 81 [0.03] &	Primary Health Care & 0 [0]\\
Gerontology & 81 [0.03] &	Andrology & 0 [0]\\
Dermatology & 78 [0.03] &	Psychology, Psychoanalysis & 0 [0]\\
Substance Abuse & 76 [0.03] \\	 
\bottomrule
\end{tabular}
\begin{tablenotes}
 \footnotesize
 \item {\it Notes:} Number of deep learning papers by WoS subject category. A document can belong to several categories. Shares in [\%].
\end{tablenotes}
 \end{threeparttable}
 }
\end{table}

\newpage

\subsection{Meta data on the estimation sample}

This Appendix provides details on the sample constructed to carry out the empirical analysis on health sciences (Section \ref{sec:empirics}). To benchmark deep learning publications, we download all the articles for the whole observation period published in the top 100 journals where research involving deep learning has been the most prominent.

\begin{footnotesize}
\begin{longtable}{p{10cm}rrr}
\caption{Sampled papers by journal and period} \\
\toprule
Journal $|$ Foundation date & 1990--1999 & 2000--2009 & 2010--2019 \\
\hline
\endfirsthead 
\caption{Sampled papers per journal and period -- continued}\\
Journal $|$ Foundation date & 1990--1999 & 2000--2009 & 2010-2019 \\
\hline
\endhead 
\multicolumn{4}{r}{Continued on next page}\\
\endfoot
\hline
\multicolumn{4}{r}{End of table.} \\
\endlastfoot
ACADEMIC RADIOLOGY $|$ 1994 & 1,250 & 2,179 & 2,265\\
ANALYTICAL AND BIOANALYTICAL CHEMISTRY $|$ 1862 & 0 & 5,235 & 8,300\\
ANNALS OF BIOMEDICAL ENGINEERING $|$ 1972 & 684 & 1,624 & 2,389\\
BASIC \& CLINICAL PHARMACOLOGY \& TOXICOLOGY $|$ 1945 & 0 & 1,944 & 8,411\\
BEHAVIORAL AND BRAIN SCIENCES $|$ 1978 & 4,162 & 3,573 & 2,273\\
BEHAVIOURAL BRAIN RESEARCH $|$ 1980 & 1,719 & 2,861 & 5,861\\
BIOLOGICAL PSYCHIATRY $|$ 1959 & 6,477 & 10,589 & 13,323\\
BIOMED RESEARCH INTERNATIONAL $|$ 2001 & 0 & 0 & 16,302\\
BIOMEDICAL ENGINEERING ONLINE $|$ 2002 & 0 & 160 & 1,309\\
BIOMEDICAL SIGNAL PROCESSING AND CONTROL $|$ 2006 & 0 & 150 & 1,398\\
BIORESOURCE TECHNOLOGY $|$ 1991 & 1,434 & 4,464 & 15,677\\
BIOSYSTEMS $|$ 1967 & 651 & 1,062 & 927\\
BMC BIOINFORMATICS $|$ 2000 & 0 & 3,455 & 6,060\\
BMC MEDICAL INFORMATICS AND DECISION MAKING $|$ 2001 & 0 & 171 & 1,350\\
BRAIN $|$ 1878 & 1,457 & 2,820 & 3,395\\
BRAIN AND LANGUAGE $|$ 1974 & 1,317 & 1,899 & 914\\
BRAIN RESEARCH $|$ 1966 & 15,725 & 11,563 & 6,503\\
CEREBRAL CORTEX $|$ 1991 & 517 & 1,877 & 3,006\\
CLINICAL NEUROPHYSIOLOGY $|$ 1949 & 286 & 3,047 & 3,493\\
COGNITIVE NEURODYNAMICS $|$ 2007 & 0 & 94 & 418\\
COGNITIVE SCIENCE $|$ 1977 & 178 & 411 & 851\\
COMBINATORIAL CHEMISTRY \& HIGH THROUGHPUT SCREENING $|$ 1998 & 47 & 766 & 849\\
COMPUTATIONAL AND MATHEMATICAL METHODS IN MEDICINE $|$ 1997 & 0 & 44 & 1,632\\
COMPUTATIONAL INTELLIGENCE AND NEUROSCIENCE $|$ 2007 & 0 & 0 & 855\\
COMPUTERIZED MEDICAL IMAGING AND GRAPHICS $|$ 1988 & 565 & 615 & 664\\
CORTEX $|$ 1964 & 562 & 1,021 & 2,213\\
CURRENT BIOLOGY $|$ 1991 & 2,744 & 7,273 & 7,714\\
CURRENT OPINION IN NEUROBIOLOGY $|$ 1991 & 527 & 1,035 & 1,400\\
EPILEPSIA $|$ 1909 & 6,736 & 17,762 & 10,780\\
EUROPEAN JOURNAL OF MEDICINAL CHEMISTRY $|$ 1966 & 1,164 & 1,990 & 7,632\\
EUROPEAN JOURNAL OF NEUROSCIENCE $|$ 1989 & 6,282 & 8,968 & 3,260\\
EXPERIMENTAL BRAIN RESEARCH $|$ 1966 & 3,062 & 4,110 & 3,517\\
FOOD CHEMISTRY $|$ 1976 & 2,077 & 6,114 & 16,416\\
FRONTIERS IN COMPUTATIONAL NEUROSCIENCE $|$ 2007 & 0 & 38 & 1,136\\
FRONTIERS IN HUMAN NEUROSCIENCE $|$ 2008 & 0 & 91 & 5,451\\
FRONTIERS IN NEUROINFORMATICS $|$ 2007 & 0 & 1 & 482\\
FRONTIERS IN NEUROSCIENCE $|$ 2009 & 0 & 114 & 4,778\\
FRONTIERS IN PSYCHOLOGY $|$ 2010 & 0 & 0 & 14,466\\
HIPPOCAMPUS $|$ 1991 & 490 & 1,007 & 1,231\\
HUMAN BRAIN MAPPING $|$ 1993 & 182 & 1,110 & 3,052\\
IEEE TRANSACTIONS ON BIOMEDICAL ENGINEERING $|$ 1964 & 1,594 & 2,528 & 3,218\\
IEEE TRANSACTIONS ON NEURAL SYSTEMS AND REHABILITATION ENGINEERING $|$ 2001 & 0 & 553 & 1,366\\
INTERNATIONAL JOURNAL OF COMPUTER ASSISTED RADIOLOGY AND SURGERY $|$ 2006 & 0 & 664 & 1,337\\
INTERNATIONAL JOURNAL OF ENVIRONMENTAL RESEARCH AND PUBLIC HEALTH $|$ 2004 & 0 & 214 & 11,117\\
INTERNATIONAL JOURNAL OF MOLECULAR SCIENCES $|$ 2000 & 0 & 804 & 18,697\\
INVESTIGATIVE OPHTHALMOLOGY \& VISUAL SCIENCE $|$ 1962 & 17,439 & 2,973 & 2,640\\
JOURNAL OF CHROMATOGRAPHY A $|$ 1958 & 7,664 & 11,861 & 9,715\\
JOURNAL OF COGNITIVE NEUROSCIENCE $|$ 1989 & 1,302 & 3,950 & 2,978\\
JOURNAL OF COMPUTATIONAL NEUROSCIENCE $|$ 1994 & 113 & 415 & 545\\
JOURNAL OF DIGITAL IMAGING $|$ 1988 & 401 & 625 & 974\\
JOURNAL OF MECHANICS IN MEDICINE AND BIOLOGY $|$ 2001 & 0 & 308 & 1,080\\
JOURNAL OF MEDICAL IMAGING AND HEALTH INFORMATICS $|$ 2011 & 0 & 0 & 1,706\\
JOURNAL OF MEDICAL SYSTEMS $|$ 1977 & 56 & 243 & 2,125\\
JOURNAL OF MEDICINAL CHEMISTRY $|$ 1959 & 5,619 & 6,783 & 7,387\\
JOURNAL OF MOLECULAR BIOLOGY $|$ 1959 & 7097 & 9691 & 4,251\\
JOURNAL OF NEURAL ENGINEERING $|$ 2004 & 0 & 331 & 1456\\
JOURNAL OF NEUROPHYSIOLOGY $|$ 1938 & 4,750 & 6,040 & 5,195\\
JOURNAL OF NEUROSCIENCE $|$ 1981 & 6,705 & 13,443 & 13,766\\
JOURNAL OF NEUROSCIENCE METHODS $|$ 1979 & 1,638 & 2,690 & 2,874\\
JOURNAL OF NUCLEAR MEDICINE $|$ 1964 & 12,218 & 11,603 & 20,672\\
JOURNAL OF PHARMACEUTICAL AND BIOMEDICAL ANALYSIS $|$ 1983 & 2,362 & 4,376 & 5,347\\
JOURNAL OF PHYSIOLOGY-PARIS $|$ 1992 & 383 & 476 & 227\\
JOURNAL OF THE ACOUSTICAL SOCIETY OF AMERICA $|$ 1929 & 7,323 & 6,801 & 7,806\\
JOURNAL OF THEORETICAL BIOLOGY $|$ 1961 & 2,371 & 3,270 & 4,203\\
JOURNAL OF UROLOGY $|$ 1917 & 12,499 & 29,396 & 39,207\\
JOURNAL OF VIBROENGINEERING $|$ 2007 & 0 & 268 & 2,562\\
MEDICAL ENGINEERING \& PHYSICS $|$ 1994 & 537 & 1,153 & 1,699\\
MEDICAL IMAGING 2018: COMPUTER-AIDED DIAGNOSIS $|$ 2018 & 0 & 0 & 136\\
MEDICAL PHYSICS $|$ 1997 & 2,268 & 14,402 & 28,629\\
MOLECULES $|$ 1996 & 210 & 1,875 & 15,389\\
NATURE NEUROSCIENCE $|$ 1998 & 414 & 2,899 & 2,727\\
NEUROIMAGE $|$ 1993 & 373 & 7,286 & 9,626\\
NEURON $|$ 1988 & 2,393 & 3,977 & 4,693\\
NEUROPSYCHOLOGIA $|$ 1963 & 1,338 & 2,465 & 3,584\\
NEUROREPORT $|$ 1990 & 5,112 & 5,152 & 2,125\\
NEUROSCIENCE $|$ 1976 & 5,846 & 7,472 & 7,491\\
NEUROSCIENCE AND BIOBEHAVIORAL REVIEWS $|$ 1977 & 617 & 772 & 2,190\\
NEUROSCIENCE LETTERS $|$ 1975 & 10,062 & 9,976 & 7,188\\
NEUROSCIENCE RESEARCH $|$ 1984 & 1,081 & 7,801 & 4,976\\
NUCLEIC ACIDS RESEARCH $|$ 1974 & 11,010 & 10,326 & 12,648\\
PERCEPTION $|$ 1972 & 4,762 & 7,581 & 7,639\\
PHYSICS IN MEDICINE AND BIOLOGY $|$ 1956 & 1,853 & 4,283 & 5,381\\
PHYSIOLOGICAL MEASUREMENT $|$ 1980 & 367 & 1,167 & 1,620\\
PLOS COMPUTATIONAL BIOLOGY $|$ 2005 & 0 & 1,149 & 5,187\\
PROTEINS-STRUCTURE FUNCTION AND BIOINFORMATICS $|$ 1986 & 437 & 2,991 & 2,190\\
PSYCHOLOGICAL REVIEW $|$ 1894 & 379 & 493 & 390\\
RADIOLOGY $|$ 1923 & 19,517 & 12,402 & 5,188\\
RADIOTHERAPY AND ONCOLOGY $|$ 1983 & 1,623 & 10,706 & 16,163\\
SCHIZOPHRENIA RESEARCH $|$ 1988 & 5,257 & 8,757 & 7,323\\
TRENDS IN COGNITIVE SCIENCES $|$ 1997 & 332 & 1,263 & 1,092\\
VISION RESEARCH $|$ 1961 & 4,295 & 3,164 & 1,890\\
2007 ANNUAL INTERNATIONAL CONFERENCE OF THE IEEE ENGINEERING IN MEDICINE AND BIOLOGY SOCIETY, VOLS 1-16 $|$ 2007 & 0 & 1,703 & 0\\
2011 ANNUAL INTERNATIONAL CONFERENCE OF THE IEEE ENGINEERING IN MEDICINE AND BIOLOGY SOCIETY (EMBC) $|$ 2011 & 0 & 0 & 2,083\\
2015 37TH ANNUAL INTERNATIONAL CONFERENCE OF THE IEEE ENGINEERING IN MEDICINE AND BIOLOGY SOCIETY (EMBC) $|$ 2015 & 0 & 0 & 2,008\\
2017 39TH ANNUAL INTERNATIONAL CONFERENCE OF THE IEEE ENGINEERING IN MEDICINE AND BIOLOGY SOCIETY (EMBC) $|$ 2017 & 0 & 0 & 1,123\\
2017 IEEE 14TH INTERNATIONAL SYMPOSIUM ON BIOMEDICAL IMAGING (ISBI 2017) $|$ 2017 & 0 & 0 & 285\\
2018 11TH INTERNATIONAL CONGRESS ON IMAGE AND SIGNAL PROCESSING, BIOMEDICAL ENGINEERING AND INFORMATICS (CISP-BMEI 2018) $|$ 2018 & 0 & 0 & 249\\
2018 IEEE 15TH INTERNATIONAL SYMPOSIUM ON BIOMEDICAL IMAGING (ISBI 2018) $|$ 2018 & 0 & 0 & 364\\
\bottomrule
\end{longtable}
\end{footnotesize}

\begin{table}[h!]\footnotesize
 \centering  
  \caption{Health sciences sample and deep learning articles}
  \label{tab:SamplePapers}
  \vspace*{1em}
	\begin{threeparttable}
\begin{tabular}{@{\extracolsep{5pt}}lrrrrrr} 
\toprule

& & \multicolumn{2}{c}{\textit{Full sample health sciences}}
 & \multicolumn{3}{c}{\textit{Sample for econometrics}} \\ 
\cline{3-4}\cline{5-7} 

\\[-1.8ex] Year & \# Journals & \# Articles & \# DL Articles & \# Articles & \# DL Articles & \\
\midrule
1990 & 44 & 14,317 & 25 &\\
1991 & 48 & 17,809 & 37 & \\
1992 & 52 & 21,029 & 87\\
1993 & 55 & 21,295 & 97\\
1994 & 57 & 24,458 & 119\\
1995 & 60 & 24,632 & 171\\
\addlinespace
1996 & 60 & 25,072 & 155\\
1997 & 62 & 24,155 & 186\\
1998 & 65 & 29,891 & 203\\
1999 & 65 & 29,254 & 226\\
2000 & 66 & 30,239 & 222\\
\addlinespace
2001 & 68 & 27,272 & 217 & 14,427 & 139\\
2002 & 70 & 31,120 & 235& 14,580 & 132\\
2003 & 70 & 31,225 & 256& 15,463 & 162\\
2004 & 72 & 34,686 & 300& 16,924 & 182\\
2005 & 72 & 35,177 & 327& 17,586 & 198\\
\addlinespace
2006 & 77 & 41,966 & 412& 20,762 & 250\\
2007 & 83 & 42,947 & 520& 23,510 & 366\\
2008 & 84 & 41,931 & 431& 23,044 & 292\\
2009 & 86 & 46,195 & 420& 23,480 & 293\\
2010 & 85 & 47,384 & 485& 25,103 & 328\\
\addlinespace
2011 & 89 & 52,550 & 554& 30,082 & 417\\
2012 & 89 & 48,763 & 559& 29,497 & 426\\
2013 & 89 & 49,814 & 500& 32,112 & 381\\
2014 & 89 & 57,045 & 586& 34,341 & 462\\
2015 & 90 & 55,277 & 701& 35,126 & 532\\
\addlinespace
2016 & 89 & 56,232 & 729\\
2017 & 90 & 57,146 & 1,114\\
2018 & 91 & 62,342 & 1,646\\
\midrule
Total &  & 1,081,223 & 11,520 & 356,037 & 4,560 \\
\bottomrule
\end{tabular}
\begin{tablenotes}
 \footnotesize
 \item {\it Notes:} The articles published in the period 1990--2000 are used to build the novelty measures for the first focal year 2001. The articles published in the period 2016--2018 are used to check whether the new combinations of referenced journals are reused in the following three years after the last focal year 2015. The discrepancy between the number of articles in the whole sample and the number in the sample used for econometric analysis is due to the presence of missing information in the variables considered.
\end{tablenotes}
 \end{threeparttable}
\end{table}

\newpage
\clearpage

\section{Re-combinatorial novelty: indicators}
\label{appendix:novelty}

This Appendix complements Section \ref{sec:DLHealth} with details on the procedure for the construction of novelty measures (Section \ref{sec:empirics}). It also reports some statistics on the most frequent combinations of Web of Science subject categories, Tables \ref{tab:wos_freq1}--\ref{tab:wos_freq2}. Codes for the variable construction and analysis are fully accessible upon request.

\subsection{Algorithm for the construction of novelty indicators}

The novelty indicators are calculated at the year-level. Let $y$ be the focal year, we compute combinations of referenced journals in scientific documents belonging to three groups:

\begin{itemize}

\item All papers published in the focal year $y$.

\item All papers published before the focal year $y$, $B_{y}$ 

\item All papers published 3 years after the focal year $y$, $A_{y}$

\end{itemize}

In our study the focal year, $y$, is moving from 2001 to 2015, while the first year for which $B_{y}$ is calculated remains fixed. We choose the year 2001 as the first focal year to guarantee a sufficiently long time window (1990--2000) over which all previous combinations of referenced journals are assessed.

Suppose a paper $P$ published in year $y$ cites three different journals $J_1$,$J_2$ and $J_3$. This gives rise to three unique combinations: $(J_1,J_2)$, $(J_1,J_3)$, and $(J_2,J_3)$.

\begin{itemize}

\item For each of these combinations, we check whether $(J_{i},J_{j}) \in B_{y}$, and if not, the pair is removed from the analysis -- i.e., the combination is simply not new.

\item If $(J_{i},J_{j}) \notin B_{y}$, we examine whether $\underset{P_{A_{y}}\in A_{y}}{\Sigma} \{(J_{i},J_{j})\ \in A_{y}\} \ge 5$. If the last statement is \textsc{false}, we remove this pair from the analysis -- i.e., the new combination is not reused in the future.\footnote{As robustness checks, we also considered different thresholds for the re-use, i.e. 3 and 10. By construction, the number of combinations considered as novel increases (decreases) significantly when the threshold is lower (higher). However, as shown in \cite{wang_2017}, the dynamics of novelty are not much affected by these alternative specifications.}

\item If $(J_{i},J_{j}) \notin B_{y}$ \& $\underset{P_{A_{y}}\in A_{y}}{\Sigma} \{(J_{i},J_{j})\ \in A_{y}\} \ge 5$, then the journal pair combination is considered new and non trivial, hence we add that pair to the set of novel combinations $N_y$.

\end{itemize}

The difficulty of making new journal combinations are not equally distributed. Journals can share `common friends' making it possible to create more or less difficult new combinations. For example, $P_{iy}$ is making for the first time the combination $(J_1,J_2)$, but $J_1$ is usually cited with $J_3$ and $J_2$ is also sometimes cited with $J_3$. 
Creating this new combination is therefore less difficult compared to two journals that do not share any `common friends'. To investigate the difficulty of citing $J_1$ and $J_2$ for the first time, we construct a co-occurrence matrix of pairs of cited journals on the 3 years preceding the focal year $y$, and compute a cosine similarity:

 $$COS_{(J_1,J_2)}= \frac{J_1.J_2}{\left\| J_1\right\| \left\| J_2 \right\|}$$ 

The difficulty of making the $(J_1,J_2)$ combination is then $1-COS_{(J_1,J_2)}$.
To construct the novelty indicator for the article $P_{iy}$, we sum up all the difficulties for pairs $\in N_y$ and apply the $log(x+1)$ transformation:

$$Novelty(P_{iy}) = log \Big[ \underset{{(J_i,J_j)\in N_y}}{\Sigma} (1-COS_{(J_i,J_j)}) + 1\Big]$$

\newpage

\subsubsection{WoS subject categories combinations}
\label{appendix:mostusedcomb}

\begin{table}[h!]\footnotesize
 \centering  
  \caption{Subject categories combinations (All Sciences)}
  \label{tab:wos_freq1}
  \vspace*{1em}
	\begin{threeparttable}
\begin{tabular}{@{\extracolsep{5pt}}lrrrrrr} 
\toprule
Combinations [Category A $|$ Category B] & \# Combinations & Share [\%]\\
\midrule
\addlinespace
\textit{\underline{DL articles / 2001--2005}} & 450 \\
\addlinespace
NEUROSCIENCE \& BEHAVIOR $|$ NEUROSCIENCE \& BEHAVIOR & 51 & 11 \\
NEUROSCIENCE \& BEHAVIOR $|$ PSYCHIATRY/PSYCHOLOGY & 49 & 11 \\
CLINICAL MEDICINE $|$ NEUROSCIENCE \& BEHAVIOR & 22 & 5 \\
BIOLOGY \& BIOCHEMISTRY $|$ COMPUTER SCIENCE & 17 & 4 \\
COMPUTER SCIENCE $|$ NEUROSCIENCE \& BEHAVIOR & 14 & 3 \\

\addlinespace
\addlinespace
\textit{\underline{Non-DL articles / 2001--2005}}& 39,018 \\
\addlinespace
NEUROSCIENCE \& BEHAVIOR $|$ NEUROSCIENCE \& BEHAVIOR & 2,618 & 7 \\
CLINICAL MEDICINE $|$ NEUROSCIENCE \& BEHAVIOR & 2,378 & 6 \\
BIOLOGY \& BIOCHEMISTRY $|$ MOLECULAR BIOLOGY \& GENETICS & 2,369 & 6 \\
CLINICAL MEDICINE $|$ CLINICAL MEDICINE & 2,036 & 5 \\
MOLECULAR BIOLOGY \& GENETICS $|$ MOLECULAR BIOLOGY \& GENETICS & 1,927 & 5 \\
\addlinespace
\midrule
\addlinespace
\textit{\underline{DL articles / 2006--2010}} & 2,266 \\
\addlinespace
NEUROSCIENCE \& BEHAVIOR $|$ NEUROSCIENCE \& BEHAVIOR & 167 & 7 \\
NEUROSCIENCE \& BEHAVIOR $|$ PSYCHIATRY/PSYCHOLOGY & 150 & 7 \\
BIOLOGY \& BIOCHEMISTRY $|$ NEUROSCIENCE \& BEHAVIOR & 108 & 5 \\
NEUROSCIENCE \& BEHAVIOR $|$ PHYSICS & 86 & 4 \\
BIOLOGY \& BIOCHEMISTRY $|$ CHEMISTRY & 81 & 4 \\

\addlinespace
\addlinespace
\textit{\underline{Non-DL articles / 2006--2010}} & 118,363 \\
\addlinespace
CLINICAL MEDICINE $|$ CLINICAL MEDICINE & 6,164 & 5 \\
NEUROSCIENCE \& BEHAVIOR $|$ NEUROSCIENCE \& BEHAVIOR & 5,444 & 5 \\
BIOLOGY \& BIOCHEMISTRY $|$ MOLECULAR BIOLOGY \& GENETICS & 4,644 & 4 \\
NEUROSCIENCE \& BEHAVIOR $|$ PSYCHIATRY/PSYCHOLOGY & 4,547 & 4 \\
CLINICAL MEDICINE $|$ NEUROSCIENCE \& BEHAVIOR & 4,389 & 4 \\

\addlinespace
\midrule
\addlinespace
\textit{\underline{DL articles / 2011--2015}} & 3,986 \\
\addlinespace
NEUROSCIENCE \& BEHAVIOR $|$ PSYCHIATRY/PSYCHOLOGY & 302 & 8 \\
COMPUTER SCIENCE $|$ ENGINEERING & 249 & 6 \\
CLINICAL MEDICINE $|$ NEUROSCIENCE \& BEHAVIOR & 200 & 5 \\
PSYCHIATRY/PSYCHOLOGY $|$ PSYCHIATRY/PSYCHOLOGY & 188 & 5 \\
ENGINEERING $|$ ENGINEERING & 181 & 5 \\

\addlinespace
\addlinespace
\textit{\underline{Non-DL articles / 2011--2015}} & 328,197 \\
\addlinespace
CLINICAL MEDICINE $|$ CLINICAL MEDICINE & 29,295 & 9 \\
BIOLOGY \& BIOCHEMISTRY $|$ CLINICAL MEDICINE & 17,817 & 5 \\
CLINICAL MEDICINE $|$ MOLECULAR BIOLOGY \& GENETICS & 15,581 & 5 \\
PSYCHIATRY/PSYCHOLOGY $|$ PSYCHIATRY/PSYCHOLOGY & 13,583 & 4 \\
CLINICAL MEDICINE $|$ NEUROSCIENCE \& BEHAVIOR & 13,027 & 4 \\
\addlinespace
\bottomrule
\end{tabular}
\begin{tablenotes}
 \footnotesize
 \item {\it Notes:} This table reports the number and share of the most frequent combinations of WoS subject categories broken down by period and DL status.  
\end{tablenotes}
 \end{threeparttable}
\end{table}

\begin{table}[h!]\footnotesize
 \centering  
  \caption{Subject categories combinations (No CS)}
  \vspace*{1em}
	\begin{threeparttable}
\begin{tabular}{@{\extracolsep{5pt}}lrrrrrr} 

\toprule
Combinations [Category A $|$ Category B] & \# Combinations & Share [\%]\\
\midrule
\addlinespace
\textit{\underline{DL articles / 2001--2005}} & 375 \\
\addlinespace
NEUROSCIENCE \& BEHAVIOR $|$ NEUROSCIENCE \& BEHAVIOR & 51 & 14 \\
NEUROSCIENCE \& BEHAVIOR $|$ PSYCHIATRY/PSYCHOLOGY & 49 & 13 \\
CLINICAL MEDICINE $|$ NEUROSCIENCE \& BEHAVIOR & 22 & 6 \\
BIOLOGY \& BIOCHEMISTRY $|$ NEUROSCIENCE \& BEHAVIOR & 13 & 3 \\
BIOLOGY \& BIOCHEMISTRY $|$ BIOLOGY \& BIOCHEMISTRY & 12 & 3 \\

\addlinespace
\addlinespace
\textit{\underline{Non-DL articles / 2001--2005}} & 37,666 \\
\addlinespace
NEUROSCIENCE \& BEHAVIOR $|$ NEUROSCIENCE \& BEHAVIOR & 2,618 & 7 \\
CLINICAL MEDICINE $|$ NEUROSCIENCE \& BEHAVIOR & 2,378 & 6 \\
BIOLOGY \& BIOCHEMISTRY $|$ MOLECULAR BIOLOGY \& GENETICS & 2,369 & 6 \\
CLINICAL MEDICINE $|$ CLINICAL MEDICINE & 2,036 & 5 \\
MOLECULAR BIOLOGY \& GENETICS $|$ MOLECULAR BIOLOGY \& GENETICS & 1,927 & 5 \\

\addlinespace
\midrule
\addlinespace
\textit{\underline{DL articles / 2006--2010}} & 1,989 \\
\addlinespace
NEUROSCIENCE \& BEHAVIOR $|$ NEUROSCIENCE \& BEHAVIOR & 167 & 8 \\
NEUROSCIENCE \& BEHAVIOR $|$ PSYCHIATRY/PSYCHOLOGY & 150 & 8 \\
BIOLOGY \& BIOCHEMISTRY $|$ NEUROSCIENCE \& BEHAVIOR & 108 & 5 \\
NEUROSCIENCE \& BEHAVIOR $|$ PHYSICS & 86 & 4 \\
BIOLOGY \& BIOCHEMISTRY $|$ CHEMISTRY & 81 & 4 \\

\addlinespace
\addlinespace
\textit{\underline{Non-DL articles / 2006--2010}} & 114,806 \\
\addlinespace
CLINICAL MEDICINE $|$ CLINICAL MEDICINE & 6,164 & 5 \\
NEUROSCIENCE \& BEHAVIOR $|$ NEUROSCIENCE \& BEHAVIOR & 5,444 & 5 \\
BIOLOGY \& BIOCHEMISTRY $|$ MOLECULAR BIOLOGY \& GENETICS & 4,644 & 4 \\
NEUROSCIENCE \& BEHAVIOR $|$ PSYCHIATRY/PSYCHOLOGY & 4,547 & 4 \\
CLINICAL MEDICINE $|$ NEUROSCIENCE \& BEHAVIOR & 4,389 & 4 \\

\addlinespace
\midrule
\addlinespace
\textit{\underline{DL articles / 2011--2015}} & 3,188 \\
\addlinespace
NEUROSCIENCE \& BEHAVIOR $|$ PSYCHIATRY/PSYCHOLOGY & 302 & 9 \\
CLINICAL MEDICINE $|$ NEUROSCIENCE \& BEHAVIOR & 200 & 6 \\
PSYCHIATRY/PSYCHOLOGY $|$ PSYCHIATRY/PSYCHOLOGY & 188 & 6 \\
ENGINEERING $|$ ENGINEERING & 181 & 6 \\
NEUROSCIENCE \& BEHAVIOR $|$ NEUROSCIENCE \& BEHAVIOR & 154 & 5 \\

\addlinespace
\addlinespace
\textit{\underline{Non-DL articles / 2011--2015}} & 319,990 \\
\addlinespace
CLINICAL MEDICINE $|$ CLINICAL MEDICINE & 29,295 & 9 \\
BIOLOGY \& BIOCHEMISTRY $|$ CLINICAL MEDICINE & 17,817 & 6 \\
CLINICAL MEDICINE $|$ MOLECULAR BIOLOGY \& GENETICS & 15,581 & 5 \\
PSYCHIATRY/PSYCHOLOGY $|$ PSYCHIATRY/PSYCHOLOGY & 13,583 & 4 \\
CLINICAL MEDICINE $|$ NEUROSCIENCE \& BEHAVIOR & 13,027 & 4 \\
\addlinespace
\bottomrule
\end{tabular}
\begin{tablenotes}
 \footnotesize
 \item {\it Notes:} This table reports the number and share of the most frequent combinations of WoS subject categories broken down by period and DL status.   
\end{tablenotes}
 \end{threeparttable}
\end{table}

\begin{table}[h!]\footnotesize
 \centering  
  \caption{Subject categories combinations (Only HS)}
  \label{tab:wos_freq2}
  \vspace*{1em}
	\begin{threeparttable}
	\begin{tabular}{@{\extracolsep{5pt}}lrrrrrr} 

\toprule
Combinations [Category A $|$ Category B] & \# Combinations & Share [\%]\\
\midrule
\addlinespace
\textit{\underline{DL articles / 2001--2005}} & 251 \\
\addlinespace
NEUROSCIENCE \& BEHAVIOR / NEUROSCIENCE \& BEHAVIOR & 51 & 20 \\
NEUROSCIENCE \& BEHAVIOR / PSYCHIATRY/PSYCHOLOGY & 49 & 20 \\
CLINICAL MEDICINE / NEUROSCIENCE \& BEHAVIOR & 22 & 9 \\
BIOLOGY \& BIOCHEMISTRY / NEUROSCIENCE \& BEHAVIOR & 13 & 5 \\
BIOLOGY \& BIOCHEMISTRY / BIOLOGY \& BIOCHEMISTRY & 12 & 5 \\

\addlinespace
\addlinespace
\textit{\underline{Non-DL articles / 2001--2005}} & 31,917 \\
\addlinespace
NEUROSCIENCE \& BEHAVIOR / NEUROSCIENCE \& BEHAVIOR & 2,618 & 8 \\
CLINICAL MEDICINE / NEUROSCIENCE \& BEHAVIOR & 2,378 & 7 \\
BIOLOGY \& BIOCHEMISTRY / MOLECULAR BIOLOGY \& GENETICS & 2,369 & 7 \\
CLINICAL MEDICINE / CLINICAL MEDICINE & 2,036 & 6 \\
MOLECULAR BIOLOGY \& GENETICS / MOLECULAR BIOLOGY \& GENETICS & 1,927 & 6 \\

\addlinespace
\midrule
\addlinespace
\textit{\underline{DL articles / 2006--2010}} & 1,293 & \\
\addlinespace
NEUROSCIENCE \& BEHAVIOR / NEUROSCIENCE \& BEHAVIOR & 167 & 13 \\
NEUROSCIENCE \& BEHAVIOR / PSYCHIATRY/PSYCHOLOGY & 150 & 12 \\
BIOLOGY \& BIOCHEMISTRY / NEUROSCIENCE \& BEHAVIOR & 108 & 8 \\
BIOLOGY \& BIOCHEMISTRY / CHEMISTRY & 81 & 6 \\
CLINICAL MEDICINE / NEUROSCIENCE \& BEHAVIOR & 68 & 5 \\

\addlinespace
\addlinespace
\textit{\underline{Non-DL articles / 2006--2010}} & 85,342 \\
\addlinespace
CLINICAL MEDICINE / CLINICAL MEDICINE & 6,164 & 7 \\
NEUROSCIENCE \& BEHAVIOR / NEUROSCIENCE \& BEHAVIOR & 5,444 & 6 \\
BIOLOGY \& BIOCHEMISTRY / MOLECULAR BIOLOGY \& GENETICS & 4,644 & 5 \\
NEUROSCIENCE \& BEHAVIOR / PSYCHIATRY/PSYCHOLOGY & 4,547 & 5 \\
CLINICAL MEDICINE / NEUROSCIENCE \& BEHAVIOR & 4,389 & 5 \\

\addlinespace
\midrule
\addlinespace
\textit{\underline{DL articles / 2011--2015}} & 1,921 \\
\addlinespace
NEUROSCIENCE \& BEHAVIOR / PSYCHIATRY/PSYCHOLOGY & 302 & 16 \\
CLINICAL MEDICINE / NEUROSCIENCE \& BEHAVIOR & 200 & 10 \\
PSYCHIATRY/PSYCHOLOGY / PSYCHIATRY/PSYCHOLOGY & 188 & 10 \\
NEUROSCIENCE \& BEHAVIOR / NEUROSCIENCE \& BEHAVIOR & 154 & 8 \\
BIOLOGY \& BIOCHEMISTRY / NEUROSCIENCE \& BEHAVIOR & 109 & 6 \\

\addlinespace
\addlinespace
\textit{\underline{Non-DL articles / 2011--2015}} & 238,226 \\
\addlinespace
CLINICAL MEDICINE / CLINICAL MEDICINE & 29,293 & 12 \\
BIOLOGY \& BIOCHEMISTRY / CLINICAL MEDICINE & 17,817 & 7 \\
CLINICAL MEDICINE / MOLECULAR BIOLOGY \& GENETICS & 15,581 & 7 \\
PSYCHIATRY/PSYCHOLOGY / PSYCHIATRY/PSYCHOLOGY & 13,583 & 6 \\
CLINICAL MEDICINE / NEUROSCIENCE \& BEHAVIOR & 13,026 & 5 \\
\addlinespace
\bottomrule
\end{tabular}
\begin{tablenotes}
 \footnotesize
 \item {\it Notes:} This table reports the number and share of the most frequent combinations of WoS subject categories broken down by period and DL status.   
\end{tablenotes}
 \end{threeparttable}
\end{table}

\newpage
\clearpage

\section{Robustness analysis: descriptive statistics and results}
\label{appendix:robcheck}

This Appendix complements Section \ref{sec:robcheck} with descriptive statistics and estimation results for regressions and matching. Tables \ref{tab:descr_no_neuro}--\ref{tab:impact_no_neuro} refer to the sample of articles that are not classified as `Neurosciences'. Tables \ref{tab:descr_no_nn}--\ref{tab:impact_no_nn} refer to the sample of articles that do not contain the terms `neural\_network' and `neural\_networks' in their title, keywords or abstract. Table \ref{tab:ATTmatch} reports the results of the matching exercises. Table 23 reports the estimates for the Multinomial Logistic regression to model the novelty/conventionality quadrant \citep{uzzi_2013, wagner_2019}. Codes for the variable construction and analysis are fully accessible upon request.
\vspace*{2em}

\subsection{Neuroscience articles excluded}

\begin{table}[h!]\footnotesize
\centering  
\caption{Descriptive statistics of the variables -- Neuroscience articles excluded}
  \begin{threeparttable}
  \label{tab:descr_no_neuro} 
  \begin{tabular}{@{\extracolsep{5pt}} lccc} 
  \toprule
  & DL Papers & Non-DL Papers & Total \\  
  \midrule
  \addlinespace
  \textit{\underline{Re-combinatorial Novelty}} & & \\
  Novelty Dummy (All Sciences) & 38.17 & 32.48 & 32.54\\
  Novelty Dummy (No CS)    & 32.65 & 31.56 & 31.57\\
  Novelty Dummy (Only HS)     & 18.77 & 23.69 & 23.64\\
  \addlinespace
  Novelty (All Sciences) & 0/0.82 (2.16) & 0/0.84 (3.41) & 0/0.84 (3.4) \\ 
                                                    
  Novelty (No CS)     & 0/0.62 (1.82) & 0/0.82 (3.38) & 0/0.81 (3.37) \\ 
                                                    
  Novelty (Only HS)   & 0/0.29 (1.17) & 0/0.53 (2.58) & 0/0.53 (2.57) \\ 
  \addlinespace
  \textit{\underline{Scientific Impact}} & & \\
  Top 5\% Cited & 7.73 & 5.59 & 5.62 \\
  Top 10\% Cited & 14.61 & 11.02 & 11.06 \\
  \# Citations (Raw Count)  & 14/31.27 (140.01) & 17/31.73 (83.93) & 17/31.72 (84.69) \\ 
  Citations (Yearly Normalized) & 1.75/3.23 (8.52) & 2/3.48 (8.46) & 2/3.48 (8.46) \\ 
  \addlinespace
  \textit{\underline{Controls}} & & \\
  \# References & 32/37.84 (25.46) & 30/33.24 (23.26) & 30/33.28 (23.29) \\
  \# Authors & 4/4.03 (2.26) & 4/5.03 (3.61) & 4/5.01 (3.6) \\
  International Collab. & 23.65 & 21.95 & 21.97 \\
  Private Collab. & 6.37 & 7.56 & 7.55 \\
  JIF & 0.86/1.33 (1.26) & 1.63/1.98 (1.51) & 1.62/1.98 (1.51) \\
  Journal Age & 22/29.16 (28.88) & 35/41.08 (32.33) & 35/40.95 (32.32) \\
  Survey   & 0.89 & 0.98 & 0.98 \\
  
  \addlinespace
  \midrule
  Time Period & [2001 -- 2015] & [2001 -- 2015] & [2001 -- 2015] \\
  \# Scientific Fields & 41 & 43 & 43 \\
  \# Journals & 54 & 54 & 54 \\
  \# Papers & 2,355(1.03\%) & 225,748(98.97\%) & 228,103(100\%) \\
  \bottomrule
  \end{tabular}
  \begin{tablenotes}
  \footnotesize
  \item {\it Notes:} Binary indicators in [\%], for continuous measures [median/mean (s.d.)]. The statistics refer to the period used for the econometric analysis.
  \end{tablenotes}
 \end{threeparttable}
\end{table}

\begin{table}[t!]\footnotesize
\centering  
\caption{Novelty profile of deep learning publications -- Neuroscience articles excluded}
\vspace*{1em}
\begin{threeparttable}
\begin{tabular}{@{\extracolsep{5pt}}lcccccc} 
\toprule
& \multicolumn{3}{c}{\textit{Tobit: Novelty}}
& \multicolumn{3}{c}{\textit{Probit: Novelty Dummy}} \\ 
\cline{2-4}\cline{5-7} 

\\[-1.8ex] & \multicolumn{1}{c}{All Sciences}  &\multicolumn{1}{c}{No CS} &\multicolumn{1}{c}{Only HS} & \multicolumn{1}{c}{All Sciences}  &\multicolumn{1}{c}{No CS}&\multicolumn{1}{c}{Only HS} \\ 
\\[-1.8ex] & (1) & (2) & (3)& (4) & (5) & (6)\\ 
\midrule 
DL & 0.030  & -0.065  & -0.225$^{***}$ & 0.046  & -0.035  & -0.181$^{***}$ \\
   & (0.052) & (0.049) & (0.066) & (0.052) & (0.049) & (0.059) \\
   &    &    &    &    &    &    \\
\# References (log) & 1.100$^{***}$ & 1.104$^{***}$ & 1.076$^{***}$ & 0.948$^{***}$ & 0.949$^{***}$ & 0.894$^{***}$ \\
   & (0.037) & (0.036) & (0.034) & (0.033) & (0.032) & (0.025) \\
   &    &    &    &    &    &    \\
\# Authors (log) & 0.124$^{***}$ & 0.129$^{***}$ & 0.167$^{***}$ & 0.131$^{***}$ & 0.135$^{***}$ & 0.166$^{***}$ \\
   & (0.021) & (0.021) & (0.026) & (0.020) & (0.020) & (0.022) \\
   &    &    &    &    &    &    \\
International Collab. & -0.036$^{***}$ & -0.041$^{***}$ & -0.074$^{***}$ & -0.035$^{***}$ & -0.039$^{***}$ & -0.068$^{***}$ \\
   & (0.012) & (0.013) & (0.013) & (0.013) & (0.013) & (0.013) \\
   &    &    &    &    &    &    \\
Private Collab. & 0.017  & 0.017  & -0.008  & 0.016  & 0.016  & -0.006  \\
   & (0.013) & (0.014) & (0.018) & (0.014) & (0.014) & (0.017) \\
   &    &    &    &    &    &    \\
JIF & 0.022  & 0.025  & 0.039  & 0.023  & 0.026  & 0.036  \\
   & (0.07) & (0.068) & (0.073) & (0.067) & (0.065) & (0.067) \\
   &    &    &    &    &    &    \\
Journal Age (log) & 0.007  & 0.031  & 0.059  & 0.015  & 0.033  & 0.057  \\
   & (0.145) & (0.143) & (0.160) & (0.135) & (0.134) & (0.144) \\
   &    &    &    &    &    &    \\
Survey & 0.126$^{***}$ & 0.119$^{***}$ & 0.077$^{*}$ & 0.123$^{***}$ & 0.115$^{***}$ & 0.074$^{*}$ \\
   & (0.041) & (0.038) & (0.042) & (0.045) & (0.041) & (0.041) \\
   &    &    &    &    &    &    \\
\midrule
Log Likelihood & 
-172,590  &  -168,967  &  -139,119  &  -115,102  &  -113,562  &  -100,187 \\
$\chi^2$ [Null Model] &  74,312$^{***}$  &  73,797$^{***}$  &  59,245$^{***}$  &  57,591$^{***}$  &  57360$^{***}$  &  49,127$^{***}$ \\
$\chi^2$ [w/o DL Model]  &  1.30   &  5.30$^{**}$  &  44.60$^{***}$  &  2.60   &  1.40   &  31.1$^{***}$ \\
\# Obs   &  228,103  &  228,103  &  228,103  &  228,103  &  228,103  &  228,103 \\
\bottomrule 
\end{tabular} 
\begin{tablenotes}
\footnotesize
\item {\it Notes:} This table reports coefficients of the effect of deep learning (\textit{DL}, dummy) on re-combinatorial novelty built by considering different knowledge landscapes. Bootstrapped (500 replications) standard errors clustered at the journal-level in parentheses: ***, ** and * indicate significance at the 1\%, 5\% and 10\% level, respectively. The effect of DL on the positive continuous novelty measure is estimated using a Tobit regression (Columns 1--3). The effect on the novelty dummy is estimated using a Probit (Columns 4--6). Each novelty measure is calculated on three different sets of journal references: `All Sciences' -- All cited journals, `No CS' -- All cited journals except for computer science journals, and `Only HS' -- Only citations to health science journals. Constant term, scientific field (WoS subject category) and time fixed effects are incorporated in all model specifications. Likelihood-ratio test are used to compare the goodness of fit of two statistical models: (i) null model against complete model; (ii) model without the $DL$ variable against the complete model.
\end{tablenotes}
\end{threeparttable}
\end{table}

\begin{table}[h]\footnotesize
 \centering
  \caption{Impact profile of deep learning publications -- Neuroscience articles excluded}
  \vspace*{1em}
    \scalebox{0.88}{
    \label{tab:impact_no_neuro}
  \vspace*{1em}
	\begin{threeparttable}
\begin{tabular}{@{\extracolsep{5pt}}llccc}
\toprule
\\[-1.8ex] & & \multicolumn{1}{c}{NegBin: \# Citations}&\multicolumn{1}{c}{Probit: Top 5\% Cited} &\multicolumn{1}{c}{Probit: Top 10\% Cited}\\
\\[-1.8ex]&  & (1) & (2) & (3)\\
\midrule

\textit{Panel A: Mean}
  & DL & 0.090  & 0.107$^{*}$ & 0.120$^{**}$ \\
   &   & (0.060) & (0.059) & (0.059) \\
   &    &    &    &    \\
   & Novelty (All Sciences) & 0.165$^{***}$ & 0.210$^{***}$ & 0.194$^{***}$ \\
   &   & (0.028) & (0.022) & (0.022) \\
   &    &    &    &    \\
   & \# References (log) & 0.470$^{***}$ & 0.367$^{***}$ & 0.416$^{***}$ \\
   &   & (0.062) & (0.103) & (0.086) \\
   &    &    &    &    \\
   & \# Authors (log) & 0.211 $^{***}$ & 0.154$^{***}$ & 0.172$^{***}$ \\
   &   & (0.032) & (0.054) & (0.050) \\
   &    &    &    &    \\
   & International Collab. & 0.068$^{***}$ & 0.093$^{***}$ & 0.089$^{***}$ \\
   &   & (0.014) & (0.016) & (0.017) \\
   &    &    &    &    \\
   & Private Collab. & -0.011  & -0.009  & -0.007  \\
   &   & (0.016) & (0.021) & (0.016) \\
   &    &    &    &    \\
   & JIF & 0.222$^{***}$ & 0.202$^{***}$ & 0.192$^{***}$ \\
   &   & (0.035) & (0.062) & (0.065) \\
   &    &    &    &    \\
   & Journal Age (log) & 0.078$^{*}$ & 0.025  & 0.045  \\
   &   & (0.044) & (0.103) & (0.111) \\
   &    &    &    &    \\
   & Survey & 0.551$^{***}$ & 0.693$^{***}$ & 0.630$^{***}$ \\
   &   & (0.050) & (0.070) & (0.060) \\
   &    &    &    &    \\
\midrule
                                                      \textit{Panel B: Dispersion}
   & DL & 0.164$^{**}$ &   &   \\
   &   & (0.075) & &\\
   &    &    &    &    \\
   & Novelty (All Sciences) & 0.097$^{***}$ &   &   \\
   &   & (0.017) & &\\
   &    &    &    &    \\
   & \# References (log) & -0.473$^{***}$ &   &   \\
   &   & (0.040) & &\\
   &    &    &    &    \\
   & \# Authors (log) & -0.199$^{***}$ &   &   \\
   &   & (0.036) & &\\
   &    &    &    &    \\
   & JIF & 0.107$^{**}$ &   &   \\
   &   & (0.054) & &\\
   &    &    &    &    \\
   & Journal Age (log) & -0.123$^{***}$ &   &   \\
   &   & (0.033) & &\\
   &    &    &    &    \\
\midrule
Log Likelihood & &
-955,206  &  -45,382  &  -72,968 \\
$\chi^2$ [Null Model] & & 193,546$^{***}$  &  7,890$^{***}$  &  12,715$^{***}$ \\
$\chi^2$ [w/o DL Model] & & 2.10   &  6.80$^{**}$  &  12.60$^{***}$ \\
\# Obs   & & 228,103  &  228,103  &  228,103 \\
\bottomrule
\end{tabular}

\begin{tablenotes}
 \footnotesize
 \item {\it Notes:} This table reports coefficients of the effect of deep learning (\textit{DL}, dummy) on scientific impact proxied by the number of received citations (Column 1) and `big hits' (Columns 2 and 3). Bootstrapped (500 replications) standard errors clustered at the journal-level in parentheses: ***, ** and * indicate significance at the 1\%, 5\% and 10\% level, respectively. The effect of DL on the citation count is estimated using a Negative Binomial regression. Estimates for the expectation and variance are reported in Panel A and B, respectively. The effects on the binary indicators is estimated using a Probit. Constant term, scientific field (WoS subject category) and time fixed effects are incorporated in all model specifications. Likelihood-ratio test are used to compare the goodness of fit of two statistical models: (i) null model against complete model; (ii) model without the $DL$ variable against the complete model.
\end{tablenotes}
 \end{threeparttable}
 }
\end{table}

\newpage
\clearpage

\subsection{Neural network(s) articles excluded}

\begin{table}[h!]\footnotesize
\centering  
\caption{Descriptive statistics of the variables -- Neural network(s) articles excluded}
  \begin{threeparttable}
  \label{tab:descr_no_nn} 
  \begin{tabular}{@{\extracolsep{5pt}} lccc} 
  \toprule
  & DL Papers & Non-DL Papers & Total \\  
  \midrule
  \addlinespace
  \textit{\underline{Re-combinatorial Novelty}} & & \\
  Novelty Dummy (All Sciences) & 37.97 & 30.05 & 30.08\\
  Novelty Dummy (No CS)    & 32.64 & 29.22 & 29.23\\
  Novelty Dummy (Only HS)     & 18.57 & 22.72 & 22.71\\
  \addlinespace
  Novelty (All Sciences) & 0/0.78 (1.92) & 0/0.74 (3.17) & 0/0.74 (3.16) \\ 
                                                    
  Novelty (No CS)     & 0/0.61 (1.67) & 0/0.72 (3.14) & 0/0.72 (3.13) \\ 
                                                    
  Novelty (Only HS)   & 0/0.26 (0.87) & 0/0.49 (2.43) & 0/0.49 (2.42) \\ 
  \addlinespace
  \textit{\underline{Scientific Impact}} & & \\
  Top 5\% Cited & 7.33 & 6.00 & 6.00 \\
  Top 10\% Cited & 14.15 & 11.7 & 11.71 \\
  \# Citations (Raw Count)  & 15/27.88 (41.87) & 17/34.95 (84.48) & 17/34.93 (84.36) \\ 
  Citations (Yearly Normalized) & 1.78/3.21 (4.95) & 2/3.73 (8.23) & 2/3.73 (8.22) \\ 
  \addlinespace
  \textit{\underline{Controls}} & & \\
  \# References & 32/36.56 (22.63) & 31/36.13 (25.60) & 31/36.14 (25.59) \\
  \# Authors & 4/4.15 (2.16) & 4/4.75 (3.28) & 4/4.75 (3.27) \\
  International Collab. & 23.06 & 22.45 & 22.45 \\
  Private Collab. & 7.58 & 6.91 & 6.92 \\
  JIF & 0.96/1.3 (1.27) & 1.57/2.37 (2.20) & 1.57/2.37 (2.20) \\
  Journal Age & 23/29.58 (28.77) & 31/37.38 (29.04) & 31/37.35 (29.04) \\
  Survey   & 1.17 & 0.83 & 0.83 \\
  
  \addlinespace
  \midrule
  Time Period & [2001 -- 2015] & [2001 -- 2015] & [2001 -- 2015] \\
  \# Scientific Fields & 45 & 48 & 48 \\
  \# Journals & 84 & 84 & 84 \\
  \# Papers & 1,201(0.37\%) & 319,755(99.63\%) & 320,956(100\%) \\
  \bottomrule
  \end{tabular}
  \begin{tablenotes}
  \footnotesize
  \item {\it Notes:} Binary indicators in [\%], for continuous measures [median/mean (s.d.)]. The statistics refer to the period used for the econometric analysis.
  \end{tablenotes}
 \end{threeparttable}
\end{table}

\begin{table}[t!]\footnotesize
\centering  
\caption{Novelty profile of deep learning publications -- Neural network(s) articles excluded}
\vspace*{1em}
\begin{threeparttable}
\begin{tabular}{@{\extracolsep{5pt}}lcccccc} 
\toprule
& \multicolumn{3}{c}{\textit{Tobit: Novelty}}
& \multicolumn{3}{c}{\textit{Probit: Novelty Dummy}} \\ 
\cline{2-4}\cline{5-7} 

\\[-1.8ex] & \multicolumn{1}{c}{All Sciences}  &\multicolumn{1}{c}{No CS} &\multicolumn{1}{c}{Only HS} & \multicolumn{1}{c}{All Sciences}  &\multicolumn{1}{c}{No CS}&\multicolumn{1}{c}{Only HS} \\ 
\\[-1.8ex] & (1) & (2) & (3)& (4) & (5) & (6)\\ 
\midrule 
DL & 0.083  & 0.003  & -0.171$^{***}$ & 0.091$^{*}$ & 0.014  & -0.137$^{**}$ \\
   & (0.051) & (0.052) & (0.061) & (0.053) & (0.057) & (0.058) \\
   &    &    &    &    &    &    \\
\# References (log) & 1.046$^{***}$ & 1.050$^{***}$ & 1.025$^{***}$ & 0.880$^{***}$ & 0.880$^{***}$ & 0.838$^{***}$ \\
   & (0.032) & (0.032) & (0.033) & (0.026) & (0.026) & (0.023) \\
   &    &    &    &    &    &    \\
\# Authors (log) & 0.186$^{***}$ & 0.194$^{***}$ & 0.241$^{***}$ & 0.191$^{***}$ & 0.197$^{***}$ & 0.233$^{***}$ \\
   & (0.023) & (0.024) & (0.027) & (0.022) & (0.022) & (0.024) \\
   &    &    &    &    &    &    \\
International Collab. & -0.058$^{***}$ & -0.064$^{***}$ & -0.095$^{***}$ & -0.055$^{***}$ & -0.061$^{***}$ & -0.086$^{***}$ \\
   & (0.010) & (0.010) & (0.010) & (0.010) & (0.010) & (0.009) \\
   &    &    &    &    &    &    \\
Private Collab. & 0.001  & 0.001  & -0.023  & 0.001  & -0.001  & -0.021  \\
   & (0.012) & (0.013) & (0.015) & (0.012) & (0.012) & (0.014) \\
   &    &    &    &    &    &    \\
JIF & -0.040$^{**}$ & -0.037$^{*}$ & -0.029  & -0.037$^{**}$ & -0.034$^{*}$ & -0.026  \\
   & (0.020) & (0.021) & (0.021) & (0.018) & (0.018) & (0.018) \\
   &    &    &    &    &    &    \\
Journal Age (log) & -0.092  & -0.077  & -0.040  & -0.069  & -0.056  & -0.026  \\
   & (0.103) & (0.106) & (0.115) & (0.094) & (0.096) & (0.101) \\
   &    &    &    &    &    &    \\
Survey & 0.204$^{***}$ & 0.195$^{***}$ & 0.160$^{***}$ & 0.192$^{***}$ & 0.184$^{***}$ & 0.146$^{***}$ \\
   & (0.042) & (0.040) & (0.043) & (0.045) & (0.042) & (0.042) \\
   &    &    &    &    &    &    \\
\midrule
Log Likelihood & 
-234,600  &  -230,021  &  -194,470  &  -160,685  &  -158,739  &  -142,454 \\
$\chi^2$ [Null Model] &  90,036$^{***}$  &  88,839$^{***}$  &  70,498$^{***}$  &  71,192$^{***}$  &  70,357$^{***}$  &  58,980$^{***}$ \\
$\chi^2$ [w/o DL Model]  &  4.70$^{*}$  &  0.02   &  12.80$^{***}$  &  5.30$^{**}$  &  0.10   &  9.4$^{***}$ \\
\# Obs   &  320,956  &  320,956  &  320,956  &  320,956  &  320,956  &  320,956 \\
\bottomrule 
\end{tabular} 
\begin{tablenotes}
\footnotesize
\item {\it Notes:} This table reports coefficients of the effect of deep learning (\textit{DL}, dummy) on re-combinatorial novelty built by considering different knowledge landscapes. Bootstrapped (500 replications) standard errors clustered at the journal-level in parentheses: ***, ** and * indicate significance at the 1\%, 5\% and 10\% level, respectively. The effect of DL on the positive continuous novelty measure is estimated using a Tobit regression (Columns 1--3). The effect on the novelty dummy is estimated using a Probit (Columns 4--6). Each novelty measure is calculated on three different sets of journal references: `All Sciences' -- All cited journals, `No CS' -- All cited journals except for computer science journals, and `Only HS' -- Only citations to health science journals. Constant term, scientific field (WoS subject category) and time fixed effects are incorporated in all model specifications. Likelihood-ratio test are used to compare the goodness of fit of two statistical models: (i) null model against complete model; (ii) model without the $DL$ variable against the complete model.
\end{tablenotes}
\end{threeparttable}
\end{table}

\begin{table}[t!]\footnotesize
 \centering
  \caption{Impact profile of deep learning publications -- Neural network(s) articles excluded}
  \vspace*{1em}
    \scalebox{0.88}{
    \label{tab:impact_no_nn}
  \vspace*{1em}
	\begin{threeparttable}
\begin{tabular}{@{\extracolsep{5pt}}llccc}
\toprule
\\[-1.8ex] & & \multicolumn{1}{c}{NegBin: \# Citations}&\multicolumn{1}{c}{Probit: Top 5\% Cited} &\multicolumn{1}{c}{Probit: Top 10\% Cited}\\
\\[-1.8ex]&  & (1) & (2) & (3)\\
\midrule

\textit{Panel A: Mean}
  & DL & 0.110$^{*}$  & 0.136$^{*}$ & 0.153$^{**}$ \\
   &   & (0.067) & (0.070) & (0.064) \\
   &    &    &    &    \\
   & Novelty (All Sciences) & 0.138$^{***}$ & 0.190$^{***}$ & 0.181$^{***}$ \\
   &   & (0.022) & (0.017) & (0.016) \\
   &    &    &    &    \\
   & \# References (log) & 0.517$^{***}$ & 0.436$^{***}$ & 0.485$^{***}$ \\
   &   & (0.061) & (0.075) & (0.063) \\
   &    &    &    &    \\
   & \# Authors (log) & 0.248$^{***}$ & 0.179$^{***}$ & 0.206$^{***}$ \\
   &   & (0.031) & (0.040) & (0.038) \\
   &    &    &    &    \\
   & International Collab. & 0.070$^{***}$ & 0.088$^{***}$ & 0.090$^{***}$ \\
   &   & (0.014) & (0.015) & (0.014) \\
   &    &    &    &    \\
   & Private Collab. & -0.034$^{**}$ & -0.031  & -0.04$^{**}$ \\
   &   & (0.017) & (0.019) & (0.016) \\
   &    &    &    &    \\
   & JIF & 0.202$^{***}$ & 0.155$^{***}$ & 0.168$^{***}$ \\
   &   & (0.022) & (0.018) & (0.019) \\
   &    &    &    &    \\
   & Journal Age (log) & 0.063$^{*}$ & -0.043  & -0.032  \\
   &   & (0.038) & (0.093) & (0.089) \\
   &    &    &    &    \\
   & Survey & 0.522$^{***}$ & 0.646$^{***}$ & 0.607$^{***}$ \\
   &   & (0.055) & (0.056) & (0.051) \\
   &    &    &    &    \\
\midrule
                                                      \textit{Panel B: Dispersion}
   & DL & 0.075  &   &   \\
   &   & (0.053) & &\\
   &    &    &    &    \\
   & Novelty (All Sciences) & 0.086$^{***}$ &   &   \\
   &   & (0.017) & &\\
   &    &    &    &    \\
   & \# References (log) & -0.488$^{***}$ &   &   \\
   &   & (0.039) & &\\
   &    &    &    &    \\
   & \# Authors (log) & -0.202$^{***}$ &   &   \\
   &   & (0.043) & &\\
   &    &    &    &    \\
   & JIF & 0.037  &   &   \\
   &   & (0.03) & &\\
   &    &    &    &    \\
   & Journal Age (log) & -0.116$^{***}$ &   &   \\
   &   & (0.032) & &\\
   &    &    &    &    \\
\midrule
Log Likelihood & &
-1,360,967  &  -63,884  &  -101,311 \\
$\chi^2$ [Null Model] & & 282,883$^{***}$  &  17,961$^{***}$  &  29,217$^{***}$ \\
$\chi^2$ [w/o DL Model]  & & 1.60   &  5.50$^{**}$  &  10.40$^{***}$ \\
\# Obs   & & 320,956  &  320,956  &  320,956 \\
\bottomrule
\end{tabular}

\begin{tablenotes}
 \footnotesize
 \footnotesize
 \item {\it Notes:} This table reports coefficients of the effect of deep learning (\textit{DL}, dummy) on scientific impact proxied by the number of received citations (Column 1) and `big hits' (Columns 2 and 3). Bootstrapped (500 replications) standard errors clustered at the journal-level in parentheses: ***, ** and * indicate significance at the 1\%, 5\% and 10\% level, respectively. The effect of DL on the citation count is estimated using a Negative Binomial regression. Estimates for the expectation and variance are reported in Panel A and B, respectively. The effects on the binary indicators is estimated using a Probit. Constant term, scientific field (WoS subject category) and time fixed effects are incorporated in all model specifications. Likelihood-ratio test are used to compare the goodness of fit of two statistical models: (i) null model against complete model; (ii) model without the $DL$ variable against the complete model.
\end{tablenotes}
 \end{threeparttable}
 }
\end{table}

\newpage

\begin{table}[t!]\footnotesize
\centering  
  \begin{threeparttable}
\caption{Novelty and impact profile -- Matching}
\vspace{2em}
\label{tab:ATTmatch}
\begin{tabular}{@{\extracolsep{10pt}}lcccc} 
\toprule
  \addlinespace
& \multicolumn{2}{c}{\textit{Exact Matching}}
& \multicolumn{2}{c}{\textit{Propensity Score Matching}} \\ 
\cline{2-3}\cline{4-5} 
  \addlinespace
 & (1) & (2) & (3)& (4)\\ 
\midrule
\addlinespace
Novelty (All Sciences) & 0.054$^{***}$ & 0.053$^{***}$ & 0.035$^{***}$ & 0.023 \\
Novelty (No CS) & 0.026$^{**}$ & 0.026$^{**}$ & 0.008  & -0.001 \\
Novelty (Only HS) & -0.005  & -0.005  & -0.025$^{**}$ & -0.033$^{***}$\\
\# Citations & 0.192$^{***}$ & 0.195$^{***}$ & 0.102$^{***}$ & 0.063$^{**}$\\
 \addlinespace
\bottomrule
\end{tabular} 
\begin{tablenotes}
  \footnotesize
  \item {\it Notes:} This table reports Average Treatment Effect on the Treated (ATT) for novelty and impact variables. ***, ** and * indicate significance at the 1\%, 5\% and 10\% level, respectively. The set of variables used for each matching is composed as follows:
  (1) Journal / WoS Categories / Publication Year;
  (2) All dummy variables in our set of control variables / Journal / WoS Categories / Publication Year; 
  (3) Number of authors (log) / Number of References (log) /  Journal / WoS Categories / Publication Year;
  (4) All Variables.
  \end{tablenotes}
    \end{threeparttable}
\end{table}

\newpage
\clearpage

\subsection{Atypical combinations in deep learning publications}

\begin{longtable}[h!]{lcccc}
\caption{Atypical profile of deep learning publications} \\
\toprule
& \multicolumn{1}{c}{Category} & \multicolumn{1}{c}{All Sciences} & \multicolumn{1}{c}{No CS} &\multicolumn{1}{c}{Only HS} \\
& & (1) & (2) & (3)\\
\endfirsthead 
\caption{Atypical profile of deep learning publications -- continued.}\\
\midrule
 & \multicolumn{1}{c}{Category} & \multicolumn{1}{c}{All Sciences}  &\multicolumn{1}{c}{No CS} &\multicolumn{1}{c}{Only HS}  \\
& & (1) & (2) & (3)\\
\hline
\endhead 
\multicolumn{5}{r}{Continued on next page}\\
\endfoot
\hline
\multicolumn{5}{r}{End of table.} \\
\endlastfoot
\midrule
DL & HC--HN & 0.008  & 0.208  & 0.308$^{**}$ \\
   & &(0.130) & (0.133) & (0.136) \\
   & &   &    &    \\
 & HC--LN &-0.041  & 0.090  & -0.049  \\
   & &(0.157) & (0.152) & (0.154) \\
   & &   &    &    \\
 & LC--LN &-0.043  & -0.086  & 0.021  \\
   & &(0.162) & (0.163) & (0.155) \\
   & &   &    &    \\
\# References (log) & HC--HN & -0.198$^{***}$ & -0.216$^{***}$ & -0.168$^{***}$ \\
   & &(0.066) & (0.065) & (0.061) \\
   & &   &    &    \\
 & HC--LN &-0.687$^{***}$ & -0.668$^{***}$ & -0.711$^{***}$ \\
   & &(0.066) & (0.064) & (0.063) \\
   & &   &    &    \\
 & LC--LN &-0.460$^{***}$ & -0.463$^{***}$ & -0.550$^{***}$ \\
   & &(0.063) & (0.060) & (0.062) \\
   & &   &    &    \\
\# Authors (log) & HC--HN &-0.392$^{***}$ & -0.393$^{***}$ & -0.433$^{***}$ \\
   & &(0.060) & (0.060) & (0.066) \\
   & &   &    &    \\
 & HC--LN & -0.557$^{***}$ & -0.597$^{***}$ & -0.603$^{***}$ \\
   & &(0.078) & (0.077) & (0.086) \\
   & &   &    &    \\
 & LC--LN &-0.260$^{***}$ & -0.254$^{***}$ & -0.299$^{***}$ \\
   & &(0.048) & (0.047) & (0.050) \\
   & &   &    &    \\
International Collab. & HC--HN &0.103$^{**}$ & 0.160$^{***}$ & 0.128$^{***}$ \\
   & &(0.042) & (0.043) & (0.044) \\
   & &   &    &    \\
 & HC--LN & 0.096$^{**}$ & 0.155$^{***}$ & 0.141$^{***}$ \\
   & &(0.041) & (0.040) & (0.043) \\
   & &   &    &    \\
 & LC--LN &-0.013  & 0.052  & 0.119$^{***}$ \\
   & &(0.047) & (0.044) & (0.044) \\
   & &   &    &    \\
Private Collab. & HC--HN & -0.050  & -0.067  & 0.045  \\
   & &(0.069) & (0.071) & (0.072) \\
   & &   &    &    \\
 & HC--LN & 0.010  & -0.108$^{*}$ & -0.093  \\
  & & (0.063) & (0.060) & (0.062) \\
   & &   &    &    \\
 & LC--LN & 0.052  & -0.016  & 0.025  \\
   & & (0.068) & (0.069) & (0.071) \\
   & &   &    &    \\
JIF & HC--HN & 0.134$^{***}$ & 0.145$^{***}$ & 0.146$^{***}$ \\
   & & (0.035) & (0.035) & (0.038) \\
   & &   &    &    \\
 & HC--LN & 0.117$^{***}$ & 0.105$^{***}$ & 0.092$^{***}$ \\
   & & (0.032) & (0.033) & (0.035) \\
   & &   &    &    \\
 & LC--LN & -0.087  & -0.114$^{*}$ & -0.116  \\
   & &(0.062) & (0.062) & (0.075) \\
   & &   &    &    \\
  Journal Age (log) & HC--HN &-0.068  & -0.064  & -0.050  \\
   & &(0.196) & (0.189) & (0.194) \\
   & &   &    &    \\
 & HC--LN &-0.207  & -0.158  & -0.178  \\
   & & (0.173) & (0.168) & (0.176) \\
   & &   &    &    \\
 & LC--LN & -0.055  & -0.089  & -0.224  \\
   & & (0.241) & (0.24) & (0.258) \\
   & &   &    &    \\
Survey & HC--HN & -0.399  & -0.294  & -0.492  \\
   & &(0.339) & (0.348) & (0.328) \\
   & &   &    &    \\
 & HC--LN & 0.458$^{**}$ & 0.096  & 0.472$^{**}$ \\
   & &(0.225) & (0.209) & (0.204) \\
   & &   &    &    \\
 & LC--LN& 0.892$^{***}$ & 0.592$^{***}$ & 0.779$^{***}$ \\
   & &(0.224) & (0.211) & (0.211) \\
\midrule
Log Likelihood & &
-374,002  &  -374,000  &  -363,855 \\
$\chi^2$ [Null Model] & & 95,913$^{***}$  &  95,488$^{***}$  &  115,891$^{***}$ \\
$\chi^2$ [w/o DL Model] & &  259$^{***}$  &  158.20$^{***}$  &  144$^{***}$ \\
\# Obs  & &  320,587  &  320,587  &  320,587 \\
\bottomrule
\end{longtable}

\vspace*{-2em}
\begin{figure}[h!]
  \begin{center}
    \includegraphics[scale=0.20]{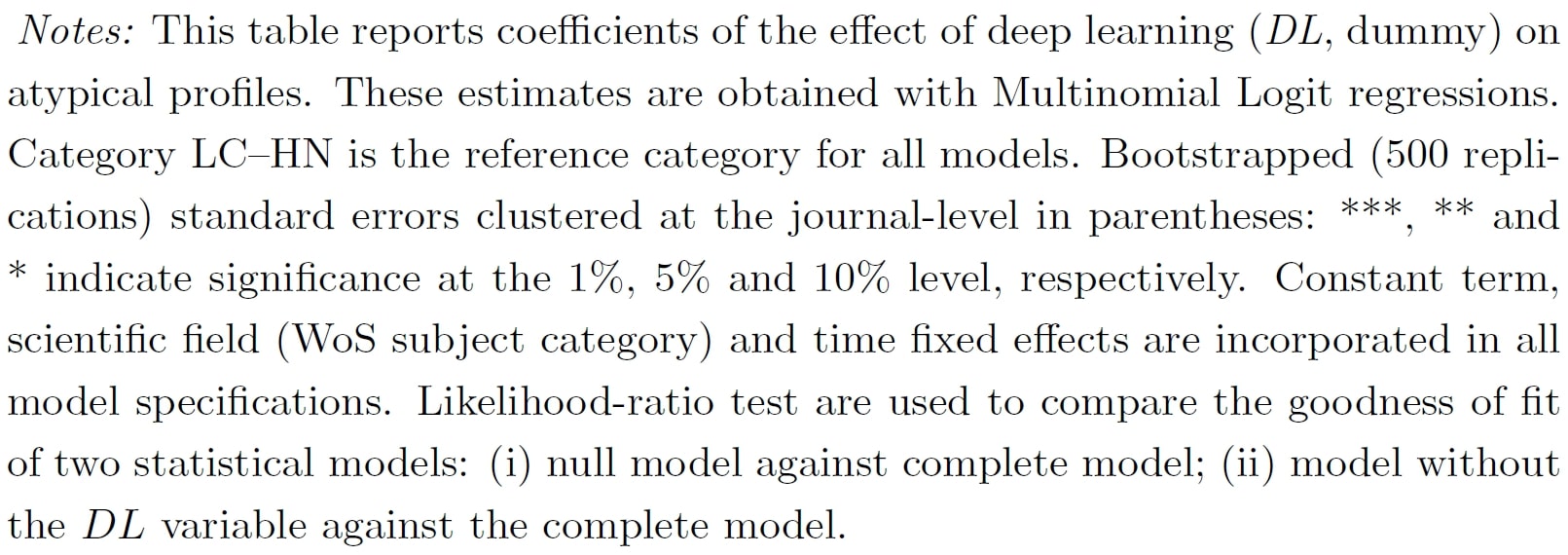}
  \end{center} 
\end{figure}

\end{document}